\def\DocumentTarget{Arxiv}

\providecommand*{\GlobalClassOptions}{}
\InputIfFileExists{preamble-local.tex}{}{}

\documentclass[\GlobalClassOptions]{llncs}
\input{preamble.tex}

\title{Aesthetic Discrimination of Graph Layouts}
\author{
  Moritz Klammler\inst{1} \and
  Tamara Mchedlidze\inst{1}\and
  Alexey Pak\inst{2}}

\institute{
  Karlsruhe Institute of Technology, 76131 Karlsruhe, Germany \and
  Fraunhofer Institute of Optronics, System Technologies and Image Exploitation,
  Fraunhoferstra{\ss}e 1, 76131 Karlsruhe, Germany \\
  \email{moritz@klammler.eu},
  \email{mched@iti.uka.de},
  \email{alexey.pak@iosb.fraunhofer.de}
}

\AtBeginDocument{

\def\NNSharedInputDims{57}
\def\NNSharedHiddenDims{15}
\def\NNSharedOutputDims{11}
\def\NNTotalAuxInputDims{2}
\def\NNTotalAuxHiddenDims{2}
\def\NNTotalCatDims{13}
\def\NNTotalOutputDims{1}
\def\NNTotalTrainableParams{1\,066}

\def\NNCorpusSizeApprox{36\,000}

\def\XValTestRuns{10}
\def\XValCountApprox{7415}
\def\XValCountMean{7414.90}
\def\XValCountStdev{620.09}
\def\XValSuccessMean{96.48}
\def\XValSuccessStdev{0.85}
\def\XValFailureMean{3.52}
\def\XValFailureStdev{0.85}
\def\XValCondPosMean{49.73}
\def\XValCondPosStdev{0.61}
\def\XValCondNegMean{50.27}
\def\XValCondNegStdev{0.61}
\def\XValPredPosMean{50.36}
\def\XValPredPosStdev{2.38}
\def\XValPredNegMean{49.64}
\def\XValPredNegStdev{2.38}
\def\XValTrueNegMean{48.20}
\def\XValTrueNegStdev{1.43}
\def\XValFalseNegMean{1.44}
\def\XValFalseNegStdev{1.15}
\def\XValTruePosMean{48.28}
\def\XValTruePosStdev{1.18}
\def\XValFalsePosMean{2.08}
\def\XValFalsePosStdev{1.43}

\HuangWeightResult[EL]{+0.4803}{0.0855}
\HuangWeightResult[CC]{+0.4679}{0.1069}
\HuangWeightResult[CR]{-0.0431}{0.0315}
\HuangWeightResult[AR]{-0.0087}{0.0072}

}

\begin{document}

\maketitle

\begin{abstract}
  This paper addresses the following basic question: given two layouts of the same graph, which one is more aesthetically pleasing? We propose a neural network-based discriminator model trained on a labeled dataset that decides which of two layouts has a higher aesthetic quality. The feature vectors used as inputs to the model are based on known graph drawing quality metrics, classical statistics, information-theoretical quantities, and two-point statistics inspired by methods of condensed matter physics. The large corpus of layout pairs used for training and testing is constructed using force-directed drawing algorithms and the layouts that naturally stem from the process of graph generation. It is further extended using data augmentation techniques. Our model demonstrates a mean prediction accuracy of $\XValSuccessMean\percent$, outperforming discriminators based on stress and on the linear combination of popular quality metrics by a small but statistically significant margin.

This paper appears in the Proceedings of the 26\textsuperscript{th} International Symposium on Graph Drawing and Network Visualization (GD~2018).

\end{abstract}


\section{Introduction}
\label{sec:intro}

What makes a drawing of a graph aesthetically pleasing? This admittedly vague question is central to the field of Graph Drawing which has over its history suggested numerous answers. Borrowing ideas from Mathematics, Physics, Arts, etc., many researchers have tried to formalize the elusive concept of aesthetics.

In particular, dozens of formulas collectively known as \emph{drawing aesthetics} (or, more precisely, \emph{quality metrics}~\cite{EadesH0K17}) have been proposed that attempt to capture in a single number how beautiful, readable and clear a drawing of an abstract graph is. Of those, simple metrics such as the number of edge crossings, minimum crossing angle, vertex distribution or angular resolution parameters, are obviously incapable \latinphrase{per se} of providing the ultimate aesthetic statement. Advanced metrics may represent, for example, the energy of a corresponding system of physical bodies~\cite{eades84,FruchtermanR91}. This approach underlies many popular graph drawing algorithms~\cite{Tamassia2013} and often leads to pleasing results in practice. However, it is known that low values of energy or stress do not always correspond to the highest degree of symmetry~\cite{Welch2017} which is an important aesthetic criterion~\cite{PurchaseCM96}.

Another direction of research aims to narrow the scope of the original question to specific application domains, focusing on the purpose of a drawing or possible user actions it may facilitate (\emph{tasks}). The target parameters -- readability and the clarity of representation -- may be assessed via user performance studies. However, even in this case such aesthetic notions as symmetry still remain important~\cite{PurchaseCM96}. In general, aesthetically pleasing designs are known to positively affect the apparent and the actual usability~\cite{Norman02,TractinskyKI00} of interfaces and induce positive mental states of users, enhancing their problem-solving abilities~\cite{Fredrickson98}.

In this work, we offer an alternative perspective on the aesthetics of graph drawings.  First, we address a slightly modified question: \enquote{Of two given drawings of the same graph, which one is more aesthetically pleasing?}. With that, we implicitly admit that \enquote{the ultimate} quality metric may not exist and one can hope for at most a (partial) ordering. Instead of a metric, we therefore search for a binary \emph{discriminator function} of graph drawings.  As limited as it is, it could be useful for practical applications such as picking the best answer out of outputs of several drawing algorithms or resolving local minima in layout optimization.

Second, like Huang et al.~\cite{Huang2013}, we believe that by combining multiple metrics computed for each drawing, one has a better chance of capturing complex aesthetic properties. We thus also consider a \enquote{meta-algorithm} that aggregates several \enquote{input} metrics into a single value. However, unlike the recipe by Huang et al., we do not specify the form of this combination \latinphrase{a priori} but let an artificial neural network \enquote{learn} it based on a sample of labeled training data. In the recent years, machine learning techniques have proven useful in such aesthetics-related tasks as assessing the appeal of 3D shapes~\cite{DevLL16} or cropping photos~\cite{Nishiyama09}. Our network architecture is based on a so-called \emph{Siamese neural network}~\cite{Bromley1994} -- a generic model specifically designed for binary functions of same-kind inputs.


Finally, we acknowledge that any simple or complex input metric may become crucial to the answer in some cases that are hard to predict \latinphrase{a priori}. We therefore implement as many input metrics as we can and relegate their ranking to the model. In addition to those known from the literature, we implement a few novel metrics inspired by statistical tools used in Condensed Matter Physics and Crystallography, which we expect to be helpful in capturing the symmetry, balance, and salient structures in large graphs. These metrics are based on so-called \emph{syndromes} -- variable-size multi-sets of numbers computed for a graph or its drawing (e.g.~vertex coordinates or pairwise distances). In order to reduce these heterogeneous multi-sets to a fixed-size \emph{feature vector} (input to the discriminator model), we perform a \emph{feature extraction} process which may involve steps such as creating histograms or performing regressions.

In our experiments, our discriminator model outperforms the known (metric-based) algorithms and achieves an average accuracy of $\XValSuccessMean\percent$ when identifying the \enquote{better} graph drawing out of a pair. The project source code including the data generation procedure is available online~\cite{GitHubRepo}.

The remainder of this paper is structured as follows. In section~\ref{sec:relwork} we briefly overview the state-of-the-art in quantifying graph layout aesthetics.  Section~\ref{sec:syndromes} discusses the used syndromes of aesthetic quality, section~\ref{sec:featex} feature extraction, and section~\ref{sec:model} the discriminator model. The dataset used in our experiments is described in section~\ref{sec:data}. The results and the comparisons with the known metrics are presented in section~\ref{sec:eval}. Section~\ref{sec:conclusion} finalizes the paper and provides an outlook for future work.


\section{Related Work}
\label{sec:relwork}

According to empirical studies, graph drawings that maximize one or several quality metrics are more aethetically pleasing and easier to read~\cite{HuangE05,Huang2013,Purchase97,PurchaseHNK12,WarePCM02}. For instance, in their seminal work, Purchase et al.~have established~\cite{PurchaseCM96} that higher numbers of edge crossings and bends as well as lower levels of symmetry negatively influence user performance in graph reading tasks.

Many graph drawing algorithms attempt to optimize multiple quality metrics. As one way to combine them, Huang et al.~\cite{Huang2013} have used a weighted sum of \enquote{simple} metrics, effects of their interactions (see Purchase~\cite{PURCHASE98} or Huang and Huang~\cite{Huang10}), and error terms to account for possible measurement errors.

In another work, Huang et al.~\cite{HuangHL16} have empirically demonstrated that their \enquote{aggregate} metric is sensitive to quality changes and is correlated with the human performance in graph comprehension tasks. They have also noticed that the dependence of aesthetic quality on input  quality metrics can be non-linear (e.g.~a quadratic relationship better describes the interplay between crossing angles and drawing quality~\cite{HuangHE08}). Our work extends this idea as we allow for arbitrary non-linear dependencies implemented by an artificial neural network.

In evolutionary graph drawing approaches, several techniques have been suggested to \enquote{train} a \emph{fitness function}\footnote{Objective function in genetic algorithms that summarizes optimization goals.} from the user's responses as a composition of several known quality metrics.  Masui~\cite{Masui1994} modeled the fitness function as a linear combination in which the weights are obtained via genetic programming from the pairs of \enquote{good} and \enquote{bad} layouts provided by users.  The so-called co-evolution was used by Barbosa and Barreto~\cite{Barbosa2001} to evolve the weights of the fitness function in parallel with a drawing population in order to match the ranking made by users.  Spönemann and others~\cite{Sponemann14} suggested two alternative techniques. In the first one, the user directly chooses the weights with a slider. In the second, they select good layouts from the current population and the weights are adjusted according to the selection.  Rosete-Suarez~\cite{Rosete-Suarez99} determined the relative importance of individual quality metrics based on user inputs.  Several machine learning-based approaches to graph drawing are described by dos Santos Vieira et al.~\cite{Raissa2015}. Recently, Kwon et al.~\cite{Kwon18} presented a novel work on topological similarity of graphs.  Their goal was to avoid expensive computations of graph layouts and their quality measures. The resulting system was able to sketch a graph in different layouts and estimate corresponding quality measures.


\section{Definitions}

In this paper we consider general simple graphs $G=(V,E)$ where $V=V(G)$ and $E=E(G)$ are the vertex and edge sets of $G$ with $\card{V}=n$ and $\card{E}=m$. A \emph{drawing} or \emph{layout} of a graph is its graphical representation where vertices are drawn as points or small circles, and the edges as straight line segments. Vertex positions in a drawing are denoted by $\vec{p^k}=(p_1^k, p_2^k)\transposed$ for $k=1,\dots,n$ and their set $P=\{\vec{p}^k\}_{k=1}^n$. Furthermore, we use $\textrm{dist}_G(u,v)$ to denote the \emph{graph-theoretical distance} -- the length of the shortest path between vertices $u$ and $v$ in $G$ -- and $\textrm{dist}_\Gamma(u,v)$ for the Euclidean distance between $u$ and $v$ in the drawing $\Gamma(G)$.

\section{Quality Syndromes of Graph Layouts}
\label{sec:syndromes}
A \emph{quality syndrome} of a layout $\Gamma$ is a multi-set of numbers sharing an interpretation that are known or suspected to correlate with the aesthetic quality (e.g.~all pairwise angles between incident edges in $\Gamma$). In the following we describe several syndromes (implemented in our code) inspired by popular quality metrics and common statistical tools. The list is by no means exhaustive, nor do we claim syndromes below as necessary or independent. Our model accepts any combination of syndromes; better choices remain to be systematically investigated.

\begin{explanation}
\explain{\enum{PRINVEC1} and \enum{PRINVEC2}} The two principal axes of the set $P$. If we define a covariance matrix $C=\{c_{ij}\}$, $c_{ij}=\frac{1}{n}\sum_{k=1}^n{(p_i^k-\overline{p_i})(p_j^k-\overline{p_j})})$, $i, j \in \{1, 2\}$, where $\overline{p_i}=\frac{1}{n}\sum_{k=1}^n{p_i^k}$ are the mean values over each dimension, then \enum{PRINVEC1} and \enum{PRINVEC2} will be its eigenvectors.

\explain{\enum{PRINCOMP1} and \enum{PRINCOMP2}} Projections of vertex positions onto $\vec{v}_1=\menum{PRINVEC1}$ and $\vec{v}_2=\menum{PRINVEC2}$, that is, ~$\{\langle\left(\vec{p}^j-\overline{\vec{p}}\right),\vec{v}_i\rangle\}_{j=1}^n$ for $i\in\{1,2\}$ where $\langle\cdot,\cdot\rangle$ denotes the scalar product.

\explain{\enum{ANGULAR}} Let $A(v)$ denote the sequence of edges incident to a vertex $v$, appearing in a clockwise order around it in $\Gamma$.  Let $\alpha(e_i,e_j)$ denote the clockwise angle between edges $e_i$ and $e_j$ incident to the same vertex.  This syndrome is then defined as $\bigcup_{v\in{}V(G)}\{\alpha(e_i,e_j)\st{}e_i,e_j\text{ are consecutive in }A(v)\}$.

\explain{\enum{EDGE\_LENGTH}} $\bigcup_{(u,v)\in{}E(G)}\{\textrm{dist}_\Gamma(u,v)\}$ is the set of edge lengths in $\Gamma$.

\explain{\enum{RDF\_GLOBAL}} $\bigcup_{u\neq{}v\in{}V(G)}\{\dist_\Gamma(u,v)\}$ contains  distances between all vertices in the drawing. The concept of a \emph{radial distribution function (RDF)}~\cite{Findenegg2015} (the distribution of \enum{RDF\_GLOBAL}) is borrowed from Statistical Physics and Crystallography and characterizes the regularity of molecular structures. In large graph layouts it captures regular, periodic and symmetric patterns in the vertex positions. \OnlyArxiv{Fig.~\ref{app:fig:rdf-global} in the Appendix shows histograms of \enum{RDF\_GLOBAL} for some graphs and layouts. Note that more regular drawings feature better-isolated peaks in the respective histograms.}

\explain{$\menum{RDF\_LOCAL}(d)$} $\bigcup_{u\neq{}v\in{}V(G)}\{\dist_\Gamma(u,v)\st\dist(u,v)\leq{}d\}$ is the set of distances between vertices such that the graph-theoretical distance between them is bounded by $d\in\IntsN$. In our implementation, we compute $\menum{RDF\_LOCAL}(2^i)$ for $i\in\{0,\ldots,\ceil{\log_2(D)}\}$ where $D$ is the diameter of $G$. $\menum{RDF\_LOCAL}(d)$ in a sense interpolates between \enum{EDGE\_LENGTH} ($d=1$) and \enum{RDF\_GLOBAL} ($d\to\infty$).

\explain{\enum{TENSION}} $\bigcup_{u\neq{}v\in{}V(G)}\{\dist_\Gamma(u,v)/\dist_G(u,v)\}$ are the ratios of Euclidean and graph-theoretical distances computed for all vertex pairs.  \enum{TENSION} is motivated by and is related to the well-known stress function~\cite{Kamada1989}.
\end{explanation}

\noindent
Note that before computing the quality syndromes, we \emph{normalize} all layouts so that the center of gravity of $V$ is at the origin and the mean edge length is fixed in order to remove the effects of scaling and translation (but not rotation).


\section{Feature Vectors}
\label{sec:featex}

The sizes of quality syndromes are in general graph- and layout-dependent. A neural network, however, requires a fixed-size input. A collection of syndromes is condensed to this \emph{feature vector} via \emph{feature extraction}. Our approach to this step relies on several auxiliary definitions. Let $S=\{x_i\}_{i=1}^p$ be a syndrome with $p$ entries. By $S^\mu$ we denote the arithmetic mean and by $S^\rho$ the root mean square of $S$.  We also define a \emph{histogram sequence} $S^\beta=\frac{1}{p}(S_1,\ldots,S_\beta)$ -- normalized counts in a histogram built over $S$ with $\beta$ bins. The \emph{entropy}~\cite{Shannon1948} of $S^\beta$ is defined as
\begin{equation}
  \label{eqn:entropy}
  \Entropy(S^\beta) = -\sum_{i=1}^{p} \log_2(S_i) S_i
  \mathendpunct{.}
\end{equation}
We expect the entropy, as a measure of disorder, to be related to the aesthetic quality of a layout and convey important information to the discriminator.

\begin{figure}[p]
  \begin{center}
  \ifpdf%
    \tikzsetnextfilename{cache/entropy-regression}%
    \input{pics/entropy-regression.pgf}%
  \else%
    \pgfimage{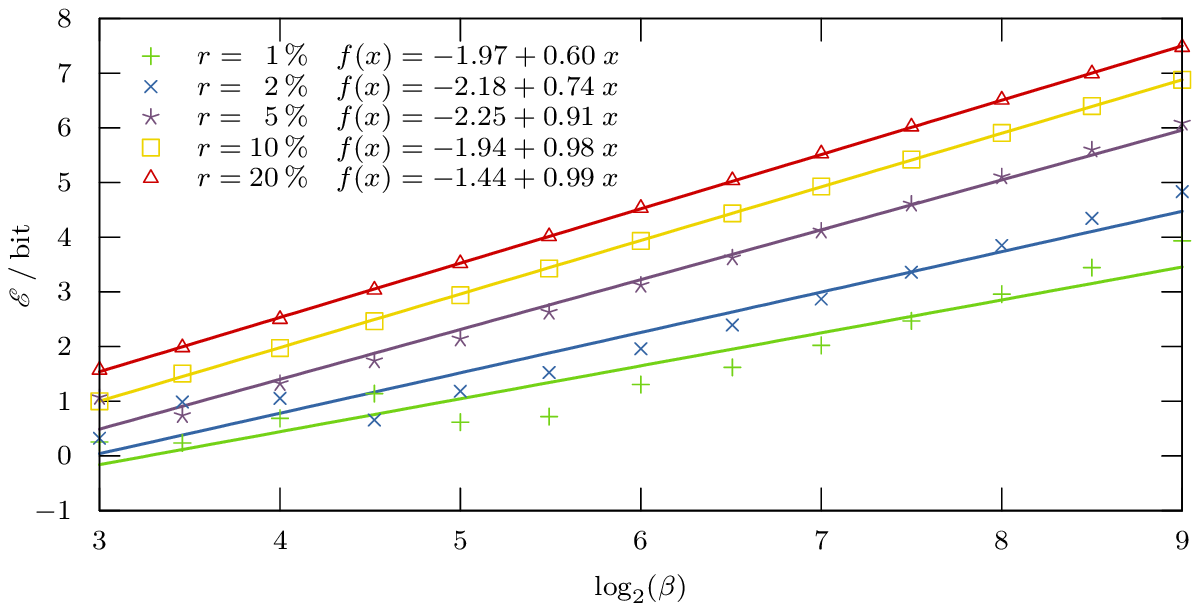}%
  \fi%

  \end{center}
  \caption{%
    Entropy $\mathcal{E}=\mathcal{E}(S^\beta)$ computed for histogram sequences $S^\beta$ defined for different numbers of histogram bins $\beta$. Different markers (colors) correspond to several layouts of a regular grid-like graph, progressively distorted according to the parameter $r$. \OnlyArxiv{See Fig.~\ref{app:fig:worsening} in the Appendix for the examples of distorted grid layout.} The dependence of $\mathcal{E}$ on $\log_2(\beta)$ is well approximated by a linear function. Both intercept and slope show a strong correlation with the levels of distortion $r$.
  }
  \label{fig:entropy-regression}
\end{figure}

The entropy $\Entropy(S^\beta)$ is sensitive to the number of bins $\beta$ (cf.~Fig.~\ref{fig:entropy-regression}). In order to avoid influencing the results via arbitrary choices of $\beta$, we compute it for $\beta=8,16,\ldots,512$. After that, we perform a linear regression of $\Entropy(S^\beta)$ as a function of $\log_2(\beta)$.  Specifically, we find $S^\eta$ and $S^\sigma$ such that $\sum_{\beta}(S^\sigma \log_2\beta+S^\eta-\Entropy(S^\beta))^2$ is minimized.  The parameters (intercept $S^\eta$ and slope $S^\sigma$) of this regression no longer depend on the histogram size and are used as feature vector components. Fig.~\ref{fig:entropy-regression} illustrates that the dependence of $\Entropy(S^\beta)$ on $\log_2(\beta)$ is indeed often close to linear and the regression provides a decent approximation.

A discrete histogram over $S$ can be generalized to a continuous \emph{sliding average}
\begin{equation}
  S^F(x) = \frac{\sum_{i=1}^{p} F(x,x_i)}{\int_{-\infty}^{+\infty} \diff{y} \sum_{i=1}^p F(y,x_i)}
  \mathendpunct{.}
\end{equation}
A natural choice for the kernel $F(x,y)$ is the Gaussian $F_\sigma(x,y)=\exp\left(-\frac{(x - y)^2}{2\sigma^2}\right)$. By analogy to Eq.~\ref{eqn:entropy}, we may now define the \emph{differential entropy}~\cite{Shannon1948} as
\begin{equation}
  \Entropy*(S^{F_\sigma}) = -\int_{-\infty}^{+\infty} \diff{x}\log_2(S^{F_\sigma}(x)) \: S^{F_\sigma}(x)
  \mathendpunct{.}
\end{equation}
This entropy via kernel function still depends on parameter $\sigma$ (the filter width). Computing
$\Entropy*(S^{F_\sigma})$ for multiple $\sigma$ values as we do for $\Entropy(S^\beta)$ is too expensive. Instead, we have found that using Scott's Normal Reference Rule~\cite{Scott1979} as a heuristic to fix $\sigma$ yields satisfactory results, and allows us to define $S^\epsilon = \Entropy*(S^{F_\sigma})$.
 
Using these definitions, for the most complex syndrome $\menum{RDF\_LOCAL}(d)$ we introduce \enum{RDF\_LOCAL} -- a $30$-tuple containing the arithmetic mean, root mean square and the differential entropy of $\menum{RDF\_LOCAL}(2^i)$ for $i\in(0,\ldots,9)$. With that\footnote{Values $i < 10$ are sufficient as no graph in our dataset has a diameter exceeding $2^9$.}, $\menum{RDF\_LOCAL} = \left(\enum{RDF\_LOCAL}(2^i)^\mu, \enum{RDF\_LOCAL}(2^i)^\rho, \enum{RDF\_LOCAL}(2^i)^\epsilon \right)_{i = 0}^9$.

Finally, we assemble the \ensuremath{\NNSharedInputDims}-dimensional\footnote{The size is one less than expected from the explanation above because we do not include the arithmetic mean for \enum{EDGE\_LENGTH} as it is constant (due to the layout normalization mentioned earlier) and therefore non-informative.} feature vector for a layout $\Gamma$ as
\begin{equation*}
  F_\mathrm{layout}(\Gamma) =
  \menum{PRINVEC1}\cup\menum{PRINVEC2}\cup\menum{RDF\_LOCAL}
  \cup\bigcup_{S}\left(S^\mu,S^\rho,S^\eta,S^\sigma\right)
\end{equation*}
where $S$ ranges over \enum{PRINCOMP1}, \enum{PRINCOMP2}, \enum{ANGULAR}, \enum{EDGE\_LENGTH}, \enum{RDF\_GLOBAL} and \enum{TENSION}.

In addition, the discriminator model receives the trivial properties of the underlying graph as the second \ensuremath{\NNTotalAuxHiddenDims}-dimensional vector $F_\mathrm{graph}(G)=(\log(n),\log(m))$.


\section{Discriminator Model}
\label{sec:model}

Feature extractors such as those introduced in the previous section reduce an arbitrary graph $G$ and its arbitrary layout $\Gamma$ to fixed-size vectors $F_\mathrm{graph}(G)$ and $F_\mathrm{layout}(\Gamma)$.  Given a graph $G$ and a pair of its alternative layouts $\Gamma_a$ and $\Gamma_b$, the discriminator function $\DM$ receives the feature vectors $\vec{v}_a=F_\mathrm{layout}(\Gamma_a)$, $\vec{v}_b=F_\mathrm{layout}(\Gamma_b)$ and $\vec{v}_G=F_\mathrm{graph}(G)$ and outputs a scalar value
\begin{equation}
  t = \DM(\vec{v}_G,\vec{v}_a,\vec{v}_b) \in [-1, 1]
  \mathendpunct{.}
\end{equation}
The interpretation is as follows: if $t<0$, then the model believes that $\Gamma_a$ is \enquote{prettier} than $\Gamma_b$; if $t>0$, then it prefers $\Gamma_b$. Its magnitude $\abs{t}$ encodes the confidence level of the decision (the higher $\abs{t}$, the more solid the answer).

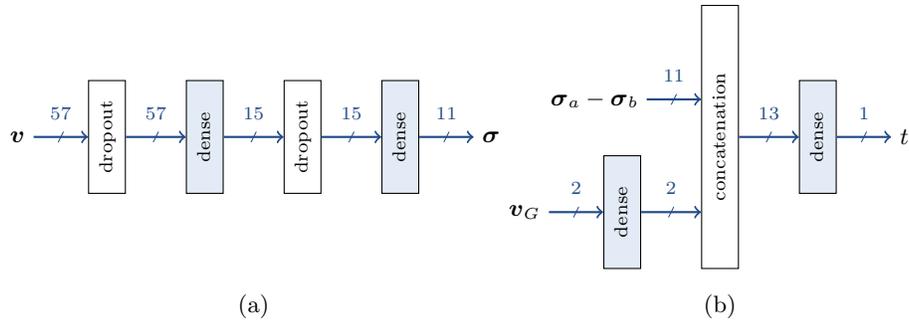
\begin{figure}
  \begin{center}

\begin{tikzpicture}
  [
    layer/.style = {
      rectangle,
      draw,
      minimum width = 15mm,
      minimum height = 4mm,
      text depth = 0.5ex,
      text height = 1.5ex,
      font = \scriptsize,
    },
    dense/.style = {
      layer,
      fill = TangoSkyBlue1!20,
    },
    operator/.style = {
      circle,
      draw,
      minimum size = 4mm,
      font = \scriptsize,
    },
    signal/.style = {
      draw = TangoSkyBlue3,
      thick,
      ->,
    },
    sigwd/.style = {
      font = \scriptsize,
      text = TangoSkyBlue3,
    },
    sigwdmark/.style = {
      draw = TangoSkyBlue3,
      very thin,
    },
    node distance = 5mm,
  ]

  \newcommand*{\NNUnknownDims}{?}
  \providecommand*{\NNSharedInputDims}{\NNUnknownDims}
  \providecommand*{\NNSharedHiddenDims}{\NNUnknownDims}
  \providecommand*{\NNSharedOutputDims}{\NNUnknownDims}
  \providecommand*{\NNTotalAuxInputDims}{\NNUnknownDims}
  \providecommand*{\NNTotalAuxHiddenDims}{\NNUnknownDims}
  \providecommand*{\NNTotalCatDims}{\NNUnknownDims}
  \providecommand*{\NNTotalOutputDims}{\NNUnknownDims}

  \coordinate (SM) at (0, 0);
  \coordinate (GM) at (7.5, -1);
  \coordinate (label) at (0, -2.25);

  \begin{scope}[shift = (SM), x = 13mm]

    \coordinate (in) at (0.25, 0);
    \coordinate (mid) at (4.75, 0);
    \node[layer, rotate = 90] (do1) at (1, 0) {dropout};
    \node[dense, rotate = 90] (l1) at (2, 0) {dense};
    \node[layer, rotate = 90] (do2) at (3, 0) {dropout};
    \node[dense, rotate = 90] (l2) at (4, 0) {dense};

    \draw[signal] (in) to node[name = a]{} (do1);
    \draw[signal] (do1) to node[name = b]{} (l1);
    \draw[signal] (l1) to node[name = c]{} (do2);
    \draw[signal] (do2) to node[name = d]{} (l2);
    \draw[signal] (l2) to node[name = e]{} (mid);

    \foreach \pos in {a, b, c, d, e} {
      \draw[sigwdmark] (\pos) +(-1pt, -2pt) -- +(+1pt, +2pt);
    }

    \node[sigwd, above = 0ex of a] {\ensuremath{\NNSharedInputDims}};
    \node[sigwd, above = 0ex of b] {\ensuremath{\NNSharedInputDims}};
    \node[sigwd, above = 0ex of c] {\ensuremath{\NNSharedHiddenDims}};
    \node[sigwd, above = 0ex of d] {\ensuremath{\NNSharedHiddenDims}};
    \node[sigwd, above = 0ex of e] {\ensuremath{\NNSharedOutputDims}};

    \node[anchor = east] at (in) {$\vec{v}$};
    \node[anchor = west] at (mid) {$\vec{\sigma}$};

    \coordinate (smlabel) at ($(in)!0.5!(mid)$);

  \end{scope}

  \begin{scope}[shift = (GM), x = 13mm]

    \coordinate (delta) at (0.75, 1.5);
    \coordinate (auxin) at (-0.25, 0);
    \coordinate (out) at (3.25, 1);

    \node[layer, rotate = 90, minimum width = 35mm] (cat) at (1.5, 1) {concatenation};
    \node[dense, rotate = 90] (aux) at (0.5, 0) {dense};
    \node[dense, rotate = 90] (l3) at (2.5, 1) {dense};

    \draw[signal] (delta) to node[name = f]{} (cat.north|-delta);
    \draw[signal] (auxin) to node[name = g]{} (aux);
    \draw[signal] (aux) to node[name = h]{} (cat.north|-aux);
    \draw[signal] (cat) to node[name = i]{} (l3);
    \draw[signal] (l3) to node[name = j]{} (out);

    \foreach \pos in {f, g, h, i, j} {
      \draw[sigwdmark] (\pos) +(-1pt, -2pt) -- +(+1pt, +2pt);
    }

    \node[sigwd, above = 0ex of f] {\ensuremath{\NNSharedOutputDims}};
    \node[sigwd, above = 0ex of g] {\ensuremath{\NNTotalAuxInputDims}};
    \node[sigwd, above = 0ex of h] {\ensuremath{\NNTotalAuxHiddenDims}};
    \node[sigwd, above = 0ex of i] {\ensuremath{\NNTotalCatDims}};
    \node[sigwd, above = 0ex of j] {\ensuremath{\NNTotalOutputDims}};

    \node[anchor = east] at (delta) {$\vec{\sigma}_a-\vec{\sigma}_b$};
    \node[anchor = east] at (auxin) {$\vec{v}_G$};
    \node[anchor = west] at (out) {$t$};

    \coordinate (gmlabel) at ($(auxin)!0.5!(out)$);

  \end{scope}

  \node at (smlabel|-label) {(a)};
  \node at (gmlabel|-label) {(b)};

\end{tikzpicture}
  \end{center}
  \caption{
    Structure of the neural networks $\SM(\vec{v})$ (a) and $\GM(\vec{\sigma}_a-\vec{\sigma}_b,\vec{v}_G)$ (b). Shaded blocks denote standard network layers, and the numbers on the arrows denote the dimensionality of the respective representations.
  }
  \label{fig:nn-structure}
\end{figure}

For the implementation of the function $\DM$ we have chosen a practically convenient and flexible model structure known as \emph{Siamese neural networks}, originally proposed by Bromley and others~\cite{Bromley1994} that is defined as
\begin{equation}
 \DM(\vec{v}_G,\vec{v}_a,\vec{v}_b) = \GM(\vec{\sigma}_a-\vec{\sigma}_b,\vec{v}_G)
\end{equation}
where $\vec{\sigma}_a=\SM(\vec{v}_a)$ and $\vec{\sigma}_b=\SM(\vec{v}_b)$.  The \emph{shared model} $\SM$ and the \emph{global model} $\GM$ are implemented as multi-layer neural networks with a simple structure shown in Fig.~\ref{fig:nn-structure}. The network was implemented using the \emph{Keras}~\cite{Keras} framework with the \emph{TensorFlow}~\cite{TensorFlow} library as back-end.

The $\SM$ network (Fig.~\ref{fig:nn-structure}(a)) consists of two \enquote{dense} (fully-connected) layers, each preceded by a \enquote{dropout} layer (discarding $50\percent$ and $25\percent$ of the signals, respectively). Dropout is a stochastic regularization technique intended to avoid overfitting that was first proposed by Srivastava and others~\cite{Srivastava2014}.

In the $\GM$ network (Fig.~\ref{fig:nn-structure}(b)), the graph-related feature vector $\vec{v}_G$ is passed through an auxiliary dense layer, and concatenated with the difference signal $(\vec{\sigma}_a-\vec{\sigma}_b)$ obtained from the output vectors of $\SM$ for the two layouts. The final dense layer produces the scalar output value. The first and the auxiliary layers use linear activation functions, the hidden layer uses $\ReLU$~\cite{Hahnloser2000} and the final layer hyperbolic tangent activation. Following the standard practice, the inputs to the network are normalized by subtracting the mean and dividing by the standard deviation of the feature computed over the complete dataset.

In total, the $\DM$ model has \ensuremath{\NNTotalTrainableParams} free parameters, trained via stochastic gradient descent-based optimization of the mean squared error (MSE) loss function.


\section{Training and Testing Data}
\label{sec:data}

For training, all machine learning methods require datasets representing the variability of possible inputs. Our $\DM$ model needs a dataset containing graphs, their layouts, and known aesthetic orderings of layout pairs. We have assembled such a dataset using two types of sources. First, we used the collections of the well-known graph archives \enum{ROME}, \enum{NORTH} and \enum{RANDDAG} which are published on \url{graphdrawing.org} as well as the NIST's \enquote{Matrix Market}~\cite{MatrixMarket}.  \OnlyArxiv{See Fig.~\ref{app:fig:archives} in the Appendix for examples.}

Second, we have generated random graphs using the algorithms listed below. As a by-product, some of them produce layouts that stem naturally from the generation logic. We refer to these as \emph{native} layouts (see~\cite{Moritz18} for details). \OnlyArxiv{Sample graphs with native layouts (where available) are shown in Fig.~\ref{app:fig:generators} in the Appendix.}

\begin{explanation}
\explain{\enum{GRID}} Regular $n\times{}m$ grids. Native layouts: regular rectangular grids.

\explain{\enum{TORUS1}} Same as \enum{GRID}, but the first and the last \enquote{rows} are connected to form a $1$-torus (a cylinder). No native layouts.

\explain{\enum{TORUS2}} Same as \enum{TORUS1}, but also the first and the last \enquote{columns} are connected to form a $2$-torus (a doughnut). No native layouts.

\explain{\enum{LINDENMAYER}} Uses a stochastic L-system~\cite{Lindenmayer1990} to derive  increasingly complex graphs by performing random replacements of individual vertices with more complicated substructures such as an $n$-ring or an $n$-clique. \OnlyArxiv{Fig.~\ref{app:fig:lindenmayer-subgens} in the Appendix shows all the implemented replacement rules.} Produces a planar native layout.

\explain{\enum{QUASI\meta{n}D} for $n\in\{3,\ldots,6\}$} Projection of a primitive cubic lattice in an $n$-di\-men\-sio\-nal space onto a $2$-dimensional plane intersecting that space at a random angle.  The native layout follows from the construction.

\explain{\enum{MOSAIC1}} Starts with a regular polygon and randomly divides faces according to a set of simple rules until the desired graph size is reached.  The rules include adding a vertex connected to all vertices of the face; subdividing each edge and adding a vertex that connects to each subdivision vertex; subdividing each edge and connecting them to a cycle. \OnlyArxiv{These operations are visualized in Fig.~\ref{app:fig:mosaic-subgens} in the Appendix.} The native layout follows from the construction.

\explain{\enum{MOSAIC2}} Applies a randomly chosen rule of \enum{MOSAIC1} to every face, with the goal of obtaining more symmetric graphs.

\explain{\enum{BOTTLE}} Constructs a graph as a three-dimensional mesh over a random solid of revolution. The native layout is an axonometric projection.
\end{explanation}

\noindent
For each graph, we have computed force-directed layouts using the FM\textsuperscript{3}~\cite{Hachul2005} and stress-minimization~\cite{Kamada1989} algorithms. We assume these and native layouts to be generally aesthetically pleasing and call them all \emph{proper} layouts of a graph.

Furthermore, we have generated \latinphrase{a priori} un-pleasing (\emph{garbage}) layouts as follows.  Given a graph $G=(V,E)$, we generate a random graph $G'=(V',E')$ with $\card{V'}=\card{V}$ and $\card{E'}=\card{E}$ and compute a force-directed layout for $G'$.  The coordinates found for the vertices $V'$ are then assigned to $V$.  We call these \enquote{phantom} layouts due to the use of a \enquote{phantom} graph $G'$. We find that phantom layouts look less artificial than purely random layouts when vertex positions are sampled from a uniform or a normal distribution. This might be due to the fact that $G$ and $G'$ have the same density and share some beneficial aspects of the force-directed method (such as mutual repelling of nodes). \OnlyArxiv{See Fig.~\ref{app:fig:layouts} in the Appendix for the examples of regular and garbage layouts.}

For training and testing of the discriminator model we need a corpus of labeled pairs -- triplets $(\Gamma_a,\Gamma_b,t)$ where $\Gamma_a$ and $\Gamma_a$ are two different layouts for the same graph and $t\in [-1,1]$ is a value indicating the relative aesthetic quality of $\Gamma_a$ and $\Gamma_b$.  A negative (positive) value for $t$ expresses that the quality of $\Gamma_a$ is superior (inferior) compared to $\Gamma_b$ and the magnitude of $t$ expresses the confidence of this prediction.  We only use pairs with sufficiently large $\abs{t}$.

As manually-labelled data were unavailable, we have fixed the values of $t$ as follows. First, we paired a proper and a garbage layout of a graph. The assumption is that the former is always more pleasing (i.e.~$t=\pm1$). Second, in order to obtain more nuanced layout pairs and to increase the amount of data, we have employed the well-known technique of \emph{data augmentation} as follows.

\paragraph{Layout Worsening:}
Given a proper layout $\Gamma$, we apply a transformation designed to gradually reduce its aesthetic quality that is modulated by some parameter $r\in[0,1]$, resulting in a transformed layout $\Gamma'_r$.  By varying the degree $r$ of the distortion, we may generate a sequence of layouts ordered by their anticipated aesthetic value: a layout with less distortion is expected to be more pleasing than a layout with more distortion when starting from a presumably decent layout.  We have implemented the following worsening techniques. \enum{PERTURB}: add Gaussian noise to each node's coordinates. \enum{FLIP\_NODES}: swap coordinates of randomly selected node pairs. \enum{FLIP\_EDGES}: same as \enum{FLIP\_NODES} but restricted to connected node pairs. \enum{MOVLSQ}: apply an affine deformation based on moving least squares suggested (although for a different purpose) by Schaefer et al.~\cite{Schaefer2006}. In essence, all vertices are shifted according to some smoothly varying coordinate mapping. \OnlyArxiv{Illustrations of these worsening algorithms can be found in Fig.~\ref{app:fig:worsening} in the Appendix}.

\paragraph{Layout Interpolation:}
As the second data augmentation technique, we linearly interpolated the positions of corresponding vertices between the proper and garbage layouts of the same graph. The resulting label $t$ is then proportional to the difference in the interpolation parameter.

\par\bigskip\noindent
In total, using all the methods described above, we have been able to collect a database of about \ensuremath{\NNCorpusSizeApprox} labeled layout pairs.


\section{Evaluation}
\label{sec:eval}

The performance of the discriminator model was evaluated using \emph{cross-validation} with $\XValTestRuns$-fold random subsampling~\cite{Kohavi1995}.  In each round, $20\percent$ of graphs (with all their layouts) were chosen randomly and were set aside for testing, and the model was trained using the remaining layout pairs. Of $N$ labeled pairs used for testing, in each round we computed the number $N_\mathrm{correct}$ of pairs for which the model properly predicted the aesthetic preference, and derived the accuracy (success rate) $A=N_\mathrm{correct}/N$.  The standard deviation of $A$ over the $\XValTestRuns$ runs was taken as the uncertainty of the results.  With the average number of test samples of $N=\XValCountApprox$, the eventual success rate was $\boldsymbol{A=(\XValSuccessMean\pm\XValSuccessStdev)\percent}$.

\subsection{Comparison With Other Metrics}

In order to assess the relative standing of the suggested method, we have implemented two known aesthetic metrics (\emph{stress} and the \emph{combined metric} by Huang et al.~\cite{HuangHL16}) and evaluated them over the same dataset. The metric values were trivially converted to the respective discriminator function outputs.

Stress $\stress$ of a layout $\Gamma$ of a simple connected graph $G=(V,E)$ was defined by Kamada and Kawai~\cite{Kamada1989} as
\begin{equation}
  \stress(\Gamma) = \sum_{i=1}^{n-1} \sum_{j=i+1}^{n}
  k_{ij} \left( \dist_\Gamma(v_i,v_j) - L \dist_G(v_i,v_j) \right)^2
  \mathendpunct{,}
  \label{eq:stress}
\end{equation}
where $L$ denotes the desirable edge length and $k_{ij}=K/\dist_G(v_i,v_j)^2$ is the strength of a \enquote{spring} attached to $v_i$ and $v_j$.  The constant $K$ is irrelevant in the context of discriminator functions and can be set to any value.

As observed by Welch and Kobourov~\cite{Welch2017}, the numeric value of stress depends on the layout scale via the constant $L$ in the Eq.~\ref{eq:stress} which complicates comparisons. Their suggested solution was for each layout to find $L$ that minimizes $\stress$ (e.g.~using binary search).  In our implementation, we applied a similar technique based on fitting and minimizing a quadratic function to the stress computed at three scales. We refer to this quantity as \enum{STRESS}.

The combined metric proposed by Huang et al.~\cite{HuangHL16} (referred to as \enum{COMB}) is a weighted average of four simpler quality metrics: the number of edge crossings (\enum{CC}), the minimum crossing angle between any two edges in the drawing (\enum{CR}), the minimum angle between two adjacent edges (\enum{AR}), and the standard deviation computed over all edge lengths (\enum{EL}).

The average is computed over the so-called $z$-scores of the above metrics. Each $z$-score is found by subtracting the mean and dividing by the standard deviation of the metric for all layouts of a given graph to be compared with each other. More formally, let $G$ be a graph and $\Gamma_1, \ldots, \Gamma_k$ be its $k$ layouts to be compared pairwise. Let $M(\Gamma_i)$ be the value of metric $M$ for $\Gamma_i$ and $\mu_M$ and $\sigma_M$ be the mean and the standard deviation of $M(\Gamma_i)$ for $i\in\{1,\ldots,k\}$. Then
\begin{equation}
  z_M^{(i)} = \frac{M(\Gamma_i) - \mu_M}{\sigma_M}
\end{equation}
is the $z$-score for metric $M$ and layout $\Gamma_i$.  The combined metric then is
\begin{equation}
  \menum{COMB}(\Gamma_j) = \sum_{M} w_M \: z_{M}^{(j)}
  \mathendpunct{.}
\end{equation}
The weights $w_M$ were found via Nelder-Mead maximization~\cite{Press2007} of the prediction accuracy over the training dataset\footnote{%
  The obtained weights are: \TextualHuangWeights.
}.

\begin{figure}[p]
  \csdef{Status0}{\textcolor{TangoScarletRed2}{\ding{55}}}%
  \csdef{Status1}{\textcolor{TangoChameleon2}{\ding{51}}}%
  \newcommand*{\ShowIt}[1]{%
  \ifpdf%
    \tikzsetnextfilename{cache/#1}%
    \tikz[x=0.25\textwidth, y=0.25\textwidth, ]{\input{pics/#1.tikz}}%
  \else%
    \pgfimage[width=0.25\textwidth]{cache/#1}%
  \fi%
}
  \newcommand*{\TellIt}[3]{%
    \begin{tabular}{l@{\quad}c}
      \enum{DISC\_MODEL} & \csuse{Status#1}\\
      \enum{STRESS}      & \csuse{Status#2}\\
      \enum{COMB}        & \csuse{Status#3}\\
    \end{tabular}
  }
  \begin{center}
    \begin{tabular}{>{\centering}m{0.35\textwidth}>{\centering}m{0.35\textwidth}@{\qquad}m{0.2\textwidth}}
      \ShowIt{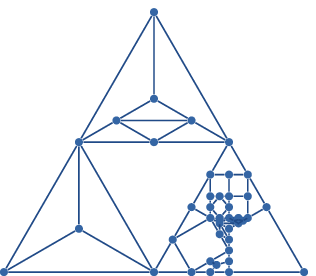} & \ShowIt{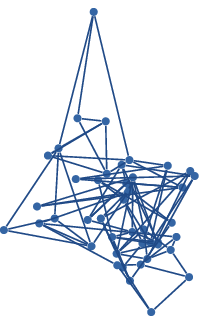} & \TellIt{1}{0}{1}\\[1ex]\\[1ex]
      \ShowIt{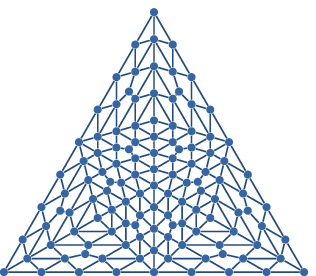} & \ShowIt{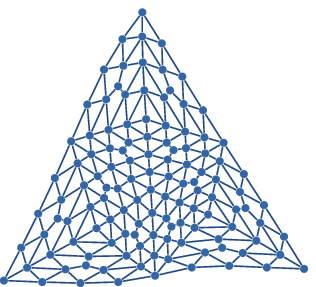} & \TellIt{1}{0}{0}\\[1ex]\\[1ex]
      \ShowIt{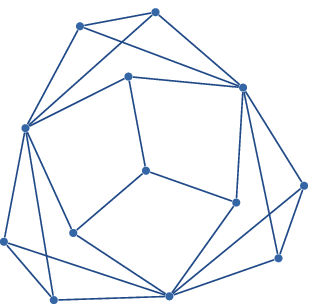} & \ShowIt{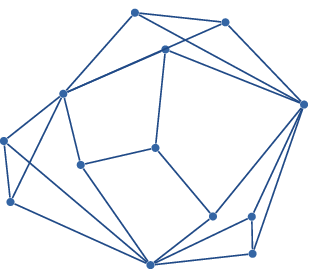} & \TellIt{1}{1}{0}\\
    \end{tabular}
  \end{center}
  \caption{%
    Examples where our discriminator model (\enum{DISC\_MODEL}) succeeds (\csuse{Status1}) and the competing metrics fail (\csuse{Status0}) to predict the answer correctly. In each row, the layout on the left is expected to be superior compared to the one on the right.
  }
  \label{fig:model-comp}
\end{figure}

The accuracy of the stress-based and the combined model-based discriminators is shown in Tab.~\ref{tab:competing-metrics}. In most cases, our model outperforms these algorithms by a comfortable margin. Fig.~\ref{fig:model-comp} provides examples of mis-predictions. By inspecting such cases, we notice that \enum{STRESS} often fails to guess the aesthetics of (almost) planar layouts that contain both very short and very long edges (such behavior may also be inferred from the definition of \enum{STRESS}). We observe that there are planar graphs, such as nested triangulations, for which this property is unavoidable in planar drawings. The mis-predictions of \enum{COMB} seem to be due to the high weight of the edge length metric \enum{EL}. Both \enum{STRESS} and \enum{COMB} are weaker than our model in capturing the absolute symmetry and regularity of layouts.

\begin{table}[p]
  \begin{center}


\CompetingMetricResult[\enum{STRESS}]{93.49}{0.86}{2.99}{1.01}
\CompetingMetricResult[\enum{COMB}]{92.76}{1.03}{3.71}{1.22}

  \end{center}
  \caption{%
    Accuracy scores for the \enum{COMB} and \enum{STRESS} model. The standard deviation in each column is estimated based on the $5$-fold cross-validation (using $20\percent$ of data for testing each time).  The \enquote{Advantage} column shows the improvement in the accuracy of our model with respect to the alternative metric.
  }
  \label{tab:competing-metrics}
\end{table}

\subsection{Significance of Individual Syndromes}

In order to estimate the influence of individual syndromes on the final result, we have tested several modifications of our model. For each syndrome, we considered the case when the feature vector contained only that syndrome. In the second case, that syndrome was removed from the original feature vector. The entries for the omitted features were set to zero. The results are shown in Tab.~\ref{tab:eval-puncture}.

\begin{table}
  \begin{center}


\def\XValTestRuns{10}
\def\XValCountApprox{7415}
\def\XValCountMean{7414.90}
\def\XValCountStdev{620.09}
\def\XValSuccessMean{96.48}
\def\XValSuccessStdev{0.85}
\def\XValFailureMean{3.52}
\def\XValFailureStdev{0.85}
\def\XValCondPosMean{49.73}
\def\XValCondPosStdev{0.61}
\def\XValCondNegMean{50.27}
\def\XValCondNegStdev{0.61}
\def\XValPredPosMean{50.36}
\def\XValPredPosStdev{2.38}
\def\XValPredNegMean{49.64}
\def\XValPredNegStdev{2.38}
\def\XValTrueNegMean{48.20}
\def\XValTrueNegStdev{1.43}
\def\XValFalseNegMean{1.44}
\def\XValFalseNegStdev{1.15}
\def\XValTruePosMean{48.28}
\def\XValTruePosStdev{1.18}
\def\XValFalsePosMean{2.08}
\def\XValFalsePosStdev{1.43}

  \end{center}
  \caption{%
    Success rates of our discriminator when a syndrome is excluded from the feature vector, and when the feature vector contains only that a syndrome. Note that \enum{RDF\_LOCAL} is a family of syndromes that are all included or excluded together. The apparent paradox of higher success rates when some syndromes are excluded can be explained by a statistical fluctuation and is well within the listed range of uncertainty.
  }
  \label{tab:eval-puncture}
\end{table}

As can be observed, the dominant contribution to the accuracy of the model is due to the RDF-based properties \enum{RDF\_LOCAL} and \enum{RDF\_GLOBAL}. The exclusion of other syndromes does not significantly change the results (they agree within the estimated uncertainty). However, the sole inclusion of these syndromes still performs better than random choice. This suggests that there is a considerable overlap between the aesthetic aspects captured by various syndromes.  Further analysis is needed to identify the nature and the magnitude of these correlations.

\section{Conclusion}
\label{sec:conclusion}

In this paper we propose a machine learning-based discriminator model that selects the more aesthetically pleasing drawing from a pair of graph layouts. Our model picks the \enquote{better} layout in more than $96\percent$ cases and outperforms known stress-based and linear combination-based models. To the best of our knowledge, this is the first application of machine learning methods to this question. Previously, such techniques have proven successful in a range of complex issues involving aesthetics, prior knowledge, and unstated rules in object recognition, industrial design, and digital arts. As our model uses a simple network architecture, investigating the performance of more complex networks is warranted.

Previous efforts were focused on determining the aesthetic quality of a layout as a weighted average of individual quality metrics.  We extend these ideas and findings in the sense that we do not assume any particular form of dependency between the overall aesthetic quality and the individual quality metrics.

Going beyond simple quality metrics, we define quality syndromes that capture arrays of information about graphs and layouts. In particular, we borrow the notion of RDF from Statistical Physics and Crystallography; RDF-based features demonstrate the strongest potential in extracting the aesthetic quality of a layout. We expect RDFs (describing the microscopic structure of materials) to be the most relevant for large graphs. It is tempting to investigate whether further tools from physics can be useful in capturing drawing aesthetics.

From multiple syndromes, we construct fixed-size feature vectors using common statistical tools. Our feature vector does not contain any information on crossings or crossing angles, nevertheless its performance is superior with respect to the weighted averages-based model which accounts for both. It would be interesting to investigate whether including these and other features further improves the performance of the neural network-based model.

In order to train and evaluate the model, we have assembled a relatively large corpus of labeled pairs of layouts, using available and generated graphs and exploiting the assumption that layouts produced by force-directed algorithms and native graph layouts are aesthetically pleasing and that disturbing them reduces the aesthetic quality. We admit that this study should ideally be repeated with human-labeled data. However, this requires that a dataset be collected with a size similar to ours, which is a challenging task. Creating such a dataset may become a critically important accomplishment in the graph drawing field.

\bibliographystyle{splncs04}
\bibliography{literature}

\ifforarxiv
\clearpage
\appendix\relax

\section*{Appendix -- Supplementary Figures and Tables}

\begin{figure}[h!]
  \begin{center}
    \begin{tabularx}{\linewidth}{
        >{\centering\arraybackslash}X
        >{\centering\arraybackslash}X
        >{\centering\arraybackslash}X
      }
  \ifpdf%
    \tikzsetnextfilename{cache/demograph-a}%
    \tikz[x=30mm, y=30mm, ]{\input{pics/demograph-a.tikz}}%
  \else%
    \pgfimage[width=30mm]{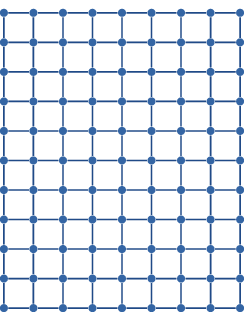}%
  \fi%
&
  \ifpdf%
    \tikzsetnextfilename{cache/demograph-b}%
    \tikz[x=30mm, y=30mm, ]{\input{pics/demograph-b.tikz}}%
  \else%
    \pgfimage[width=30mm]{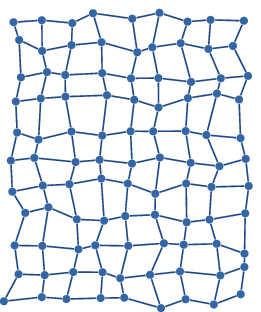}%
  \fi%
&
  \ifpdf%
    \tikzsetnextfilename{cache/demograph-c}%
    \tikz[x=30mm, y=30mm, ]{\input{pics/demograph-c.tikz}}%
  \else%
    \pgfimage[width=30mm]{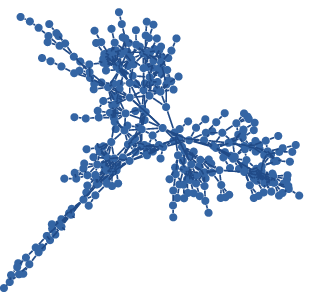}%
  \fi%
\\[3ex]
  \ifpdf%
    \tikzsetnextfilename{cache/rdf-global-a}%
    \input{pics/rdf-global-a.pgf}%
  \else%
    \pgfimage{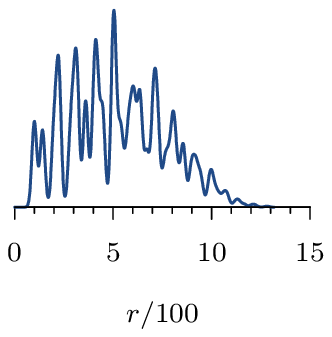}%
  \fi%
&
  \ifpdf%
    \tikzsetnextfilename{cache/rdf-global-b}%
    \input{pics/rdf-global-b.pgf}%
  \else%
    \pgfimage{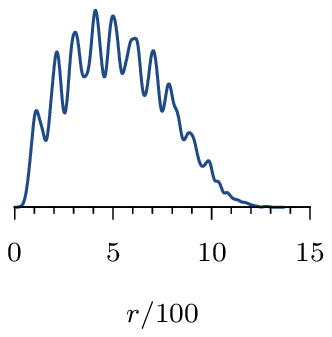}%
  \fi%
&
  \ifpdf%
    \tikzsetnextfilename{cache/rdf-global-c}%
    \input{pics/rdf-global-c.pgf}%
  \else%
    \pgfimage{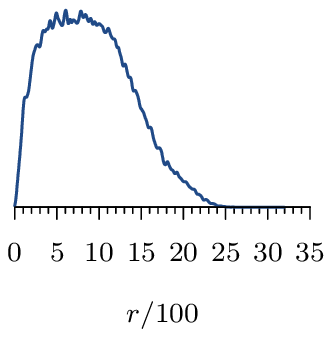}%
  \fi%
\\[2ex]
  \ifpdf%
    \tikzsetnextfilename{cache/angular-a}%
    \begin{tikzpicture}[gnuplot]
\tikzset{every node/.append style={font={\fontsize{8.0pt}{9.6pt}\selectfont}}}
\gpcolor{color=gp lt color border}
\gpsetlinetype{gp lt border}
\gpsetdashtype{gp dt solid}
\gpsetlinewidth{1.00}
\draw[gp path] (0.639,1.414)--(0.639,1.293);
\node[gp node center] at (0.639,0.956) {0.0};
\draw[gp path] (1.321,1.414)--(1.321,1.293);
\node[gp node center] at (1.321,0.956) {0.5};
\draw[gp path] (2.003,1.414)--(2.003,1.293);
\node[gp node center] at (2.003,0.956) {1.0};
\draw[gp path] (2.684,1.414)--(2.684,1.293);
\node[gp node center] at (2.684,0.956) {1.5};
\draw[gp path] (3.366,1.414)--(3.366,1.293);
\node[gp node center] at (3.366,0.956) {2.0};
\draw[gp path] (0.503,1.414)--(3.503,1.414);
\node[gp node center] at (2.002,0.335) {$\phi / \pi$};
\gpcolor{rgb color={0.125,0.290,0.529}}
\gpsetlinewidth{2.00}
\draw[gp path] (0.574,1.414)--(0.579,1.414)--(0.585,1.414)--(0.590,1.414)--(0.596,1.414)%
  --(0.601,1.414)--(0.608,1.414)--(0.613,1.414)--(0.619,1.414)--(0.624,1.414)--(0.630,1.414)%
  --(0.635,1.414)--(0.641,1.414)--(0.646,1.414)--(0.652,1.414)--(0.657,1.414)--(0.663,1.414)%
  --(0.668,1.414)--(0.675,1.414)--(0.681,1.414)--(0.686,1.414)--(0.691,1.414)--(0.697,1.414)%
  --(0.702,1.414)--(0.708,1.414)--(0.713,1.414)--(0.719,1.414)--(0.724,1.414)--(0.730,1.414)%
  --(0.735,1.414)--(0.742,1.414)--(0.748,1.414)--(0.753,1.414)--(0.759,1.414)--(0.764,1.414)%
  --(0.770,1.414)--(0.775,1.414)--(0.780,1.414)--(0.786,1.414)--(0.791,1.414)--(0.797,1.414)%
  --(0.802,1.414)--(0.809,1.414)--(0.815,1.414)--(0.820,1.414)--(0.826,1.414)--(0.831,1.414)%
  --(0.837,1.414)--(0.842,1.414)--(0.848,1.414)--(0.853,1.414)--(0.859,1.414)--(0.864,1.414)%
  --(0.869,1.414)--(0.876,1.414)--(0.882,1.414)--(0.887,1.414)--(0.893,1.414)--(0.898,1.414)%
  --(0.904,1.414)--(0.909,1.414)--(0.915,1.414)--(0.920,1.414)--(0.926,1.414)--(0.931,1.414)%
  --(0.937,1.414)--(0.943,1.414)--(0.949,1.414)--(0.954,1.414)--(0.960,1.414)--(0.965,1.414)%
  --(0.971,1.414)--(0.976,1.414)--(0.982,1.414)--(0.987,1.414)--(0.993,1.414)--(0.998,1.414)%
  --(1.005,1.414)--(1.010,1.414)--(1.016,1.414)--(1.021,1.414)--(1.027,1.414)--(1.032,1.414)%
  --(1.038,1.414)--(1.043,1.414)--(1.049,1.414)--(1.054,1.414)--(1.060,1.414)--(1.065,1.414)%
  --(1.072,1.414)--(1.078,1.414)--(1.083,1.414)--(1.089,1.414)--(1.094,1.414)--(1.099,1.414)%
  --(1.105,1.414)--(1.110,1.414)--(1.116,1.414)--(1.121,1.414)--(1.127,1.414)--(1.132,1.414)%
  --(1.139,1.414)--(1.145,1.414)--(1.150,1.414)--(1.156,1.414)--(1.161,1.414)--(1.167,1.414)%
  --(1.172,1.414)--(1.178,1.414)--(1.183,1.414)--(1.188,1.414)--(1.194,1.414)--(1.199,1.414)%
  --(1.206,1.414)--(1.212,1.414)--(1.217,1.414)--(1.223,1.414)--(1.228,1.414)--(1.234,1.414)%
  --(1.239,1.416)--(1.245,1.418)--(1.250,1.424)--(1.256,1.437)--(1.261,1.463)--(1.267,1.509)%
  --(1.273,1.589)--(1.279,1.714)--(1.284,1.893)--(1.290,2.132)--(1.295,2.419)--(1.301,2.732)%
  --(1.306,3.030)--(1.312,3.268)--(1.317,3.404)--(1.323,3.414)--(1.328,3.295)--(1.334,3.069)%
  --(1.340,2.775)--(1.346,2.463)--(1.351,2.171)--(1.357,1.924)--(1.362,1.735)--(1.368,1.605)%
  --(1.373,1.519)--(1.379,1.468)--(1.384,1.441)--(1.390,1.426)--(1.395,1.420)--(1.401,1.416)%
  --(1.408,1.414)--(1.413,1.414)--(1.419,1.414)--(1.424,1.414)--(1.429,1.414)--(1.435,1.414)%
  --(1.440,1.414)--(1.446,1.414)--(1.451,1.414)--(1.457,1.414)--(1.462,1.414)--(1.468,1.414)%
  --(1.475,1.414)--(1.480,1.414)--(1.486,1.414)--(1.491,1.414)--(1.497,1.414)--(1.502,1.414)%
  --(1.508,1.414)--(1.513,1.414)--(1.518,1.414)--(1.524,1.414)--(1.529,1.414)--(1.535,1.414)%
  --(1.542,1.414)--(1.547,1.414)--(1.553,1.414)--(1.558,1.414)--(1.564,1.414)--(1.569,1.414)%
  --(1.575,1.414)--(1.580,1.414)--(1.586,1.414)--(1.591,1.414)--(1.597,1.414)--(1.602,1.414)%
  --(1.609,1.414)--(1.614,1.414)--(1.620,1.414)--(1.625,1.414)--(1.631,1.414)--(1.636,1.414)%
  --(1.642,1.414)--(1.647,1.414)--(1.653,1.414)--(1.658,1.414)--(1.664,1.414)--(1.670,1.414)%
  --(1.676,1.414)--(1.681,1.414)--(1.687,1.414)--(1.692,1.414)--(1.698,1.414)--(1.703,1.414)%
  --(1.709,1.414)--(1.714,1.414)--(1.720,1.414)--(1.725,1.414)--(1.731,1.414)--(1.738,1.414)%
  --(1.743,1.414)--(1.749,1.414)--(1.754,1.414)--(1.759,1.414)--(1.765,1.414)--(1.770,1.414)%
  --(1.776,1.414)--(1.781,1.414)--(1.787,1.414)--(1.792,1.414)--(1.798,1.414)--(1.805,1.414)%
  --(1.810,1.414)--(1.816,1.414)--(1.821,1.414)--(1.827,1.414)--(1.832,1.414)--(1.838,1.414)%
  --(1.843,1.414)--(1.848,1.414)--(1.854,1.414)--(1.859,1.414)--(1.865,1.414)--(1.872,1.414)%
  --(1.877,1.414)--(1.883,1.414)--(1.888,1.414)--(1.894,1.414)--(1.899,1.414)--(1.905,1.414)%
  --(1.910,1.414)--(1.916,1.414)--(1.921,1.414)--(1.927,1.414)--(1.932,1.416)--(1.939,1.416)%
  --(1.944,1.420)--(1.950,1.424)--(1.955,1.431)--(1.961,1.445)--(1.966,1.463)--(1.972,1.488)%
  --(1.977,1.517)--(1.983,1.548)--(1.988,1.577)--(1.994,1.601)--(1.999,1.612)--(2.006,1.612)%
  --(2.011,1.601)--(2.017,1.577)--(2.022,1.548)--(2.028,1.517)--(2.033,1.488)--(2.039,1.463)%
  --(2.044,1.445)--(2.050,1.431)--(2.055,1.424)--(2.061,1.420)--(2.066,1.416)--(2.073,1.416)%
  --(2.079,1.414)--(2.084,1.414)--(2.089,1.414)--(2.095,1.414)--(2.100,1.414)--(2.106,1.414)%
  --(2.111,1.414)--(2.117,1.414)--(2.122,1.414)--(2.128,1.414)--(2.133,1.414)--(2.140,1.414)%
  --(2.146,1.414)--(2.151,1.414)--(2.157,1.414)--(2.162,1.414)--(2.168,1.414)--(2.173,1.414)%
  --(2.178,1.414)--(2.184,1.414)--(2.189,1.414)--(2.195,1.414)--(2.200,1.414)--(2.207,1.414)%
  --(2.213,1.414)--(2.218,1.414)--(2.224,1.414)--(2.229,1.414)--(2.235,1.414)--(2.240,1.414)%
  --(2.246,1.414)--(2.251,1.414)--(2.257,1.414)--(2.262,1.414)--(2.267,1.414)--(2.274,1.414)%
  --(2.280,1.414)--(2.285,1.414)--(2.291,1.414)--(2.296,1.414)--(2.302,1.414)--(2.307,1.414)%
  --(2.313,1.414)--(2.318,1.414)--(2.324,1.414)--(2.329,1.414)--(2.335,1.414)--(2.341,1.414)%
  --(2.347,1.414)--(2.352,1.414)--(2.358,1.414)--(2.363,1.414)--(2.369,1.414)--(2.374,1.414)%
  --(2.380,1.414)--(2.385,1.414)--(2.391,1.414)--(2.396,1.414)--(2.403,1.414)--(2.408,1.414)%
  --(2.414,1.414)--(2.419,1.414)--(2.425,1.414)--(2.430,1.414)--(2.436,1.414)--(2.441,1.414)%
  --(2.447,1.414)--(2.452,1.414)--(2.458,1.414)--(2.463,1.414)--(2.470,1.414)--(2.476,1.414)%
  --(2.481,1.414)--(2.487,1.414)--(2.492,1.414)--(2.497,1.414)--(2.503,1.414)--(2.508,1.414)%
  --(2.514,1.414)--(2.519,1.414)--(2.525,1.414)--(2.530,1.414)--(2.537,1.414)--(2.543,1.414)%
  --(2.548,1.414)--(2.554,1.414)--(2.559,1.414)--(2.565,1.414)--(2.570,1.414)--(2.576,1.414)%
  --(2.581,1.414)--(2.586,1.414)--(2.592,1.414)--(2.597,1.414)--(2.604,1.414)--(2.610,1.414)%
  --(2.615,1.414)--(2.621,1.414)--(2.626,1.414)--(2.632,1.416)--(2.637,1.416)--(2.643,1.418)%
  --(2.648,1.420)--(2.654,1.424)--(2.659,1.427)--(2.665,1.431)--(2.671,1.435)--(2.677,1.437)%
  --(2.682,1.439)--(2.688,1.439)--(2.693,1.437)--(2.699,1.433)--(2.704,1.429)--(2.710,1.426)%
  --(2.715,1.424)--(2.721,1.420)--(2.726,1.418)--(2.732,1.416)--(2.738,1.416)--(2.744,1.414)%
  --(2.749,1.414)--(2.755,1.414)--(2.760,1.414)--(2.766,1.414)--(2.771,1.414)--(2.777,1.414)%
  --(2.782,1.414)--(2.788,1.414)--(2.793,1.414)--(2.799,1.414)--(2.806,1.414)--(2.811,1.414)%
  --(2.817,1.414)--(2.822,1.414)--(2.827,1.414)--(2.833,1.414)--(2.838,1.414)--(2.844,1.414)%
  --(2.849,1.414)--(2.855,1.414)--(2.860,1.414)--(2.866,1.414)--(2.873,1.414)--(2.878,1.414)%
  --(2.884,1.414)--(2.889,1.414)--(2.895,1.414)--(2.900,1.414)--(2.906,1.414)--(2.911,1.414)%
  --(2.916,1.414)--(2.922,1.414)--(2.927,1.414)--(2.933,1.414)--(2.940,1.414)--(2.945,1.414)%
  --(2.951,1.414)--(2.956,1.414)--(2.962,1.414)--(2.967,1.414)--(2.973,1.414)--(2.978,1.414)%
  --(2.984,1.414)--(2.989,1.414)--(2.995,1.414)--(3.000,1.414)--(3.007,1.414)--(3.012,1.414)%
  --(3.018,1.414)--(3.023,1.414)--(3.029,1.414)--(3.034,1.414)--(3.040,1.414)--(3.045,1.414)%
  --(3.051,1.414)--(3.056,1.414)--(3.062,1.414)--(3.068,1.414)--(3.074,1.414)--(3.079,1.414)%
  --(3.085,1.414)--(3.090,1.414)--(3.096,1.414)--(3.101,1.414)--(3.107,1.414)--(3.112,1.414)%
  --(3.118,1.414)--(3.123,1.414)--(3.129,1.414)--(3.136,1.414)--(3.141,1.414)--(3.147,1.414)%
  --(3.152,1.414)--(3.157,1.414)--(3.163,1.414)--(3.168,1.414)--(3.174,1.414)--(3.179,1.414)%
  --(3.185,1.414)--(3.190,1.414)--(3.196,1.414)--(3.203,1.414)--(3.208,1.414)--(3.214,1.414)%
  --(3.219,1.414)--(3.225,1.414)--(3.230,1.414)--(3.236,1.414)--(3.241,1.414)--(3.246,1.414)%
  --(3.252,1.414)--(3.257,1.414)--(3.263,1.414)--(3.270,1.414)--(3.275,1.414)--(3.281,1.414)%
  --(3.286,1.414)--(3.292,1.414)--(3.297,1.414)--(3.303,1.414)--(3.308,1.414)--(3.314,1.414)%
  --(3.319,1.414)--(3.325,1.414)--(3.330,1.414)--(3.337,1.414)--(3.342,1.414)--(3.348,1.414)%
  --(3.353,1.414)--(3.359,1.414)--(3.364,1.414)--(3.370,1.414)--(3.375,1.414)--(3.381,1.414)%
  --(3.386,1.414)--(3.392,1.414)--(3.397,1.414)--(3.404,1.414)--(3.409,1.414)--(3.415,1.414)%
  --(3.420,1.414)--(3.426,1.414)--(3.431,1.414);
\gpcolor{color=gp lt color border}
\gpsetlinewidth{1.00}
\draw[gp path] (0.503,1.414)--(3.503,1.414);
\gpdefrectangularnode{gp plot 1}{\pgfpoint{0.503cm}{1.414cm}}{\pgfpoint{3.503cm}{3.414cm}}
\end{tikzpicture}
%
  \else%
    \pgfimage{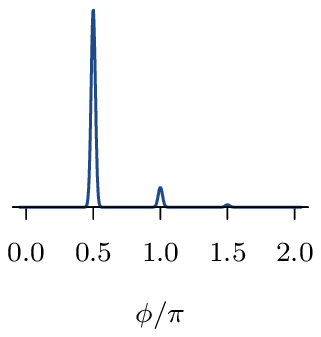}%
  \fi%
&
  \ifpdf%
    \tikzsetnextfilename{cache/angular-b}%
    \begin{tikzpicture}[gnuplot]
\tikzset{every node/.append style={font={\fontsize{8.0pt}{9.6pt}\selectfont}}}
\gpcolor{color=gp lt color border}
\gpsetlinetype{gp lt border}
\gpsetdashtype{gp dt solid}
\gpsetlinewidth{1.00}
\draw[gp path] (0.639,1.414)--(0.639,1.293);
\node[gp node center] at (0.639,0.956) {0.0};
\draw[gp path] (1.321,1.414)--(1.321,1.293);
\node[gp node center] at (1.321,0.956) {0.5};
\draw[gp path] (2.003,1.414)--(2.003,1.293);
\node[gp node center] at (2.003,0.956) {1.0};
\draw[gp path] (2.684,1.414)--(2.684,1.293);
\node[gp node center] at (2.684,0.956) {1.5};
\draw[gp path] (3.366,1.414)--(3.366,1.293);
\node[gp node center] at (3.366,0.956) {2.0};
\draw[gp path] (0.503,1.414)--(3.503,1.414);
\node[gp node center] at (2.002,0.335) {$\phi / \pi$};
\gpcolor{rgb color={0.125,0.290,0.529}}
\gpsetlinewidth{2.00}
\draw[gp path] (0.574,1.414)--(0.579,1.414)--(0.585,1.414)--(0.590,1.414)--(0.596,1.414)%
  --(0.601,1.414)--(0.608,1.414)--(0.613,1.414)--(0.619,1.414)--(0.624,1.414)--(0.630,1.414)%
  --(0.635,1.414)--(0.641,1.414)--(0.646,1.414)--(0.652,1.414)--(0.657,1.414)--(0.663,1.414)%
  --(0.668,1.414)--(0.675,1.414)--(0.681,1.414)--(0.686,1.414)--(0.691,1.414)--(0.697,1.414)%
  --(0.702,1.414)--(0.708,1.414)--(0.713,1.414)--(0.719,1.414)--(0.724,1.414)--(0.730,1.414)%
  --(0.735,1.414)--(0.742,1.414)--(0.748,1.414)--(0.753,1.414)--(0.759,1.414)--(0.764,1.414)%
  --(0.770,1.414)--(0.775,1.414)--(0.780,1.414)--(0.786,1.414)--(0.791,1.414)--(0.797,1.414)%
  --(0.802,1.414)--(0.809,1.414)--(0.815,1.414)--(0.820,1.414)--(0.826,1.414)--(0.831,1.414)%
  --(0.837,1.414)--(0.842,1.414)--(0.848,1.414)--(0.853,1.414)--(0.859,1.414)--(0.864,1.414)%
  --(0.869,1.414)--(0.876,1.414)--(0.882,1.414)--(0.887,1.414)--(0.893,1.414)--(0.898,1.414)%
  --(0.904,1.414)--(0.909,1.414)--(0.915,1.414)--(0.920,1.414)--(0.926,1.414)--(0.931,1.414)%
  --(0.937,1.414)--(0.943,1.414)--(0.949,1.414)--(0.954,1.414)--(0.960,1.414)--(0.965,1.414)%
  --(0.971,1.414)--(0.976,1.414)--(0.982,1.414)--(0.987,1.414)--(0.993,1.416)--(0.998,1.416)%
  --(1.005,1.420)--(1.010,1.424)--(1.016,1.429)--(1.021,1.441)--(1.027,1.455)--(1.032,1.472)%
  --(1.038,1.496)--(1.043,1.519)--(1.049,1.542)--(1.054,1.568)--(1.060,1.589)--(1.065,1.609)%
  --(1.072,1.630)--(1.078,1.653)--(1.083,1.681)--(1.089,1.716)--(1.094,1.759)--(1.099,1.809)%
  --(1.105,1.866)--(1.110,1.924)--(1.116,1.984)--(1.121,2.043)--(1.127,2.101)--(1.132,2.158)%
  --(1.139,2.212)--(1.145,2.265)--(1.150,2.317)--(1.156,2.370)--(1.161,2.423)--(1.167,2.477)%
  --(1.172,2.532)--(1.178,2.590)--(1.183,2.647)--(1.188,2.703)--(1.194,2.756)--(1.199,2.806)%
  --(1.206,2.851)--(1.212,2.896)--(1.217,2.939)--(1.223,2.989)--(1.228,3.046)--(1.234,3.112)%
  --(1.239,3.186)--(1.245,3.262)--(1.250,3.332)--(1.256,3.385)--(1.261,3.414)--(1.267,3.414)%
  --(1.273,3.383)--(1.279,3.322)--(1.284,3.242)--(1.290,3.151)--(1.295,3.059)--(1.301,2.976)%
  --(1.306,2.907)--(1.312,2.861)--(1.317,2.839)--(1.323,2.837)--(1.328,2.855)--(1.334,2.880)%
  --(1.340,2.904)--(1.346,2.915)--(1.351,2.909)--(1.357,2.880)--(1.362,2.830)--(1.368,2.761)%
  --(1.373,2.684)--(1.379,2.606)--(1.384,2.536)--(1.390,2.475)--(1.395,2.430)--(1.401,2.401)%
  --(1.408,2.388)--(1.413,2.388)--(1.419,2.399)--(1.424,2.419)--(1.429,2.444)--(1.435,2.473)%
  --(1.440,2.500)--(1.446,2.524)--(1.451,2.536)--(1.457,2.534)--(1.462,2.512)--(1.468,2.471)%
  --(1.475,2.413)--(1.480,2.341)--(1.486,2.263)--(1.491,2.189)--(1.497,2.123)--(1.502,2.068)%
  --(1.508,2.031)--(1.513,2.010)--(1.518,2.000)--(1.524,2.000)--(1.529,2.006)--(1.535,2.014)%
  --(1.542,2.019)--(1.547,2.019)--(1.553,2.012)--(1.558,1.994)--(1.564,1.965)--(1.569,1.928)%
  --(1.575,1.885)--(1.580,1.840)--(1.586,1.796)--(1.591,1.755)--(1.597,1.720)--(1.602,1.688)%
  --(1.609,1.659)--(1.614,1.636)--(1.620,1.612)--(1.625,1.589)--(1.631,1.568)--(1.636,1.544)%
  --(1.642,1.523)--(1.647,1.503)--(1.653,1.486)--(1.658,1.470)--(1.664,1.455)--(1.670,1.445)%
  --(1.676,1.435)--(1.681,1.429)--(1.687,1.427)--(1.692,1.427)--(1.698,1.429)--(1.703,1.433)%
  --(1.709,1.439)--(1.714,1.445)--(1.720,1.451)--(1.725,1.455)--(1.731,1.461)--(1.738,1.466)%
  --(1.743,1.470)--(1.749,1.476)--(1.754,1.480)--(1.759,1.484)--(1.765,1.486)--(1.770,1.486)%
  --(1.776,1.482)--(1.781,1.476)--(1.787,1.470)--(1.792,1.464)--(1.798,1.463)--(1.805,1.464)%
  --(1.810,1.468)--(1.816,1.478)--(1.821,1.490)--(1.827,1.500)--(1.832,1.509)--(1.838,1.519)%
  --(1.843,1.525)--(1.848,1.529)--(1.854,1.535)--(1.859,1.540)--(1.865,1.548)--(1.872,1.556)%
  --(1.877,1.566)--(1.883,1.574)--(1.888,1.577)--(1.894,1.577)--(1.899,1.574)--(1.905,1.566)%
  --(1.910,1.554)--(1.916,1.540)--(1.921,1.527)--(1.927,1.513)--(1.932,1.501)--(1.939,1.490)%
  --(1.944,1.482)--(1.950,1.476)--(1.955,1.472)--(1.961,1.470)--(1.966,1.470)--(1.972,1.472)%
  --(1.977,1.476)--(1.983,1.482)--(1.988,1.486)--(1.994,1.490)--(1.999,1.492)--(2.006,1.492)%
  --(2.011,1.492)--(2.017,1.496)--(2.022,1.500)--(2.028,1.509)--(2.033,1.523)--(2.039,1.538)%
  --(2.044,1.552)--(2.050,1.564)--(2.055,1.570)--(2.061,1.568)--(2.066,1.564)--(2.073,1.556)%
  --(2.079,1.546)--(2.084,1.538)--(2.089,1.533)--(2.095,1.531)--(2.100,1.529)--(2.106,1.527)%
  --(2.111,1.523)--(2.117,1.515)--(2.122,1.503)--(2.128,1.490)--(2.133,1.474)--(2.140,1.461)%
  --(2.146,1.447)--(2.151,1.439)--(2.157,1.433)--(2.162,1.431)--(2.168,1.433)--(2.173,1.437)%
  --(2.178,1.445)--(2.184,1.455)--(2.189,1.466)--(2.195,1.480)--(2.200,1.494)--(2.207,1.505)%
  --(2.213,1.517)--(2.218,1.525)--(2.224,1.533)--(2.229,1.537)--(2.235,1.537)--(2.240,1.537)%
  --(2.246,1.533)--(2.251,1.527)--(2.257,1.519)--(2.262,1.509)--(2.267,1.500)--(2.274,1.488)%
  --(2.280,1.474)--(2.285,1.463)--(2.291,1.451)--(2.296,1.441)--(2.302,1.433)--(2.307,1.427)%
  --(2.313,1.424)--(2.318,1.424)--(2.324,1.424)--(2.329,1.427)--(2.335,1.431)--(2.341,1.435)%
  --(2.347,1.441)--(2.352,1.445)--(2.358,1.447)--(2.363,1.447)--(2.369,1.445)--(2.374,1.441)%
  --(2.380,1.437)--(2.385,1.431)--(2.391,1.427)--(2.396,1.422)--(2.403,1.420)--(2.408,1.418)%
  --(2.414,1.418)--(2.419,1.418)--(2.425,1.420)--(2.430,1.422)--(2.436,1.426)--(2.441,1.431)%
  --(2.447,1.435)--(2.452,1.441)--(2.458,1.445)--(2.463,1.447)--(2.470,1.447)--(2.476,1.445)%
  --(2.481,1.441)--(2.487,1.435)--(2.492,1.431)--(2.497,1.426)--(2.503,1.422)--(2.508,1.420)%
  --(2.514,1.416)--(2.519,1.416)--(2.525,1.414)--(2.530,1.414)--(2.537,1.414)--(2.543,1.414)%
  --(2.548,1.414)--(2.554,1.414)--(2.559,1.414)--(2.565,1.414)--(2.570,1.414)--(2.576,1.414)%
  --(2.581,1.414)--(2.586,1.414)--(2.592,1.414)--(2.597,1.414)--(2.604,1.414)--(2.610,1.414)%
  --(2.615,1.414)--(2.621,1.414)--(2.626,1.414)--(2.632,1.414)--(2.637,1.414)--(2.643,1.414)%
  --(2.648,1.414)--(2.654,1.414)--(2.659,1.414)--(2.665,1.414)--(2.671,1.414)--(2.677,1.414)%
  --(2.682,1.414)--(2.688,1.414)--(2.693,1.414)--(2.699,1.414)--(2.704,1.414)--(2.710,1.414)%
  --(2.715,1.414)--(2.721,1.414)--(2.726,1.414)--(2.732,1.416)--(2.738,1.416)--(2.744,1.418)%
  --(2.749,1.422)--(2.755,1.426)--(2.760,1.429)--(2.766,1.435)--(2.771,1.439)--(2.777,1.443)%
  --(2.782,1.447)--(2.788,1.447)--(2.793,1.447)--(2.799,1.443)--(2.806,1.441)--(2.811,1.437)%
  --(2.817,1.437)--(2.822,1.437)--(2.827,1.437)--(2.833,1.441)--(2.838,1.443)--(2.844,1.447)%
  --(2.849,1.447)--(2.855,1.447)--(2.860,1.443)--(2.866,1.439)--(2.873,1.433)--(2.878,1.429)%
  --(2.884,1.426)--(2.889,1.422)--(2.895,1.420)--(2.900,1.420)--(2.906,1.420)--(2.911,1.424)%
  --(2.916,1.427)--(2.922,1.431)--(2.927,1.437)--(2.933,1.441)--(2.940,1.445)--(2.945,1.447)%
  --(2.951,1.447)--(2.956,1.445)--(2.962,1.441)--(2.967,1.435)--(2.973,1.429)--(2.978,1.426)%
  --(2.984,1.422)--(2.989,1.418)--(2.995,1.416)--(3.000,1.416)--(3.007,1.414)--(3.012,1.414)%
  --(3.018,1.414)--(3.023,1.414)--(3.029,1.414)--(3.034,1.414)--(3.040,1.414)--(3.045,1.414)%
  --(3.051,1.414)--(3.056,1.414)--(3.062,1.414)--(3.068,1.414)--(3.074,1.414)--(3.079,1.414)%
  --(3.085,1.414)--(3.090,1.414)--(3.096,1.414)--(3.101,1.414)--(3.107,1.414)--(3.112,1.414)%
  --(3.118,1.414)--(3.123,1.414)--(3.129,1.414)--(3.136,1.414)--(3.141,1.414)--(3.147,1.414)%
  --(3.152,1.414)--(3.157,1.414)--(3.163,1.414)--(3.168,1.414)--(3.174,1.414)--(3.179,1.414)%
  --(3.185,1.414)--(3.190,1.414)--(3.196,1.414)--(3.203,1.414)--(3.208,1.414)--(3.214,1.414)%
  --(3.219,1.414)--(3.225,1.414)--(3.230,1.414)--(3.236,1.414)--(3.241,1.414)--(3.246,1.414)%
  --(3.252,1.414)--(3.257,1.414)--(3.263,1.414)--(3.270,1.414)--(3.275,1.414)--(3.281,1.414)%
  --(3.286,1.414)--(3.292,1.414)--(3.297,1.414)--(3.303,1.414)--(3.308,1.414)--(3.314,1.414)%
  --(3.319,1.414)--(3.325,1.414)--(3.330,1.414)--(3.337,1.414)--(3.342,1.414)--(3.348,1.414)%
  --(3.353,1.414)--(3.359,1.414)--(3.364,1.414)--(3.370,1.414)--(3.375,1.414)--(3.381,1.414)%
  --(3.386,1.414)--(3.392,1.414)--(3.397,1.414)--(3.404,1.414)--(3.409,1.414)--(3.415,1.414)%
  --(3.420,1.414)--(3.426,1.414)--(3.431,1.414);
\gpcolor{color=gp lt color border}
\gpsetlinewidth{1.00}
\draw[gp path] (0.503,1.414)--(3.503,1.414);
\gpdefrectangularnode{gp plot 1}{\pgfpoint{0.503cm}{1.414cm}}{\pgfpoint{3.503cm}{3.414cm}}
\end{tikzpicture}
%
  \else%
    \pgfimage{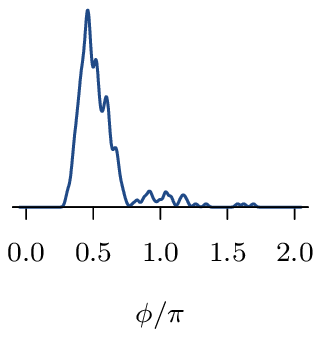}%
  \fi%
&
  \ifpdf%
    \tikzsetnextfilename{cache/angular-c}%
    \begin{tikzpicture}[gnuplot]
\tikzset{every node/.append style={font={\fontsize{8.0pt}{9.6pt}\selectfont}}}
\gpcolor{color=gp lt color border}
\gpsetlinetype{gp lt border}
\gpsetdashtype{gp dt solid}
\gpsetlinewidth{1.00}
\draw[gp path] (0.639,1.414)--(0.639,1.293);
\node[gp node center] at (0.639,0.956) {0.0};
\draw[gp path] (1.321,1.414)--(1.321,1.293);
\node[gp node center] at (1.321,0.956) {0.5};
\draw[gp path] (2.003,1.414)--(2.003,1.293);
\node[gp node center] at (2.003,0.956) {1.0};
\draw[gp path] (2.684,1.414)--(2.684,1.293);
\node[gp node center] at (2.684,0.956) {1.5};
\draw[gp path] (3.366,1.414)--(3.366,1.293);
\node[gp node center] at (3.366,0.956) {2.0};
\draw[gp path] (0.503,1.414)--(3.503,1.414);
\node[gp node center] at (2.002,0.335) {$\phi / \pi$};
\gpcolor{rgb color={0.125,0.290,0.529}}
\gpsetlinewidth{2.00}
\draw[gp path] (0.574,1.416)--(0.579,1.418)--(0.585,1.424)--(0.590,1.431)--(0.596,1.447)%
  --(0.601,1.474)--(0.608,1.515)--(0.613,1.577)--(0.619,1.663)--(0.624,1.772)--(0.630,1.908)%
  --(0.635,2.064)--(0.641,2.234)--(0.646,2.409)--(0.652,2.576)--(0.657,2.730)--(0.663,2.865)%
  --(0.668,2.972)--(0.675,3.056)--(0.681,3.116)--(0.686,3.161)--(0.691,3.196)--(0.697,3.227)%
  --(0.702,3.262)--(0.708,3.299)--(0.713,3.338)--(0.719,3.373)--(0.724,3.400)--(0.730,3.414)%
  --(0.735,3.408)--(0.742,3.383)--(0.748,3.342)--(0.753,3.289)--(0.759,3.231)--(0.764,3.170)%
  --(0.770,3.116)--(0.775,3.067)--(0.780,3.024)--(0.786,2.985)--(0.791,2.950)--(0.797,2.917)%
  --(0.802,2.882)--(0.809,2.849)--(0.815,2.816)--(0.820,2.785)--(0.826,2.756)--(0.831,2.728)%
  --(0.837,2.701)--(0.842,2.674)--(0.848,2.645)--(0.853,2.611)--(0.859,2.573)--(0.864,2.534)%
  --(0.869,2.491)--(0.876,2.452)--(0.882,2.417)--(0.887,2.389)--(0.893,2.372)--(0.898,2.364)%
  --(0.904,2.364)--(0.909,2.368)--(0.915,2.374)--(0.920,2.380)--(0.926,2.380)--(0.931,2.374)%
  --(0.937,2.364)--(0.943,2.351)--(0.949,2.335)--(0.954,2.319)--(0.960,2.306)--(0.965,2.296)%
  --(0.971,2.290)--(0.976,2.290)--(0.982,2.298)--(0.987,2.314)--(0.993,2.337)--(0.998,2.364)%
  --(1.005,2.393)--(1.010,2.419)--(1.016,2.432)--(1.021,2.436)--(1.027,2.428)--(1.032,2.411)%
  --(1.038,2.388)--(1.043,2.360)--(1.049,2.331)--(1.054,2.304)--(1.060,2.278)--(1.065,2.257)%
  --(1.072,2.236)--(1.078,2.216)--(1.083,2.199)--(1.089,2.183)--(1.094,2.166)--(1.099,2.146)%
  --(1.105,2.125)--(1.110,2.101)--(1.116,2.074)--(1.121,2.047)--(1.127,2.018)--(1.132,1.994)%
  --(1.139,1.973)--(1.145,1.957)--(1.150,1.947)--(1.156,1.940)--(1.161,1.934)--(1.167,1.928)%
  --(1.172,1.918)--(1.178,1.907)--(1.183,1.891)--(1.188,1.875)--(1.194,1.860)--(1.199,1.850)%
  --(1.206,1.846)--(1.212,1.846)--(1.217,1.852)--(1.223,1.860)--(1.228,1.868)--(1.234,1.875)%
  --(1.239,1.879)--(1.245,1.877)--(1.250,1.873)--(1.256,1.866)--(1.261,1.854)--(1.267,1.838)%
  --(1.273,1.823)--(1.279,1.807)--(1.284,1.792)--(1.290,1.778)--(1.295,1.768)--(1.301,1.762)%
  --(1.306,1.760)--(1.312,1.762)--(1.317,1.768)--(1.323,1.778)--(1.328,1.792)--(1.334,1.805)%
  --(1.340,1.817)--(1.346,1.825)--(1.351,1.829)--(1.357,1.825)--(1.362,1.817)--(1.368,1.801)%
  --(1.373,1.786)--(1.379,1.768)--(1.384,1.755)--(1.390,1.745)--(1.395,1.741)--(1.401,1.745)%
  --(1.408,1.751)--(1.413,1.759)--(1.419,1.764)--(1.424,1.766)--(1.429,1.764)--(1.435,1.755)%
  --(1.440,1.739)--(1.446,1.718)--(1.451,1.698)--(1.457,1.679)--(1.462,1.665)--(1.468,1.657)%
  --(1.475,1.657)--(1.480,1.661)--(1.486,1.667)--(1.491,1.675)--(1.497,1.681)--(1.502,1.685)%
  --(1.508,1.688)--(1.513,1.688)--(1.518,1.690)--(1.524,1.692)--(1.529,1.698)--(1.535,1.702)%
  --(1.542,1.708)--(1.547,1.710)--(1.553,1.710)--(1.558,1.706)--(1.564,1.696)--(1.569,1.686)%
  --(1.575,1.673)--(1.580,1.661)--(1.586,1.649)--(1.591,1.640)--(1.597,1.630)--(1.602,1.622)%
  --(1.609,1.614)--(1.614,1.611)--(1.620,1.607)--(1.625,1.607)--(1.631,1.607)--(1.636,1.611)%
  --(1.642,1.614)--(1.647,1.618)--(1.653,1.618)--(1.658,1.618)--(1.664,1.616)--(1.670,1.612)%
  --(1.676,1.609)--(1.681,1.605)--(1.687,1.601)--(1.692,1.599)--(1.698,1.601)--(1.703,1.607)%
  --(1.709,1.614)--(1.714,1.628)--(1.720,1.642)--(1.725,1.655)--(1.731,1.667)--(1.738,1.675)%
  --(1.743,1.679)--(1.749,1.675)--(1.754,1.669)--(1.759,1.659)--(1.765,1.649)--(1.770,1.640)%
  --(1.776,1.632)--(1.781,1.628)--(1.787,1.626)--(1.792,1.626)--(1.798,1.628)--(1.805,1.632)%
  --(1.810,1.638)--(1.816,1.646)--(1.821,1.655)--(1.827,1.667)--(1.832,1.683)--(1.838,1.702)%
  --(1.843,1.722)--(1.848,1.739)--(1.854,1.753)--(1.859,1.759)--(1.865,1.757)--(1.872,1.745)%
  --(1.877,1.729)--(1.883,1.710)--(1.888,1.690)--(1.894,1.677)--(1.899,1.669)--(1.905,1.669)%
  --(1.910,1.677)--(1.916,1.690)--(1.921,1.710)--(1.927,1.729)--(1.932,1.751)--(1.939,1.772)%
  --(1.944,1.794)--(1.950,1.811)--(1.955,1.827)--(1.961,1.840)--(1.966,1.852)--(1.972,1.860)%
  --(1.977,1.866)--(1.983,1.868)--(1.988,1.870)--(1.994,1.868)--(1.999,1.860)--(2.006,1.850)%
  --(2.011,1.838)--(2.017,1.821)--(2.022,1.803)--(2.028,1.782)--(2.033,1.762)--(2.039,1.743)%
  --(2.044,1.723)--(2.050,1.704)--(2.055,1.686)--(2.061,1.669)--(2.066,1.653)--(2.073,1.636)%
  --(2.079,1.620)--(2.084,1.605)--(2.089,1.589)--(2.095,1.574)--(2.100,1.560)--(2.106,1.546)%
  --(2.111,1.537)--(2.117,1.529)--(2.122,1.525)--(2.128,1.521)--(2.133,1.517)--(2.140,1.513)%
  --(2.146,1.507)--(2.151,1.500)--(2.157,1.492)--(2.162,1.482)--(2.168,1.474)--(2.173,1.468)%
  --(2.178,1.463)--(2.184,1.461)--(2.189,1.461)--(2.195,1.463)--(2.200,1.464)--(2.207,1.466)%
  --(2.213,1.466)--(2.218,1.466)--(2.224,1.463)--(2.229,1.461)--(2.235,1.457)--(2.240,1.453)%
  --(2.246,1.451)--(2.251,1.447)--(2.257,1.447)--(2.262,1.447)--(2.267,1.447)--(2.274,1.447)%
  --(2.280,1.445)--(2.285,1.445)--(2.291,1.443)--(2.296,1.443)--(2.302,1.441)--(2.307,1.439)%
  --(2.313,1.439)--(2.318,1.439)--(2.324,1.439)--(2.329,1.439)--(2.335,1.439)--(2.341,1.439)%
  --(2.347,1.439)--(2.352,1.437)--(2.358,1.437)--(2.363,1.437)--(2.369,1.437)--(2.374,1.437)%
  --(2.380,1.439)--(2.385,1.439)--(2.391,1.441)--(2.396,1.443)--(2.403,1.443)--(2.408,1.443)%
  --(2.414,1.443)--(2.419,1.443)--(2.425,1.443)--(2.430,1.443)--(2.436,1.445)--(2.441,1.447)%
  --(2.447,1.451)--(2.452,1.453)--(2.458,1.453)--(2.463,1.453)--(2.470,1.453)--(2.476,1.451)%
  --(2.481,1.447)--(2.487,1.445)--(2.492,1.443)--(2.497,1.439)--(2.503,1.437)--(2.508,1.437)%
  --(2.514,1.435)--(2.519,1.433)--(2.525,1.431)--(2.530,1.429)--(2.537,1.429)--(2.543,1.427)%
  --(2.548,1.427)--(2.554,1.427)--(2.559,1.427)--(2.565,1.431)--(2.570,1.435)--(2.576,1.439)%
  --(2.581,1.443)--(2.586,1.449)--(2.592,1.451)--(2.597,1.453)--(2.604,1.453)--(2.610,1.453)%
  --(2.615,1.449)--(2.621,1.447)--(2.626,1.443)--(2.632,1.437)--(2.637,1.433)--(2.643,1.429)%
  --(2.648,1.426)--(2.654,1.424)--(2.659,1.420)--(2.665,1.420)--(2.671,1.420)--(2.677,1.420)%
  --(2.682,1.422)--(2.688,1.422)--(2.693,1.424)--(2.699,1.426)--(2.704,1.426)--(2.710,1.426)%
  --(2.715,1.426)--(2.721,1.424)--(2.726,1.422)--(2.732,1.420)--(2.738,1.420)--(2.744,1.418)%
  --(2.749,1.416)--(2.755,1.416)--(2.760,1.416)--(2.766,1.416)--(2.771,1.416)--(2.777,1.416)%
  --(2.782,1.418)--(2.788,1.420)--(2.793,1.422)--(2.799,1.426)--(2.806,1.429)--(2.811,1.433)%
  --(2.817,1.437)--(2.822,1.439)--(2.827,1.441)--(2.833,1.443)--(2.838,1.443)--(2.844,1.441)%
  --(2.849,1.439)--(2.855,1.437)--(2.860,1.435)--(2.866,1.431)--(2.873,1.429)--(2.878,1.426)%
  --(2.884,1.424)--(2.889,1.422)--(2.895,1.422)--(2.900,1.422)--(2.906,1.422)--(2.911,1.424)%
  --(2.916,1.426)--(2.922,1.426)--(2.927,1.427)--(2.933,1.426)--(2.940,1.426)--(2.945,1.424)%
  --(2.951,1.422)--(2.956,1.420)--(2.962,1.418)--(2.967,1.418)--(2.973,1.416)--(2.978,1.416)%
  --(2.984,1.414)--(2.989,1.414)--(2.995,1.414)--(3.000,1.414)--(3.007,1.414)--(3.012,1.414)%
  --(3.018,1.414)--(3.023,1.414)--(3.029,1.414)--(3.034,1.414)--(3.040,1.414)--(3.045,1.414)%
  --(3.051,1.414)--(3.056,1.414)--(3.062,1.414)--(3.068,1.414)--(3.074,1.414)--(3.079,1.414)%
  --(3.085,1.414)--(3.090,1.414)--(3.096,1.414)--(3.101,1.414)--(3.107,1.414)--(3.112,1.414)%
  --(3.118,1.414)--(3.123,1.414)--(3.129,1.414)--(3.136,1.414)--(3.141,1.414)--(3.147,1.414)%
  --(3.152,1.414)--(3.157,1.414)--(3.163,1.414)--(3.168,1.414)--(3.174,1.414)--(3.179,1.414)%
  --(3.185,1.414)--(3.190,1.414)--(3.196,1.414)--(3.203,1.414)--(3.208,1.414)--(3.214,1.414)%
  --(3.219,1.414)--(3.225,1.414)--(3.230,1.414)--(3.236,1.414)--(3.241,1.414)--(3.246,1.414)%
  --(3.252,1.414)--(3.257,1.414)--(3.263,1.414)--(3.270,1.414)--(3.275,1.414)--(3.281,1.414)%
  --(3.286,1.416)--(3.292,1.418)--(3.297,1.422)--(3.303,1.431)--(3.308,1.447)--(3.314,1.478)%
  --(3.319,1.529)--(3.325,1.605)--(3.330,1.714)--(3.337,1.854)--(3.342,2.019)--(3.348,2.195)%
  --(3.353,2.354)--(3.359,2.475)--(3.364,2.536)--(3.370,2.522)--(3.375,2.440)--(3.381,2.304)%
  --(3.386,2.134)--(3.392,1.961)--(3.397,1.803)--(3.404,1.673)--(3.409,1.575)--(3.415,1.507)%
  --(3.420,1.464)--(3.426,1.439)--(3.431,1.426);
\gpcolor{color=gp lt color border}
\gpsetlinewidth{1.00}
\draw[gp path] (0.503,1.414)--(3.503,1.414);
\gpdefrectangularnode{gp plot 1}{\pgfpoint{0.503cm}{1.414cm}}{\pgfpoint{3.503cm}{3.414cm}}
\end{tikzpicture}
%
  \else%
    \pgfimage{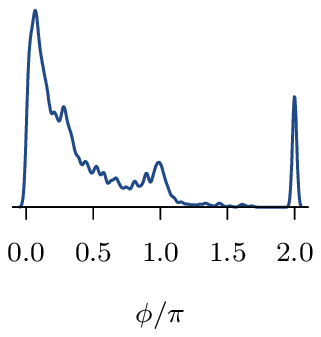}%
  \fi%

    \end{tabularx}
  \end{center}
  \caption{%
    Illustration of the syndromes \enum{RDF\_GLOBAL} and \enum{ANGULAR}.  Upper row, from left to right: proper, distorted layouts of a regular grid, and a force-directed layout of an irregular graph (``power grid'').  Central row: smoothed relative frequency distributions for the \enum{RDF\_GLOBAL} syndromes computed for the respective layouts in the upper row.  The isolated peaks in the leftmost distribution correspond to characteristic distances in the lattice.  In the central plot, these peaks are widened due to random distortion. In the rightmost plot, no regular structure can be identified.  Lower row: smoothed relative frequency distributions for the \enum{ANGULAR} syndromes.  The leftmost plot clearly shows the dominance of angles proportional to $\pi/2$.  In the rightmost plot, the distinctive peak at $\phi = 2\pi$ corresponds to the large number of degree one vertices.
  }
  \label{app:fig:rdf-global}
\end{figure}
\clearpage

\begin{figure}[bp]
  \begin{center}
    \begin{tabularx}{\linewidth}{%
        >{\centering\arraybackslash}X
        >{\centering\arraybackslash}X
        >{\centering\arraybackslash}X
      }
  \ifpdf%
    \tikzsetnextfilename{cache/rome}%
    \tikz[x=0.25\textwidth, y=0.25\textwidth, ]{
\iftikzgraphpreamble
\def\aspectratio{0.8291505574}
\else
\node[vertex] (v0) at (-0.2249597601, 0.1755465674) {};
\node[vertex] (v1) at (-0.3128785238, 0.3747195899) {};
\node[vertex] (v2) at (0.0251341400, 0.0626333651) {};
\node[vertex] (v3) at (-0.2294678345, 0.3008657788) {};
\node[vertex] (v4) at (-0.0030771204, 0.1755966572) {};
\node[vertex] (v5) at (0.1874172949, -0.2776345612) {};
\node[vertex] (v6) at (0.1591593418, 0.2670967483) {};
\node[vertex] (v7) at (0.0220391531, -0.1063396213) {};
\node[vertex] (v8) at (0.3281186770, 0.1017215067) {};
\node[vertex] (v9) at (-0.1089206471, -0.0384654868) {};
\node[vertex] (v10) at (-0.2218040647, -0.0913788712) {};
\node[vertex] (v11) at (-0.1373829051, 0.0129602542) {};
\node[vertex] (v12) at (0.1062305304, -0.1821399357) {};
\node[vertex] (v13) at (0.1959669457, 0.3819496502) {};
\node[vertex] (v14) at (0.1617776494, -0.0285769796) {};
\node[vertex] (v15) at (-0.2235142423, -0.1891521666) {};
\node[vertex] (v16) at (-0.0250906175, -0.3294657217) {};
\node[vertex] (v17) at (0.4354216085, 0.1282771669) {};
\node[vertex] (v18) at (-0.3436766874, -0.2175390360) {};
\node[vertex] (v19) at (0.0663180718, -0.2952567886) {};
\node[vertex] (v20) at (0.5359835344, 0.1628055307) {};
\node[vertex] (v21) at (0.0431198367, 0.2989571536) {};
\node[vertex] (v22) at (-0.0401301676, 0.1363458838) {};
\node[vertex] (v23) at (-0.0310820924, 0.0398319652) {};
\node[vertex] (v24) at (-0.1304832765, -0.0884404294) {};
\node[vertex] (v25) at (-0.2182895940, -0.3084478351) {};
\node[vertex] (v26) at (-0.4640164656, 0.1809614485) {};
\node[vertex] (v27) at (0.2736339100, -0.3565424012) {};
\node[vertex] (v28) at (-0.1204783184, 0.0896916772) {};
\node[vertex] (v29) at (0.0945765221, -0.0466130010) {};
\node[vertex] (v30) at (0.0554783515, 0.1760989728) {};
\node[vertex] (v31) at (0.2240256581, -0.2076929644) {};
\node[vertex] (v32) at (0.1498558327, 0.1122566682) {};
\node[vertex] (v33) at (0.2613631952, 0.1749444009) {};
\node[vertex] (v34) at (-0.3506735519, 0.1598369617) {};
\node[vertex] (v35) at (-0.0129810886, -0.2038624303) {};
\node[vertex] (v36) at (-0.2401639558, -0.0231825855) {};
\node[vertex] (v37) at (0.1705068615, -0.0977601726) {};
\node[vertex] (v38) at (0.2239430352, 0.0617581138) {};
\node[vertex] (v39) at (0.1105981011, 0.1325877201) {};
\node[vertex] (v40) at (-0.2928217166, 0.2640144638) {};
\node[vertex] (v41) at (-0.0012545099, -0.4472009073) {};
\node[vertex] (v42) at (-0.1301496628, -0.3739274051) {};
\node[vertex] (v43) at (0.0326285522, -0.0618389445) {};
\draw[edge] (v0) -- (v3);
\draw[edge] (v3) -- (v1);
\draw[edge] (v2) -- (v4);
\draw[edge] (v5) -- (v12);
\draw[edge] (v12) -- (v7);
\draw[edge] (v13) -- (v6);
\draw[edge] (v7) -- (v14);
\draw[edge] (v9) -- (v10);
\draw[edge] (v9) -- (v11);
\draw[edge] (v15) -- (v18);
\draw[edge] (v19) -- (v16);
\draw[edge] (v17) -- (v20);
\draw[edge] (v27) -- (v5);
\draw[edge] (v0) -- (v28);
\draw[edge] (v28) -- (v2);
\draw[edge] (v2) -- (v22);
\draw[edge] (v23) -- (v11);
\draw[edge] (v11) -- (v24);
\draw[edge] (v24) -- (v15);
\draw[edge] (v15) -- (v25);
\draw[edge] (v7) -- (v9);
\draw[edge] (v12) -- (v29);
\draw[edge] (v29) -- (v2);
\draw[edge] (v30) -- (v21);
\draw[edge] (v9) -- (v24);
\draw[edge] (v31) -- (v12);
\draw[edge] (v32) -- (v2);
\draw[edge] (v33) -- (v8);
\draw[edge] (v33) -- (v32);
\draw[edge] (v26) -- (v34);
\draw[edge] (v34) -- (v0);
\draw[edge] (v35) -- (v16);
\draw[edge] (v9) -- (v36);
\draw[edge] (v37) -- (v14);
\draw[edge] (v28) -- (v9);
\draw[edge] (v14) -- (v38);
\draw[edge] (v38) -- (v8);
\draw[edge] (v39) -- (v30);
\draw[edge] (v39) -- (v38);
\draw[edge] (v39) -- (v6);
\draw[edge] (v39) -- (v23);
\draw[edge] (v8) -- (v17);
\draw[edge] (v19) -- (v12);
\draw[edge] (v24) -- (v23);
\draw[edge] (v0) -- (v40);
\draw[edge] (v40) -- (v1);
\draw[edge] (v41) -- (v16);
\draw[edge] (v2) -- (v30);
\draw[edge] (v25) -- (v42);
\draw[edge] (v42) -- (v16);
\draw[edge] (v43) -- (v37);
\draw[edge] (v43) -- (v2);
\draw[edge] (v43) -- (v35);
\fi}%
  \else%
    \pgfimage[width=0.25\textwidth]{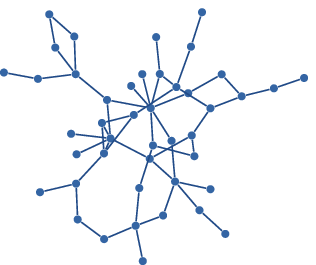}%
  \fi%
&
  \ifpdf%
    \tikzsetnextfilename{cache/north}%
    \tikz[x=0.25\textwidth, y=0.25\textwidth, ]{
\iftikzgraphpreamble
\def\aspectratio{1.0190868096}
\else
\node[vertex] (v0) at (-0.4906353367, 0.1080136901) {};
\node[vertex] (v1) at (0.0000000000, 0.0000000000) {};
\node[vertex] (v2) at (0.4604211796, 0.2010033042) {};
\node[vertex] (v3) at (-0.3334434714, 0.3757731261) {};
\node[vertex] (v4) at (-0.2543417809, -0.4332438682) {};
\node[vertex] (v5) at (-0.4604211796, -0.2010033042) {};
\node[vertex] (v6) at (0.3334434714, -0.3757731261) {};
\node[vertex] (v7) at (0.0488875332, -0.5000000000) {};
\node[vertex] (v8) at (-0.0488875332, 0.5000000000) {};
\node[vertex] (v9) at (0.4906353367, -0.1080136901) {};
\node[vertex] (v10) at (0.2543417809, 0.4332438682) {};
\draw[edge] (v0) -- (v1);
\draw[edge] (v1) -- (v2);
\draw[edge] (v1) -- (v3);
\draw[edge] (v1) -- (v4);
\draw[edge] (v1) -- (v5);
\draw[edge] (v1) -- (v6);
\draw[edge] (v1) -- (v7);
\draw[edge] (v1) -- (v8);
\draw[edge] (v1) -- (v9);
\draw[edge] (v1) -- (v10);
\fi}%
  \else%
    \pgfimage[width=0.25\textwidth]{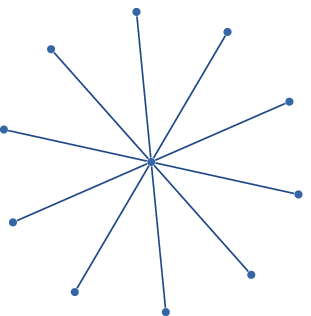}%
  \fi%
&
  \ifpdf%
    \tikzsetnextfilename{cache/randdag}%
    \tikz[x=0.25\textwidth, y=0.25\textwidth, ]{
\iftikzgraphpreamble
\def\aspectratio{0.8758730479}
\else
\node[vertex] (v0) at (0.0819623630, 0.1556251343) {};
\node[vertex] (v1) at (-0.0709662388, 0.1672688527) {};
\node[vertex] (v2) at (0.1143537431, 0.2001433753) {};
\node[vertex] (v3) at (0.1762052051, 0.1462640981) {};
\node[vertex] (v4) at (0.0100901887, 0.2489249353) {};
\node[vertex] (v5) at (0.0070286220, 0.0466374938) {};
\node[vertex] (v6) at (-0.1324055234, -0.0623077901) {};
\node[vertex] (v7) at (0.0051649662, -0.0888308028) {};
\node[vertex] (v8) at (-0.0505153781, 0.0859667930) {};
\node[vertex] (v9) at (0.2981676518, 0.0259168252) {};
\node[vertex] (v10) at (0.1903532528, -0.1335941441) {};
\node[vertex] (v11) at (0.2076348009, 0.2946794167) {};
\node[vertex] (v12) at (0.0428938971, -0.1494405946) {};
\node[vertex] (v13) at (0.1793897805, -0.1967732374) {};
\node[vertex] (v14) at (0.2898578278, -0.2849699410) {};
\node[vertex] (v15) at (0.0719812520, 0.0124425134) {};
\node[vertex] (v16) at (-0.0678709596, -0.1287361425) {};
\node[vertex] (v17) at (-0.1395466560, -0.2459044633) {};
\node[vertex] (v18) at (-0.0641132063, -0.2797651334) {};
\node[vertex] (v19) at (-0.0418512272, -0.3581734243) {};
\node[vertex] (v20) at (0.1522689118, -0.2976676566) {};
\node[vertex] (v21) at (0.4314429544, -0.0692712686) {};
\node[vertex] (v22) at (0.3374682637, -0.2116415821) {};
\node[vertex] (v23) at (0.2338242354, 0.0013763506) {};
\node[vertex] (v24) at (-0.2172839999, 0.2462805780) {};
\node[vertex] (v25) at (-0.3220203776, 0.0454606025) {};
\node[vertex] (v26) at (-0.2041528132, -0.1587334045) {};
\node[vertex] (v27) at (0.0806360666, -0.2504601145) {};
\node[vertex] (v28) at (0.1527436120, -0.4103530998) {};
\node[vertex] (v29) at (0.0147515110, -0.2837986715) {};
\node[vertex] (v30) at (-0.3082185263, -0.2576900007) {};
\node[vertex] (v31) at (-0.4007120601, -0.0948843715) {};
\node[vertex] (v32) at (-0.2607771650, 0.0547119744) {};
\node[vertex] (v33) at (-0.1905142273, 0.0541030783) {};
\node[vertex] (v34) at (-0.3395758864, 0.1902176832) {};
\node[vertex] (v35) at (0.2620820176, -0.0786848703) {};
\node[vertex] (v36) at (-0.2619443286, 0.1893178539) {};
\node[vertex] (v37) at (-0.4188916353, 0.0617958249) {};
\node[vertex] (v38) at (-0.3301911605, -0.1045515537) {};
\node[vertex] (v39) at (0.1255212451, 0.3764970497) {};
\node[vertex] (v40) at (0.2203680738, 0.4622976789) {};
\node[vertex] (v41) at (-0.0168220541, 0.1996223483) {};
\node[vertex] (v42) at (0.5483796128, 0.0356367637) {};
\node[vertex] (v43) at (0.4798455489, 0.1890596487) {};
\node[vertex] (v44) at (0.3087076414, 0.2213836809) {};
\node[vertex] (v45) at (-0.2184007288, 0.4152496863) {};
\node[vertex] (v46) at (-0.0637692549, 0.4117974256) {};
\node[vertex] (v47) at (0.3192152420, 0.4103992979) {};
\node[vertex] (v48) at (-0.2328559060, -0.4008609890) {};
\node[vertex] (v49) at (-0.4030805524, -0.4135753689) {};
\node[vertex] (v50) at (-0.4516203872, -0.2486607713) {};
\node[vertex] (v51) at (-0.1342382346, 0.2602524329) {};
\draw[edge] (v3) -- (v0);
\draw[edge] (v4) -- (v0);
\draw[edge] (v5) -- (v0);
\draw[edge] (v3) -- (v1);
\draw[edge] (v1) -- (v4);
\draw[edge] (v5) -- (v1);
\draw[edge] (v3) -- (v2);
\draw[edge] (v2) -- (v4);
\draw[edge] (v5) -- (v2);
\draw[edge] (v6) -- (v5);
\draw[edge] (v7) -- (v6);
\draw[edge] (v8) -- (v7);
\draw[edge] (v8) -- (v2);
\draw[edge] (v3) -- (v9);
\draw[edge] (v10) -- (v9);
\draw[edge] (v10) -- (v7);
\draw[edge] (v11) -- (v2);
\draw[edge] (v3) -- (v11);
\draw[edge] (v12) -- (v5);
\draw[edge] (v13) -- (v12);
\draw[edge] (v13) -- (v14);
\draw[edge] (v14) -- (v10);
\draw[edge] (v15) -- (v4);
\draw[edge] (v15) -- (v16);
\draw[edge] (v16) -- (v6);
\draw[edge] (v16) -- (v17);
\draw[edge] (v12) -- (v17);
\draw[edge] (v8) -- (v1);
\draw[edge] (v18) -- (v7);
\draw[edge] (v19) -- (v18);
\draw[edge] (v20) -- (v19);
\draw[edge] (v12) -- (v20);
\draw[edge] (v9) -- (v21);
\draw[edge] (v21) -- (v22);
\draw[edge] (v20) -- (v22);
\draw[edge] (v3) -- (v23);
\draw[edge] (v23) -- (v13);
\draw[edge] (v24) -- (v1);
\draw[edge] (v25) -- (v24);
\draw[edge] (v26) -- (v25);
\draw[edge] (v26) -- (v6);
\draw[edge] (v27) -- (v10);
\draw[edge] (v28) -- (v27);
\draw[edge] (v20) -- (v28);
\draw[edge] (v13) -- (v29);
\draw[edge] (v29) -- (v16);
\draw[edge] (v27) -- (v16);
\draw[edge] (v17) -- (v30);
\draw[edge] (v30) -- (v31);
\draw[edge] (v31) -- (v32);
\draw[edge] (v32) -- (v8);
\draw[edge] (v33) -- (v16);
\draw[edge] (v34) -- (v33);
\draw[edge] (v24) -- (v34);
\draw[edge] (v22) -- (v35);
\draw[edge] (v35) -- (v15);
\draw[edge] (v19) -- (v26);
\draw[edge] (v36) -- (v1);
\draw[edge] (v37) -- (v36);
\draw[edge] (v38) -- (v37);
\draw[edge] (v38) -- (v6);
\draw[edge] (v39) -- (v2);
\draw[edge] (v40) -- (v39);
\draw[edge] (v11) -- (v40);
\draw[edge] (v5) -- (v41);
\draw[edge] (v41) -- (v0);
\draw[edge] (v21) -- (v42);
\draw[edge] (v42) -- (v43);
\draw[edge] (v43) -- (v44);
\draw[edge] (v44) -- (v2);
\draw[edge] (v24) -- (v45);
\draw[edge] (v45) -- (v46);
\draw[edge] (v46) -- (v4);
\draw[edge] (v11) -- (v47);
\draw[edge] (v47) -- (v40);
\draw[edge] (v17) -- (v48);
\draw[edge] (v48) -- (v49);
\draw[edge] (v49) -- (v50);
\draw[edge] (v50) -- (v38);
\draw[edge] (v51) -- (v4);
\draw[edge] (v33) -- (v51);
\fi}%
  \else%
    \pgfimage[width=0.25\textwidth]{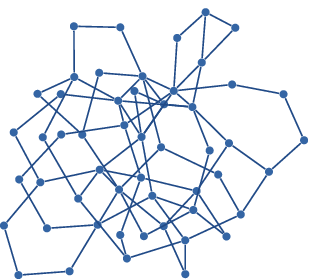}%
  \fi%
\\[2ex]
      \enum{ROME} & \enum{NORTH} & \enum{RANDDAG}
    \end{tabularx}
    \par\vspace{1cm}
    \begin{tabularx}{\linewidth}{%
        >{\centering\arraybackslash}X
        >{\centering\arraybackslash}X
      }
  \ifpdf%
    \tikzsetnextfilename{cache/bcspwr}%
    \tikz[x=0.35\textwidth, y=0.35\textwidth, ]{\input{pics/bcspwr.tikz}}%
  \else%
    \pgfimage[width=0.35\textwidth]{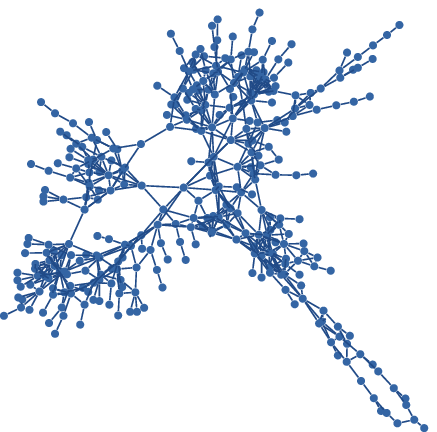}%
  \fi%
&
  \ifpdf%
    \tikzsetnextfilename{cache/grenoble}%
    \tikz[x=0.35\textwidth, y=0.35\textwidth, ]{\input{pics/grenoble.tikz}}%
  \else%
    \pgfimage[width=0.35\textwidth]{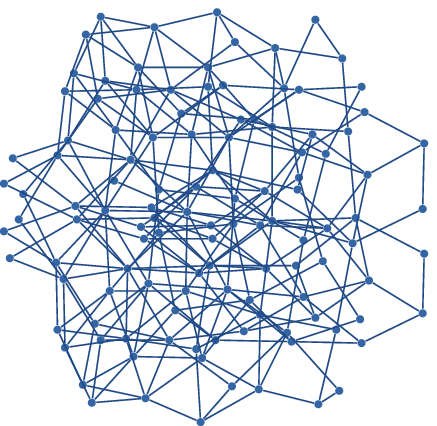}%
  \fi%
\\[2ex]
      \enum{BCSPWR} & \enum{GRENOBLE}
    \end{tabularx}
    \par\vspace{1cm}
    \begin{tabularx}{\linewidth}{%
        >{\centering\arraybackslash}X
        >{\centering\arraybackslash}X
      }
  \ifpdf%
    \tikzsetnextfilename{cache/psadmit}%
    \tikz[x=0.35\textwidth, y=0.35\textwidth, ]{\input{pics/psadmit.tikz}}%
  \else%
    \pgfimage[width=0.35\textwidth]{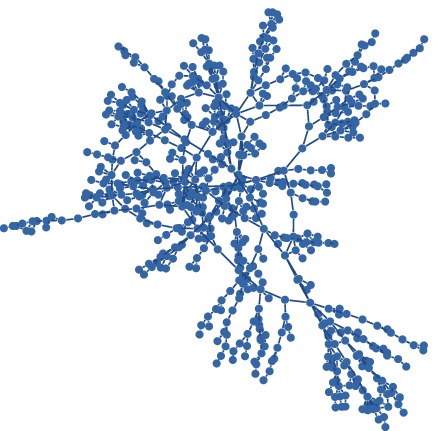}%
  \fi%
&
  \ifpdf%
    \tikzsetnextfilename{cache/smtape}%
    \tikz[x=0.35\textwidth, y=0.35\textwidth, ]{
\iftikzgraphpreamble
\def\aspectratio{0.9747747748}
\else
\node[vertex] (v0) at (-0.1426195426, 0.0073458073) {};
\node[vertex] (v1) at (-0.1065835066, -0.0845460845) {};
\node[vertex] (v2) at (-0.1011781012, -0.0971586972) {};
\node[vertex] (v3) at (-0.0579348579, -0.0142758143) {};
\node[vertex] (v4) at (-0.1191961192, -0.0124740125) {};
\node[vertex] (v5) at (-0.0921690922, -0.1800415800) {};
\node[vertex] (v6) at (-0.1534303534, -0.1710325710) {};
\node[vertex] (v7) at (-0.1696465696, -0.0106722107) {};
\node[vertex] (v8) at (0.0159390159, -0.0809424809) {};
\node[vertex] (v9) at (0.0033264033, -0.0989604990) {};
\node[vertex] (v10) at (-0.2218988219, -0.1386001386) {};
\node[vertex] (v11) at (-0.2759528760, 0.1532917533) {};
\node[vertex] (v12) at (-0.2218988219, 0.0866250866) {};
\node[vertex] (v13) at (-0.2363132363, 0.1785169785) {};
\node[vertex] (v14) at (-0.1984753985, -0.3422037422) {};
\node[vertex] (v15) at (-0.1696465696, -0.2539154539) {};
\node[vertex] (v16) at (-0.1480249480, -0.2611226611) {};
\node[vertex] (v17) at (-0.2399168399, -0.2142758143) {};
\node[vertex] (v18) at (-0.1480249480, -0.3584199584) {};
\node[vertex] (v19) at (-0.0471240471, 0.0704088704) {};
\node[vertex] (v20) at (-0.1552321552, -0.1259875260) {};
\node[vertex] (v21) at (-0.0705474705, -0.1404019404) {};
\node[vertex] (v22) at (-0.1606375606, -0.2340956341) {};
\node[vertex] (v23) at (0.2591822592, 0.1839223839) {};
\node[vertex] (v24) at (0.1762993763, 0.1586971587) {};
\node[vertex] (v25) at (0.0862092862, 0.0217602218) {};
\node[vertex] (v26) at (0.2591822592, 0.2956340956) {};
\node[vertex] (v27) at (0.4501732502, 0.3082467082) {};
\node[vertex] (v28) at (0.3582813583, 0.2974358974) {};
\node[vertex] (v29) at (0.2501732502, 0.5010395010) {};
\node[vertex] (v30) at (0.2970200970, 0.4199584200) {};
\node[vertex] (v31) at (0.3204435204, 0.4037422037) {};
\node[vertex] (v32) at (0.3456687457, 0.4812196812) {};
\node[vertex] (v33) at (0.3871101871, 0.4415800416) {};
\node[vertex] (v34) at (0.3546777547, 0.5172557173) {};
\node[vertex] (v35) at (0.2483714484, 0.3713097713) {};
\node[vertex] (v36) at (0.4303534304, 0.5064449064) {};
\node[vertex] (v37) at (0.3961191961, 0.5767151767) {};
\node[vertex] (v38) at (0.3204435204, 0.6073458073) {};
\node[vertex] (v39) at (0.3762993763, 0.6109494109) {};
\node[vertex] (v40) at (0.4411642412, 0.5659043659) {};
\node[vertex] (v41) at (0.3979209979, 0.6091476091) {};
\node[vertex] (v42) at (0.1943173943, -0.1313929314) {};
\node[vertex] (v43) at (0.1006237006, -0.1548163548) {};
\node[vertex] (v44) at (0.0790020790, -0.1728343728) {};
\node[vertex] (v45) at (0.1366597367, -0.2286902287) {};
\node[vertex] (v46) at (0.1690921691, -0.1836451836) {};
\node[vertex] (v47) at (0.1582813583, -0.2647262647) {};
\node[vertex] (v48) at (0.2249480249, -0.2232848233) {};
\node[vertex] (v49) at (0.2537768538, -0.2719334719) {};
\node[vertex] (v50) at (0.2303534304, -0.3223839224) {};
\node[vertex] (v51) at (0.1726957727, -0.3638253638) {};
\node[vertex] (v52) at (0.2159390159, -0.3404019404) {};
\node[vertex] (v53) at (0.1456687457, -0.3620235620) {};
\node[vertex] (v54) at (-0.3354123354, -0.1836451836) {};
\node[vertex] (v55) at (-0.3300069300, -0.1998613999) {};
\node[vertex] (v56) at (-0.4128898129, -0.2052668053) {};
\node[vertex] (v57) at (-0.3840609841, -0.2611226611) {};
\node[vertex] (v58) at (-0.4471240471, -0.2178794179) {};
\node[vertex] (v59) at (-0.4561330561, -0.2827442827) {};
\node[vertex] (v60) at (-0.5209979210, -0.2503118503) {};
\node[vertex] (v61) at (-0.5282051282, -0.1782397782) {};
\node[vertex] (v62) at (-0.5498267498, -0.2449064449) {};
\node[vertex] (v63) at (-0.5119889120, -0.2917532918) {};
\node[vertex] (v64) at (-0.5426195426, -0.2196812197) {};
\draw[edge] (v0) -- (v1);
\draw[edge] (v1) -- (v2);
\draw[edge] (v1) -- (v3);
\draw[edge] (v1) -- (v4);
\draw[edge] (v1) -- (v5);
\draw[edge] (v1) -- (v6);
\draw[edge] (v1) -- (v7);
\draw[edge] (v1) -- (v8);
\draw[edge] (v1) -- (v9);
\draw[edge] (v1) -- (v10);
\draw[edge] (v2) -- (v3);
\draw[edge] (v2) -- (v4);
\draw[edge] (v2) -- (v5);
\draw[edge] (v2) -- (v6);
\draw[edge] (v2) -- (v7);
\draw[edge] (v2) -- (v8);
\draw[edge] (v2) -- (v9);
\draw[edge] (v2) -- (v10);
\draw[edge] (v11) -- (v12);
\draw[edge] (v12) -- (v13);
\draw[edge] (v7) -- (v12);
\draw[edge] (v14) -- (v15);
\draw[edge] (v15) -- (v16);
\draw[edge] (v5) -- (v15);
\draw[edge] (v6) -- (v15);
\draw[edge] (v15) -- (v17);
\draw[edge] (v16) -- (v18);
\draw[edge] (v5) -- (v16);
\draw[edge] (v6) -- (v16);
\draw[edge] (v16) -- (v17);
\draw[edge] (v3) -- (v4);
\draw[edge] (v3) -- (v19);
\draw[edge] (v20) -- (v21);
\draw[edge] (v5) -- (v21);
\draw[edge] (v6) -- (v21);
\draw[edge] (v5) -- (v6);
\draw[edge] (v5) -- (v9);
\draw[edge] (v6) -- (v22);
\draw[edge] (v6) -- (v17);
\draw[edge] (v5) -- (v22);
\draw[edge] (v23) -- (v24);
\draw[edge] (v24) -- (v25);
\draw[edge] (v24) -- (v26);
\draw[edge] (v27) -- (v28);
\draw[edge] (v26) -- (v28);
\draw[edge] (v8) -- (v25);
\draw[edge] (v9) -- (v25);
\draw[edge] (v29) -- (v30);
\draw[edge] (v30) -- (v31);
\draw[edge] (v30) -- (v32);
\draw[edge] (v30) -- (v33);
\draw[edge] (v30) -- (v34);
\draw[edge] (v26) -- (v30);
\draw[edge] (v30) -- (v35);
\draw[edge] (v31) -- (v32);
\draw[edge] (v31) -- (v33);
\draw[edge] (v31) -- (v34);
\draw[edge] (v26) -- (v31);
\draw[edge] (v31) -- (v35);
\draw[edge] (v32) -- (v33);
\draw[edge] (v32) -- (v36);
\draw[edge] (v32) -- (v34);
\draw[edge] (v33) -- (v36);
\draw[edge] (v33) -- (v34);
\draw[edge] (v36) -- (v37);
\draw[edge] (v34) -- (v36);
\draw[edge] (v37) -- (v38);
\draw[edge] (v34) -- (v37);
\draw[edge] (v38) -- (v39);
\draw[edge] (v34) -- (v38);
\draw[edge] (v39) -- (v40);
\draw[edge] (v34) -- (v39);
\draw[edge] (v40) -- (v41);
\draw[edge] (v34) -- (v40);
\draw[edge] (v34) -- (v41);
\draw[edge] (v26) -- (v35);
\draw[edge] (v42) -- (v43);
\draw[edge] (v43) -- (v44);
\draw[edge] (v43) -- (v45);
\draw[edge] (v43) -- (v46);
\draw[edge] (v43) -- (v47);
\draw[edge] (v8) -- (v43);
\draw[edge] (v9) -- (v43);
\draw[edge] (v44) -- (v45);
\draw[edge] (v44) -- (v46);
\draw[edge] (v44) -- (v47);
\draw[edge] (v8) -- (v44);
\draw[edge] (v9) -- (v44);
\draw[edge] (v45) -- (v46);
\draw[edge] (v45) -- (v48);
\draw[edge] (v45) -- (v47);
\draw[edge] (v46) -- (v48);
\draw[edge] (v46) -- (v47);
\draw[edge] (v48) -- (v49);
\draw[edge] (v47) -- (v48);
\draw[edge] (v49) -- (v50);
\draw[edge] (v47) -- (v49);
\draw[edge] (v50) -- (v51);
\draw[edge] (v47) -- (v50);
\draw[edge] (v51) -- (v52);
\draw[edge] (v47) -- (v51);
\draw[edge] (v52) -- (v53);
\draw[edge] (v47) -- (v52);
\draw[edge] (v47) -- (v53);
\draw[edge] (v8) -- (v9);
\draw[edge] (v9) -- (v21);
\draw[edge] (v10) -- (v17);
\draw[edge] (v10) -- (v54);
\draw[edge] (v17) -- (v22);
\draw[edge] (v17) -- (v55);
\draw[edge] (v10) -- (v55);
\draw[edge] (v54) -- (v55);
\draw[edge] (v55) -- (v56);
\draw[edge] (v55) -- (v57);
\draw[edge] (v55) -- (v58);
\draw[edge] (v17) -- (v54);
\draw[edge] (v54) -- (v56);
\draw[edge] (v54) -- (v57);
\draw[edge] (v54) -- (v58);
\draw[edge] (v56) -- (v57);
\draw[edge] (v56) -- (v59);
\draw[edge] (v56) -- (v58);
\draw[edge] (v57) -- (v59);
\draw[edge] (v57) -- (v58);
\draw[edge] (v59) -- (v60);
\draw[edge] (v58) -- (v59);
\draw[edge] (v60) -- (v61);
\draw[edge] (v58) -- (v60);
\draw[edge] (v61) -- (v62);
\draw[edge] (v58) -- (v61);
\draw[edge] (v62) -- (v63);
\draw[edge] (v58) -- (v62);
\draw[edge] (v63) -- (v64);
\draw[edge] (v58) -- (v63);
\draw[edge] (v58) -- (v64);
\fi}%
  \else%
    \pgfimage[width=0.35\textwidth]{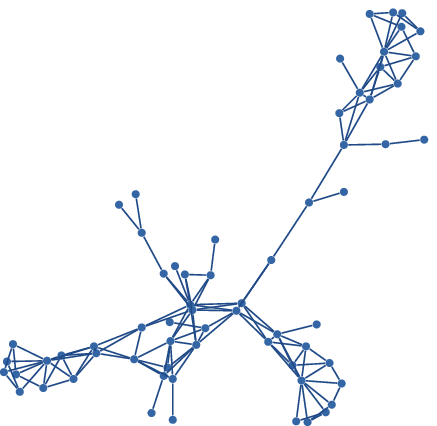}%
  \fi%
\\[2ex]
      \enum{PSADMIT} & \enum{SMTAPE}
    \end{tabularx}
  \end{center}
  \caption{%
    Examples of imported graphs.  The \enum{BCSPWR}, \enum{GRENOBLE}, \enum{PSADMIT} and \enum{SMTAPE} graphs come from the respective datasets in the Harwell-Boeing collection in NIST's \enquote{Matrix Market}~\cite{MatrixMarket}.  All graphs are visualized using the FM\textsuperscript{3} algorithm.
  }
  \label{app:fig:archives}
\end{figure}

\begin{figure}[bp]
  \begin{center}
    \begin{tabularx}{\linewidth}{%
        >{\centering\arraybackslash}X
        >{\centering\arraybackslash}X
        >{\centering\arraybackslash}X
        >{\centering\arraybackslash}X
      }
  \ifpdf%
    \tikzsetnextfilename{cache/grid}%
    \tikz[x=0.2\textwidth, y=0.2\textwidth, ]{
\iftikzgraphpreamble
\def\aspectratio{0.7000000000}
\else
\node[vertex] (v0) at (-0.5000000000, -0.3500000000) {};
\node[vertex] (v1) at (-0.4000000000, -0.3500000000) {};
\node[vertex] (v2) at (-0.3000000000, -0.3500000000) {};
\node[vertex] (v3) at (-0.2000000000, -0.3500000000) {};
\node[vertex] (v4) at (-0.1000000000, -0.3500000000) {};
\node[vertex] (v5) at (0.0000000000, -0.3500000000) {};
\node[vertex] (v6) at (0.1000000000, -0.3500000000) {};
\node[vertex] (v7) at (0.2000000000, -0.3500000000) {};
\node[vertex] (v8) at (0.3000000000, -0.3500000000) {};
\node[vertex] (v9) at (0.4000000000, -0.3500000000) {};
\node[vertex] (v10) at (0.5000000000, -0.3500000000) {};
\node[vertex] (v11) at (-0.5000000000, -0.2500000000) {};
\node[vertex] (v12) at (-0.4000000000, -0.2500000000) {};
\node[vertex] (v13) at (-0.3000000000, -0.2500000000) {};
\node[vertex] (v14) at (-0.2000000000, -0.2500000000) {};
\node[vertex] (v15) at (-0.1000000000, -0.2500000000) {};
\node[vertex] (v16) at (0.0000000000, -0.2500000000) {};
\node[vertex] (v17) at (0.1000000000, -0.2500000000) {};
\node[vertex] (v18) at (0.2000000000, -0.2500000000) {};
\node[vertex] (v19) at (0.3000000000, -0.2500000000) {};
\node[vertex] (v20) at (0.4000000000, -0.2500000000) {};
\node[vertex] (v21) at (0.5000000000, -0.2500000000) {};
\node[vertex] (v22) at (-0.5000000000, -0.1500000000) {};
\node[vertex] (v23) at (-0.4000000000, -0.1500000000) {};
\node[vertex] (v24) at (-0.3000000000, -0.1500000000) {};
\node[vertex] (v25) at (-0.2000000000, -0.1500000000) {};
\node[vertex] (v26) at (-0.1000000000, -0.1500000000) {};
\node[vertex] (v27) at (0.0000000000, -0.1500000000) {};
\node[vertex] (v28) at (0.1000000000, -0.1500000000) {};
\node[vertex] (v29) at (0.2000000000, -0.1500000000) {};
\node[vertex] (v30) at (0.3000000000, -0.1500000000) {};
\node[vertex] (v31) at (0.4000000000, -0.1500000000) {};
\node[vertex] (v32) at (0.5000000000, -0.1500000000) {};
\node[vertex] (v33) at (-0.5000000000, -0.0500000000) {};
\node[vertex] (v34) at (-0.4000000000, -0.0500000000) {};
\node[vertex] (v35) at (-0.3000000000, -0.0500000000) {};
\node[vertex] (v36) at (-0.2000000000, -0.0500000000) {};
\node[vertex] (v37) at (-0.1000000000, -0.0500000000) {};
\node[vertex] (v38) at (0.0000000000, -0.0500000000) {};
\node[vertex] (v39) at (0.1000000000, -0.0500000000) {};
\node[vertex] (v40) at (0.2000000000, -0.0500000000) {};
\node[vertex] (v41) at (0.3000000000, -0.0500000000) {};
\node[vertex] (v42) at (0.4000000000, -0.0500000000) {};
\node[vertex] (v43) at (0.5000000000, -0.0500000000) {};
\node[vertex] (v44) at (-0.5000000000, 0.0500000000) {};
\node[vertex] (v45) at (-0.4000000000, 0.0500000000) {};
\node[vertex] (v46) at (-0.3000000000, 0.0500000000) {};
\node[vertex] (v47) at (-0.2000000000, 0.0500000000) {};
\node[vertex] (v48) at (-0.1000000000, 0.0500000000) {};
\node[vertex] (v49) at (0.0000000000, 0.0500000000) {};
\node[vertex] (v50) at (0.1000000000, 0.0500000000) {};
\node[vertex] (v51) at (0.2000000000, 0.0500000000) {};
\node[vertex] (v52) at (0.3000000000, 0.0500000000) {};
\node[vertex] (v53) at (0.4000000000, 0.0500000000) {};
\node[vertex] (v54) at (0.5000000000, 0.0500000000) {};
\node[vertex] (v55) at (-0.5000000000, 0.1500000000) {};
\node[vertex] (v56) at (-0.4000000000, 0.1500000000) {};
\node[vertex] (v57) at (-0.3000000000, 0.1500000000) {};
\node[vertex] (v58) at (-0.2000000000, 0.1500000000) {};
\node[vertex] (v59) at (-0.1000000000, 0.1500000000) {};
\node[vertex] (v60) at (0.0000000000, 0.1500000000) {};
\node[vertex] (v61) at (0.1000000000, 0.1500000000) {};
\node[vertex] (v62) at (0.2000000000, 0.1500000000) {};
\node[vertex] (v63) at (0.3000000000, 0.1500000000) {};
\node[vertex] (v64) at (0.4000000000, 0.1500000000) {};
\node[vertex] (v65) at (0.5000000000, 0.1500000000) {};
\node[vertex] (v66) at (-0.5000000000, 0.2500000000) {};
\node[vertex] (v67) at (-0.4000000000, 0.2500000000) {};
\node[vertex] (v68) at (-0.3000000000, 0.2500000000) {};
\node[vertex] (v69) at (-0.2000000000, 0.2500000000) {};
\node[vertex] (v70) at (-0.1000000000, 0.2500000000) {};
\node[vertex] (v71) at (0.0000000000, 0.2500000000) {};
\node[vertex] (v72) at (0.1000000000, 0.2500000000) {};
\node[vertex] (v73) at (0.2000000000, 0.2500000000) {};
\node[vertex] (v74) at (0.3000000000, 0.2500000000) {};
\node[vertex] (v75) at (0.4000000000, 0.2500000000) {};
\node[vertex] (v76) at (0.5000000000, 0.2500000000) {};
\node[vertex] (v77) at (-0.5000000000, 0.3500000000) {};
\node[vertex] (v78) at (-0.4000000000, 0.3500000000) {};
\node[vertex] (v79) at (-0.3000000000, 0.3500000000) {};
\node[vertex] (v80) at (-0.2000000000, 0.3500000000) {};
\node[vertex] (v81) at (-0.1000000000, 0.3500000000) {};
\node[vertex] (v82) at (0.0000000000, 0.3500000000) {};
\node[vertex] (v83) at (0.1000000000, 0.3500000000) {};
\node[vertex] (v84) at (0.2000000000, 0.3500000000) {};
\node[vertex] (v85) at (0.3000000000, 0.3500000000) {};
\node[vertex] (v86) at (0.4000000000, 0.3500000000) {};
\node[vertex] (v87) at (0.5000000000, 0.3500000000) {};
\draw[edge] (v0) -- (v1);
\draw[edge] (v1) -- (v2);
\draw[edge] (v2) -- (v3);
\draw[edge] (v3) -- (v4);
\draw[edge] (v4) -- (v5);
\draw[edge] (v5) -- (v6);
\draw[edge] (v6) -- (v7);
\draw[edge] (v7) -- (v8);
\draw[edge] (v8) -- (v9);
\draw[edge] (v9) -- (v10);
\draw[edge] (v0) -- (v11);
\draw[edge] (v1) -- (v12);
\draw[edge] (v2) -- (v13);
\draw[edge] (v3) -- (v14);
\draw[edge] (v4) -- (v15);
\draw[edge] (v5) -- (v16);
\draw[edge] (v6) -- (v17);
\draw[edge] (v7) -- (v18);
\draw[edge] (v8) -- (v19);
\draw[edge] (v9) -- (v20);
\draw[edge] (v10) -- (v21);
\draw[edge] (v11) -- (v12);
\draw[edge] (v12) -- (v13);
\draw[edge] (v13) -- (v14);
\draw[edge] (v14) -- (v15);
\draw[edge] (v15) -- (v16);
\draw[edge] (v16) -- (v17);
\draw[edge] (v17) -- (v18);
\draw[edge] (v18) -- (v19);
\draw[edge] (v19) -- (v20);
\draw[edge] (v20) -- (v21);
\draw[edge] (v11) -- (v22);
\draw[edge] (v12) -- (v23);
\draw[edge] (v13) -- (v24);
\draw[edge] (v14) -- (v25);
\draw[edge] (v15) -- (v26);
\draw[edge] (v16) -- (v27);
\draw[edge] (v17) -- (v28);
\draw[edge] (v18) -- (v29);
\draw[edge] (v19) -- (v30);
\draw[edge] (v20) -- (v31);
\draw[edge] (v21) -- (v32);
\draw[edge] (v22) -- (v23);
\draw[edge] (v23) -- (v24);
\draw[edge] (v24) -- (v25);
\draw[edge] (v25) -- (v26);
\draw[edge] (v26) -- (v27);
\draw[edge] (v27) -- (v28);
\draw[edge] (v28) -- (v29);
\draw[edge] (v29) -- (v30);
\draw[edge] (v30) -- (v31);
\draw[edge] (v31) -- (v32);
\draw[edge] (v22) -- (v33);
\draw[edge] (v23) -- (v34);
\draw[edge] (v24) -- (v35);
\draw[edge] (v25) -- (v36);
\draw[edge] (v26) -- (v37);
\draw[edge] (v27) -- (v38);
\draw[edge] (v28) -- (v39);
\draw[edge] (v29) -- (v40);
\draw[edge] (v30) -- (v41);
\draw[edge] (v31) -- (v42);
\draw[edge] (v32) -- (v43);
\draw[edge] (v33) -- (v34);
\draw[edge] (v34) -- (v35);
\draw[edge] (v35) -- (v36);
\draw[edge] (v36) -- (v37);
\draw[edge] (v37) -- (v38);
\draw[edge] (v38) -- (v39);
\draw[edge] (v39) -- (v40);
\draw[edge] (v40) -- (v41);
\draw[edge] (v41) -- (v42);
\draw[edge] (v42) -- (v43);
\draw[edge] (v33) -- (v44);
\draw[edge] (v34) -- (v45);
\draw[edge] (v35) -- (v46);
\draw[edge] (v36) -- (v47);
\draw[edge] (v37) -- (v48);
\draw[edge] (v38) -- (v49);
\draw[edge] (v39) -- (v50);
\draw[edge] (v40) -- (v51);
\draw[edge] (v41) -- (v52);
\draw[edge] (v42) -- (v53);
\draw[edge] (v43) -- (v54);
\draw[edge] (v44) -- (v45);
\draw[edge] (v45) -- (v46);
\draw[edge] (v46) -- (v47);
\draw[edge] (v47) -- (v48);
\draw[edge] (v48) -- (v49);
\draw[edge] (v49) -- (v50);
\draw[edge] (v50) -- (v51);
\draw[edge] (v51) -- (v52);
\draw[edge] (v52) -- (v53);
\draw[edge] (v53) -- (v54);
\draw[edge] (v44) -- (v55);
\draw[edge] (v45) -- (v56);
\draw[edge] (v46) -- (v57);
\draw[edge] (v47) -- (v58);
\draw[edge] (v48) -- (v59);
\draw[edge] (v49) -- (v60);
\draw[edge] (v50) -- (v61);
\draw[edge] (v51) -- (v62);
\draw[edge] (v52) -- (v63);
\draw[edge] (v53) -- (v64);
\draw[edge] (v54) -- (v65);
\draw[edge] (v55) -- (v56);
\draw[edge] (v56) -- (v57);
\draw[edge] (v57) -- (v58);
\draw[edge] (v58) -- (v59);
\draw[edge] (v59) -- (v60);
\draw[edge] (v60) -- (v61);
\draw[edge] (v61) -- (v62);
\draw[edge] (v62) -- (v63);
\draw[edge] (v63) -- (v64);
\draw[edge] (v64) -- (v65);
\draw[edge] (v55) -- (v66);
\draw[edge] (v56) -- (v67);
\draw[edge] (v57) -- (v68);
\draw[edge] (v58) -- (v69);
\draw[edge] (v59) -- (v70);
\draw[edge] (v60) -- (v71);
\draw[edge] (v61) -- (v72);
\draw[edge] (v62) -- (v73);
\draw[edge] (v63) -- (v74);
\draw[edge] (v64) -- (v75);
\draw[edge] (v65) -- (v76);
\draw[edge] (v66) -- (v67);
\draw[edge] (v67) -- (v68);
\draw[edge] (v68) -- (v69);
\draw[edge] (v69) -- (v70);
\draw[edge] (v70) -- (v71);
\draw[edge] (v71) -- (v72);
\draw[edge] (v72) -- (v73);
\draw[edge] (v73) -- (v74);
\draw[edge] (v74) -- (v75);
\draw[edge] (v75) -- (v76);
\draw[edge] (v66) -- (v77);
\draw[edge] (v67) -- (v78);
\draw[edge] (v68) -- (v79);
\draw[edge] (v69) -- (v80);
\draw[edge] (v70) -- (v81);
\draw[edge] (v71) -- (v82);
\draw[edge] (v72) -- (v83);
\draw[edge] (v73) -- (v84);
\draw[edge] (v74) -- (v85);
\draw[edge] (v75) -- (v86);
\draw[edge] (v76) -- (v87);
\draw[edge] (v77) -- (v78);
\draw[edge] (v78) -- (v79);
\draw[edge] (v79) -- (v80);
\draw[edge] (v80) -- (v81);
\draw[edge] (v81) -- (v82);
\draw[edge] (v82) -- (v83);
\draw[edge] (v83) -- (v84);
\draw[edge] (v84) -- (v85);
\draw[edge] (v85) -- (v86);
\draw[edge] (v86) -- (v87);
\fi}%
  \else%
    \pgfimage[width=0.2\textwidth]{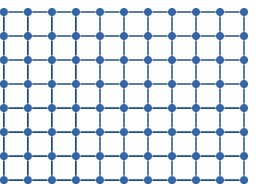}%
  \fi%
&
  \ifpdf%
    \tikzsetnextfilename{cache/torus1}%
    \tikz[x=0.2\textwidth, y=0.2\textwidth, ]{
\iftikzgraphpreamble
\def\aspectratio{1.4782318251}
\else
\node[vertex] (v0) at (-0.1508456200, -0.4205765746) {};
\node[vertex] (v1) at (-0.2770588307, -0.4087990813) {};
\node[vertex] (v2) at (-0.3370742583, -0.4300457216) {};
\node[vertex] (v3) at (-0.2215288838, -0.4771009791) {};
\node[vertex] (v4) at (-0.0873116298, -0.4949542423) {};
\node[vertex] (v5) at (0.0491289885, -0.5001892698) {};
\node[vertex] (v6) at (0.1843196863, -0.4926728548) {};
\node[vertex] (v7) at (0.3031311861, -0.4546099033) {};
\node[vertex] (v8) at (0.2449211066, -0.4288262806) {};
\node[vertex] (v9) at (0.1181766520, -0.4308984009) {};
\node[vertex] (v10) at (-0.0163995224, -0.4274326652) {};
\node[vertex] (v11) at (-0.1455192854, -0.2901395998) {};
\node[vertex] (v12) at (-0.2730806416, -0.2831194307) {};
\node[vertex] (v13) at (-0.3328668453, -0.3085990164) {};
\node[vertex] (v14) at (-0.2153861865, -0.3484469032) {};
\node[vertex] (v15) at (-0.0813873607, -0.3668374123) {};
\node[vertex] (v16) at (0.0530390209, -0.3719951592) {};
\node[vertex] (v17) at (0.1880531006, -0.3639263169) {};
\node[vertex] (v18) at (0.3082420158, -0.3331976757) {};
\node[vertex] (v19) at (0.2505846788, -0.3032114238) {};
\node[vertex] (v20) at (0.1228605667, -0.3004367946) {};
\node[vertex] (v21) at (-0.0113705691, -0.2963622326) {};
\node[vertex] (v22) at (-0.1417161113, -0.1582758139) {};
\node[vertex] (v23) at (-0.2704299917, -0.1563888634) {};
\node[vertex] (v24) at (-0.3299444229, -0.1846767393) {};
\node[vertex] (v25) at (-0.2127916909, -0.2169608302) {};
\node[vertex] (v26) at (-0.0769338174, -0.2330776296) {};
\node[vertex] (v27) at (0.0588477687, -0.2382873665) {};
\node[vertex] (v28) at (0.1955412194, -0.2326279633) {};
\node[vertex] (v29) at (0.3148236236, -0.2094156901) {};
\node[vertex] (v30) at (0.2576523549, -0.1766504141) {};
\node[vertex] (v31) at (0.1291724512, -0.1686692957) {};
\node[vertex] (v32) at (-0.0062546297, -0.1630242788) {};
\node[vertex] (v33) at (-0.1373053754, -0.0267143706) {};
\node[vertex] (v34) at (-0.2666758873, -0.0293490182) {};
\node[vertex] (v35) at (-0.3264195661, -0.0600639183) {};
\node[vertex] (v36) at (-0.2096852114, -0.0866011426) {};
\node[vertex] (v37) at (-0.0725119012, -0.0996478500) {};
\node[vertex] (v38) at (0.0646627425, -0.1049110252) {};
\node[vertex] (v39) at (0.2024325256, -0.1024134534) {};
\node[vertex] (v40) at (0.3208574400, -0.0848990712) {};
\node[vertex] (v41) at (0.2636435084, -0.0496964547) {};
\node[vertex] (v42) at (0.1348553700, -0.0371566932) {};
\node[vertex] (v43) at (-0.0011457715, -0.0298705847) {};
\node[vertex] (v44) at (-0.1318245988, 0.1050857189) {};
\node[vertex] (v45) at (-0.2613459783, 0.0981341528) {};
\node[vertex] (v46) at (-0.3217530942, 0.0645000630) {};
\node[vertex] (v47) at (-0.2053008478, 0.0427104572) {};
\node[vertex] (v48) at (-0.0676596407, 0.0328787817) {};
\node[vertex] (v49) at (0.0699794259, 0.0275977987) {};
\node[vertex] (v50) at (0.2079693833, 0.0268539623) {};
\node[vertex] (v51) at (0.3257492393, 0.0396563019) {};
\node[vertex] (v52) at (0.2680974603, 0.0778202831) {};
\node[vertex] (v53) at (0.1394896705, 0.0946758496) {};
\node[vertex] (v54) at (0.0039704779, 0.1034758971) {};
\node[vertex] (v55) at (-0.1253184720, 0.2378197187) {};
\node[vertex] (v56) at (-0.2543817578, 0.2263216368) {};
\node[vertex] (v57) at (-0.3160219122, 0.1885186340) {};
\node[vertex] (v58) at (-0.1996618757, 0.1715097016) {};
\node[vertex] (v59) at (-0.0623184864, 0.1651202763) {};
\node[vertex] (v60) at (0.0747868101, 0.1598597837) {};
\node[vertex] (v61) at (0.2122160638, 0.1557066586) {};
\node[vertex] (v62) at (0.3295376668, 0.1637494348) {};
\node[vertex] (v63) at (0.2709758813, 0.2061644920) {};
\node[vertex] (v64) at (0.1431731970, 0.2275181280) {};
\node[vertex] (v65) at (0.0091170640, 0.2376130397) {};
\node[vertex] (v66) at (-0.1186446953, 0.3712415935) {};
\node[vertex] (v67) at (-0.2463499394, 0.3553607867) {};
\node[vertex] (v68) at (-0.3100591779, 0.3112261184) {};
\node[vertex] (v69) at (-0.1934158137, 0.3002150002) {};
\node[vertex] (v70) at (-0.0567744151, 0.2974568221) {};
\node[vertex] (v71) at (0.0793991767, 0.2922320852) {};
\node[vertex] (v72) at (0.2158502589, 0.2845121983) {};
\node[vertex] (v73) at (0.3329947770, 0.2865530683) {};
\node[vertex] (v74) at (0.2728551170, 0.3354396650) {};
\node[vertex] (v75) at (0.1467422869, 0.3610591128) {};
\node[vertex] (v76) at (0.0142653610, 0.3717945954) {};
\node[vertex] (v77) at (-0.1150140456, 0.4998107302) {};
\node[vertex] (v78) at (-0.2412633480, 0.4829987856) {};
\node[vertex] (v79) at (-0.3073180844, 0.4306124857) {};
\node[vertex] (v80) at (-0.1896809623, 0.4277980662) {};
\node[vertex] (v81) at (-0.0523402763, 0.4276004338) {};
\node[vertex] (v82) at (0.0849502118, 0.4223328428) {};
\node[vertex] (v83) at (0.2219023355, 0.4120063670) {};
\node[vertex] (v84) at (0.3394096213, 0.4057984871) {};
\node[vertex] (v85) at (0.2775635998, 0.4630921347) {};
\node[vertex] (v86) at (0.1529737812, 0.4895284551) {};
\node[vertex] (v87) at (0.0191765584, 0.4997958005) {};
\draw[edge] (v0) -- (v1);
\draw[edge] (v1) -- (v2);
\draw[edge] (v2) -- (v3);
\draw[edge] (v3) -- (v4);
\draw[edge] (v4) -- (v5);
\draw[edge] (v5) -- (v6);
\draw[edge] (v6) -- (v7);
\draw[edge] (v7) -- (v8);
\draw[edge] (v8) -- (v9);
\draw[edge] (v9) -- (v10);
\draw[edge] (v10) -- (v0);
\draw[edge] (v0) -- (v11);
\draw[edge] (v1) -- (v12);
\draw[edge] (v2) -- (v13);
\draw[edge] (v3) -- (v14);
\draw[edge] (v4) -- (v15);
\draw[edge] (v5) -- (v16);
\draw[edge] (v6) -- (v17);
\draw[edge] (v7) -- (v18);
\draw[edge] (v8) -- (v19);
\draw[edge] (v9) -- (v20);
\draw[edge] (v10) -- (v21);
\draw[edge] (v11) -- (v12);
\draw[edge] (v12) -- (v13);
\draw[edge] (v13) -- (v14);
\draw[edge] (v14) -- (v15);
\draw[edge] (v15) -- (v16);
\draw[edge] (v16) -- (v17);
\draw[edge] (v17) -- (v18);
\draw[edge] (v18) -- (v19);
\draw[edge] (v19) -- (v20);
\draw[edge] (v20) -- (v21);
\draw[edge] (v21) -- (v11);
\draw[edge] (v11) -- (v22);
\draw[edge] (v12) -- (v23);
\draw[edge] (v13) -- (v24);
\draw[edge] (v14) -- (v25);
\draw[edge] (v15) -- (v26);
\draw[edge] (v16) -- (v27);
\draw[edge] (v17) -- (v28);
\draw[edge] (v18) -- (v29);
\draw[edge] (v19) -- (v30);
\draw[edge] (v20) -- (v31);
\draw[edge] (v21) -- (v32);
\draw[edge] (v22) -- (v23);
\draw[edge] (v23) -- (v24);
\draw[edge] (v24) -- (v25);
\draw[edge] (v25) -- (v26);
\draw[edge] (v26) -- (v27);
\draw[edge] (v27) -- (v28);
\draw[edge] (v28) -- (v29);
\draw[edge] (v29) -- (v30);
\draw[edge] (v30) -- (v31);
\draw[edge] (v31) -- (v32);
\draw[edge] (v32) -- (v22);
\draw[edge] (v22) -- (v33);
\draw[edge] (v23) -- (v34);
\draw[edge] (v24) -- (v35);
\draw[edge] (v25) -- (v36);
\draw[edge] (v26) -- (v37);
\draw[edge] (v27) -- (v38);
\draw[edge] (v28) -- (v39);
\draw[edge] (v29) -- (v40);
\draw[edge] (v30) -- (v41);
\draw[edge] (v31) -- (v42);
\draw[edge] (v32) -- (v43);
\draw[edge] (v33) -- (v34);
\draw[edge] (v34) -- (v35);
\draw[edge] (v35) -- (v36);
\draw[edge] (v36) -- (v37);
\draw[edge] (v37) -- (v38);
\draw[edge] (v38) -- (v39);
\draw[edge] (v39) -- (v40);
\draw[edge] (v40) -- (v41);
\draw[edge] (v41) -- (v42);
\draw[edge] (v42) -- (v43);
\draw[edge] (v43) -- (v33);
\draw[edge] (v33) -- (v44);
\draw[edge] (v34) -- (v45);
\draw[edge] (v35) -- (v46);
\draw[edge] (v36) -- (v47);
\draw[edge] (v37) -- (v48);
\draw[edge] (v38) -- (v49);
\draw[edge] (v39) -- (v50);
\draw[edge] (v40) -- (v51);
\draw[edge] (v41) -- (v52);
\draw[edge] (v42) -- (v53);
\draw[edge] (v43) -- (v54);
\draw[edge] (v44) -- (v45);
\draw[edge] (v45) -- (v46);
\draw[edge] (v46) -- (v47);
\draw[edge] (v47) -- (v48);
\draw[edge] (v48) -- (v49);
\draw[edge] (v49) -- (v50);
\draw[edge] (v50) -- (v51);
\draw[edge] (v51) -- (v52);
\draw[edge] (v52) -- (v53);
\draw[edge] (v53) -- (v54);
\draw[edge] (v54) -- (v44);
\draw[edge] (v44) -- (v55);
\draw[edge] (v45) -- (v56);
\draw[edge] (v46) -- (v57);
\draw[edge] (v47) -- (v58);
\draw[edge] (v48) -- (v59);
\draw[edge] (v49) -- (v60);
\draw[edge] (v50) -- (v61);
\draw[edge] (v51) -- (v62);
\draw[edge] (v52) -- (v63);
\draw[edge] (v53) -- (v64);
\draw[edge] (v54) -- (v65);
\draw[edge] (v55) -- (v56);
\draw[edge] (v56) -- (v57);
\draw[edge] (v57) -- (v58);
\draw[edge] (v58) -- (v59);
\draw[edge] (v59) -- (v60);
\draw[edge] (v60) -- (v61);
\draw[edge] (v61) -- (v62);
\draw[edge] (v62) -- (v63);
\draw[edge] (v63) -- (v64);
\draw[edge] (v64) -- (v65);
\draw[edge] (v65) -- (v55);
\draw[edge] (v55) -- (v66);
\draw[edge] (v56) -- (v67);
\draw[edge] (v57) -- (v68);
\draw[edge] (v58) -- (v69);
\draw[edge] (v59) -- (v70);
\draw[edge] (v60) -- (v71);
\draw[edge] (v61) -- (v72);
\draw[edge] (v62) -- (v73);
\draw[edge] (v63) -- (v74);
\draw[edge] (v64) -- (v75);
\draw[edge] (v65) -- (v76);
\draw[edge] (v66) -- (v67);
\draw[edge] (v67) -- (v68);
\draw[edge] (v68) -- (v69);
\draw[edge] (v69) -- (v70);
\draw[edge] (v70) -- (v71);
\draw[edge] (v71) -- (v72);
\draw[edge] (v72) -- (v73);
\draw[edge] (v73) -- (v74);
\draw[edge] (v74) -- (v75);
\draw[edge] (v75) -- (v76);
\draw[edge] (v76) -- (v66);
\draw[edge] (v66) -- (v77);
\draw[edge] (v67) -- (v78);
\draw[edge] (v68) -- (v79);
\draw[edge] (v69) -- (v80);
\draw[edge] (v70) -- (v81);
\draw[edge] (v71) -- (v82);
\draw[edge] (v72) -- (v83);
\draw[edge] (v73) -- (v84);
\draw[edge] (v74) -- (v85);
\draw[edge] (v75) -- (v86);
\draw[edge] (v76) -- (v87);
\draw[edge] (v77) -- (v78);
\draw[edge] (v78) -- (v79);
\draw[edge] (v79) -- (v80);
\draw[edge] (v80) -- (v81);
\draw[edge] (v81) -- (v82);
\draw[edge] (v82) -- (v83);
\draw[edge] (v83) -- (v84);
\draw[edge] (v84) -- (v85);
\draw[edge] (v85) -- (v86);
\draw[edge] (v86) -- (v87);
\draw[edge] (v87) -- (v77);
\fi}%
  \else%
    \pgfimage[width=0.2\textwidth]{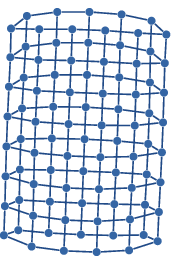}%
  \fi%
&
  \ifpdf%
    \tikzsetnextfilename{cache/torus2}%
    \tikz[x=0.2\textwidth, y=0.2\textwidth, ]{\input{pics/torus2.tikz}}%
  \else%
    \pgfimage[width=0.2\textwidth]{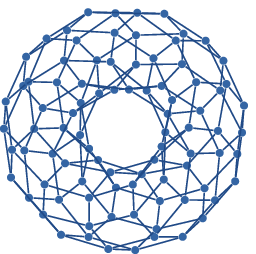}%
  \fi%
&
  \ifpdf%
    \tikzsetnextfilename{cache/bottle}%
    \tikz[x=0.2\textwidth, y=0.2\textwidth, ]{\input{pics/bottle.tikz}}%
  \else%
    \pgfimage[width=0.2\textwidth]{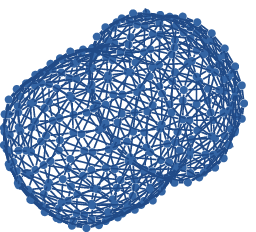}%
  \fi%
\\[2ex]
      \enum{GRID} & \enum{TORUS1} & \enum{TORUS2} & \enum{BOTTLE}
    \end{tabularx}
    \par\vspace{1cm}
    \begin{tabularx}{\linewidth}{%
        >{\centering\arraybackslash}X
        >{\centering\arraybackslash}X
        >{\centering\arraybackslash}X
        >{\centering\arraybackslash}X
      }
  \ifpdf%
    \tikzsetnextfilename{cache/quasi3d}%
    \tikz[x=0.2\textwidth, y=0.2\textwidth, ]{
\iftikzgraphpreamble
\def\aspectratio{0.4992978101}
\else
\node[vertex] (v0) at (-0.3379153860, -0.0734476984) {};
\node[vertex] (v1) at (-0.2667241249, -0.0724132923) {};
\node[vertex] (v2) at (-0.4091066471, -0.0744821046) {};
\node[vertex] (v3) at (-0.2659724689, -0.0260747490) {};
\node[vertex] (v4) at (-0.4098583032, -0.1208206479) {};
\node[vertex] (v5) at (-0.3946308766, 0.0118590055) {};
\node[vertex] (v6) at (-0.2811998954, -0.1587544024) {};
\node[vertex] (v7) at (-0.3226879594, 0.0592319549) {};
\node[vertex] (v8) at (-0.2514966984, 0.0602663611) {};
\node[vertex] (v9) at (-0.3938792205, 0.0581975488) {};
\node[vertex] (v10) at (-0.2507450423, 0.1066049044) {};
\node[vertex] (v11) at (-0.3794034500, 0.1445386589) {};
\node[vertex] (v12) at (-0.3082121890, 0.1455730650) {};
\node[vertex] (v13) at (-0.4505947111, 0.1435042527) {};
\node[vertex] (v14) at (-0.3074605329, 0.1919116083) {};
\node[vertex] (v15) at (-0.4513463672, 0.0971657094) {};
\node[vertex] (v16) at (-0.4361189406, 0.2298453628) {};
\node[vertex] (v17) at (-0.2362692718, 0.1929460145) {};
\node[vertex] (v18) at (-0.3786517939, 0.1908772022) {};
\node[vertex] (v19) at (-0.3641760234, 0.2772183123) {};
\node[vertex] (v20) at (-0.2929847624, 0.2782527184) {};
\node[vertex] (v21) at (-0.4353672845, 0.2761839061) {};
\node[vertex] (v22) at (-0.1947812078, -0.0250403428) {};
\node[vertex] (v23) at (-0.1235899467, -0.0240059367) {};
\node[vertex] (v24) at (-0.1228382907, 0.0223326066) {};
\node[vertex] (v25) at (-0.1380657172, -0.1103470467) {};
\node[vertex] (v26) at (-0.1795537812, 0.1076393106) {};
\node[vertex] (v27) at (-0.1083625202, 0.1086737167) {};
\node[vertex] (v28) at (-0.1076108641, 0.1550122600) {};
\node[vertex] (v29) at (-0.1650780107, 0.1939804207) {};
\node[vertex] (v30) at (-0.0668744562, -0.1093126406) {};
\node[vertex] (v31) at (-0.2092569783, -0.1113814529) {};
\node[vertex] (v32) at (-0.0661228001, -0.0629740973) {};
\node[vertex] (v33) at (-0.2100086344, -0.1577199962) {};
\node[vertex] (v34) at (-0.0813502267, -0.1956537507) {};
\node[vertex] (v35) at (-0.0516470296, 0.0233670128) {};
\node[vertex] (v36) at (-0.1940295517, 0.0212982005) {};
\node[vertex] (v37) at (-0.0508953735, 0.0697055561) {};
\node[vertex] (v38) at (0.0050684610, -0.0619396911) {};
\node[vertex] (v39) at (0.0762597220, -0.0609052850) {};
\node[vertex] (v40) at (0.0770113781, -0.0145667417) {};
\node[vertex] (v41) at (0.0617839516, -0.1472463951) {};
\node[vertex] (v42) at (0.1482026392, -0.0135323355) {};
\node[vertex] (v43) at (0.0058201171, -0.0156011479) {};
\node[vertex] (v44) at (0.1489542953, 0.0328062077) {};
\node[vertex] (v45) at (0.0202958876, 0.0707399622) {};
\node[vertex] (v46) at (0.1337268687, -0.0998734456) {};
\node[vertex] (v47) at (0.0914871486, 0.0717743684) {};
\node[vertex] (v48) at (0.0922388047, 0.1181129117) {};
\node[vertex] (v49) at (-0.0364196030, 0.1560466662) {};
\node[vertex] (v50) at (0.2049181298, -0.0988390395) {};
\node[vertex] (v51) at (0.0625356076, -0.1009078518) {};
\node[vertex] (v52) at (0.2056697858, -0.0525004962) {};
\node[vertex] (v53) at (0.1904423593, -0.1851801496) {};
\node[vertex] (v54) at (0.1634300658, 0.1191473178) {};
\node[vertex] (v55) at (0.0210475436, 0.1170785055) {};
\node[vertex] (v56) at (0.2761093908, -0.0978046333) {};
\node[vertex] (v57) at (0.2768610469, -0.0514660900) {};
\node[vertex] (v58) at (0.1329752126, -0.1462119889) {};
\node[vertex] (v59) at (0.2616336203, -0.1841457434) {};
\node[vertex] (v60) at (0.2201455563, 0.0338406139) {};
\node[vertex] (v61) at (0.2913368174, 0.0348750201) {};
\node[vertex] (v62) at (0.4047677985, -0.1357383878) {};
\node[vertex] (v63) at (0.4759590596, -0.1347039816) {};
\node[vertex] (v64) at (0.3335765375, -0.1367727939) {};
\node[vertex] (v65) at (0.4767107157, -0.0883654383) {};
\node[vertex] (v66) at (0.3328248814, -0.1831113372) {};
\node[vertex] (v67) at (0.3480523080, -0.0504316839) {};
\node[vertex] (v68) at (0.4614832891, -0.2210450917) {};
\node[vertex] (v69) at (0.4199952251, -0.0030587344) {};
\node[vertex] (v70) at (0.4911864862, -0.0020243282) {};
\node[vertex] (v71) at (0.3488039641, -0.0040931406) {};
\node[vertex] (v72) at (0.4344709956, 0.0832823757) {};
\node[vertex] (v73) at (0.5479019768, -0.0873310322) {};
\node[vertex] (v74) at (0.4055194546, -0.0893998445) {};
\node[vertex] (v75) at (0.5486536328, -0.0409924889) {};
\node[vertex] (v76) at (0.5334262063, -0.1736721423) {};
\draw[edge] (v2) -- (v0);
\draw[edge] (v9) -- (v7);
\draw[edge] (v9) -- (v13);
\draw[edge] (v13) -- (v11);
\draw[edge] (v13) -- (v18);
\draw[edge] (v18) -- (v14);
\draw[edge] (v18) -- (v21);
\draw[edge] (v21) -- (v19);
\draw[edge] (v4) -- (v0);
\draw[edge] (v6) -- (v33);
\draw[edge] (v6) -- (v31);
\draw[edge] (v6) -- (v0);
\draw[edge] (v0) -- (v1);
\draw[edge] (v0) -- (v3);
\draw[edge] (v0) -- (v5);
\draw[edge] (v5) -- (v7);
\draw[edge] (v5) -- (v15);
\draw[edge] (v15) -- (v11);
\draw[edge] (v31) -- (v25);
\draw[edge] (v31) -- (v3);
\draw[edge] (v3) -- (v22);
\draw[edge] (v3) -- (v36);
\draw[edge] (v3) -- (v7);
\draw[edge] (v7) -- (v8);
\draw[edge] (v7) -- (v10);
\draw[edge] (v7) -- (v11);
\draw[edge] (v11) -- (v12);
\draw[edge] (v11) -- (v14);
\draw[edge] (v11) -- (v16);
\draw[edge] (v16) -- (v19);
\draw[edge] (v36) -- (v24);
\draw[edge] (v36) -- (v10);
\draw[edge] (v10) -- (v26);
\draw[edge] (v10) -- (v14);
\draw[edge] (v14) -- (v17);
\draw[edge] (v14) -- (v19);
\draw[edge] (v19) -- (v20);
\draw[edge] (v33) -- (v25);
\draw[edge] (v33) -- (v1);
\draw[edge] (v1) -- (v22);
\draw[edge] (v34) -- (v25);
\draw[edge] (v25) -- (v30);
\draw[edge] (v25) -- (v32);
\draw[edge] (v25) -- (v22);
\draw[edge] (v22) -- (v23);
\draw[edge] (v22) -- (v24);
\draw[edge] (v22) -- (v8);
\draw[edge] (v8) -- (v26);
\draw[edge] (v8) -- (v12);
\draw[edge] (v12) -- (v17);
\draw[edge] (v32) -- (v38);
\draw[edge] (v32) -- (v43);
\draw[edge] (v32) -- (v24);
\draw[edge] (v24) -- (v35);
\draw[edge] (v24) -- (v37);
\draw[edge] (v24) -- (v26);
\draw[edge] (v26) -- (v27);
\draw[edge] (v26) -- (v28);
\draw[edge] (v26) -- (v17);
\draw[edge] (v17) -- (v29);
\draw[edge] (v17) -- (v20);
\draw[edge] (v51) -- (v46);
\draw[edge] (v51) -- (v43);
\draw[edge] (v43) -- (v40);
\draw[edge] (v43) -- (v37);
\draw[edge] (v37) -- (v45);
\draw[edge] (v37) -- (v55);
\draw[edge] (v37) -- (v28);
\draw[edge] (v28) -- (v49);
\draw[edge] (v55) -- (v48);
\draw[edge] (v30) -- (v38);
\draw[edge] (v30) -- (v23);
\draw[edge] (v23) -- (v35);
\draw[edge] (v41) -- (v58);
\draw[edge] (v41) -- (v46);
\draw[edge] (v41) -- (v38);
\draw[edge] (v38) -- (v39);
\draw[edge] (v38) -- (v40);
\draw[edge] (v38) -- (v35);
\draw[edge] (v35) -- (v45);
\draw[edge] (v35) -- (v27);
\draw[edge] (v27) -- (v49);
\draw[edge] (v27) -- (v29);
\draw[edge] (v53) -- (v59);
\draw[edge] (v53) -- (v46);
\draw[edge] (v46) -- (v50);
\draw[edge] (v46) -- (v52);
\draw[edge] (v46) -- (v40);
\draw[edge] (v40) -- (v42);
\draw[edge] (v40) -- (v44);
\draw[edge] (v40) -- (v45);
\draw[edge] (v45) -- (v47);
\draw[edge] (v45) -- (v48);
\draw[edge] (v45) -- (v49);
\draw[edge] (v52) -- (v57);
\draw[edge] (v52) -- (v44);
\draw[edge] (v44) -- (v60);
\draw[edge] (v44) -- (v48);
\draw[edge] (v48) -- (v54);
\draw[edge] (v58) -- (v50);
\draw[edge] (v58) -- (v39);
\draw[edge] (v39) -- (v42);
\draw[edge] (v59) -- (v66);
\draw[edge] (v59) -- (v64);
\draw[edge] (v59) -- (v50);
\draw[edge] (v50) -- (v56);
\draw[edge] (v50) -- (v57);
\draw[edge] (v50) -- (v42);
\draw[edge] (v42) -- (v60);
\draw[edge] (v42) -- (v47);
\draw[edge] (v47) -- (v54);
\draw[edge] (v64) -- (v62);
\draw[edge] (v64) -- (v74);
\draw[edge] (v64) -- (v57);
\draw[edge] (v57) -- (v67);
\draw[edge] (v57) -- (v71);
\draw[edge] (v57) -- (v60);
\draw[edge] (v60) -- (v61);
\draw[edge] (v60) -- (v54);
\draw[edge] (v74) -- (v65);
\draw[edge] (v74) -- (v71);
\draw[edge] (v71) -- (v69);
\draw[edge] (v66) -- (v62);
\draw[edge] (v66) -- (v56);
\draw[edge] (v56) -- (v67);
\draw[edge] (v68) -- (v76);
\draw[edge] (v68) -- (v62);
\draw[edge] (v62) -- (v63);
\draw[edge] (v62) -- (v65);
\draw[edge] (v62) -- (v67);
\draw[edge] (v67) -- (v69);
\draw[edge] (v67) -- (v61);
\draw[edge] (v76) -- (v65);
\draw[edge] (v65) -- (v73);
\draw[edge] (v65) -- (v75);
\draw[edge] (v65) -- (v69);
\draw[edge] (v69) -- (v70);
\draw[edge] (v63) -- (v73);
\draw[edge] (v73) -- (v70);
\draw[edge] (v70) -- (v72);
\fi}%
  \else%
    \pgfimage[width=0.2\textwidth]{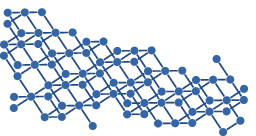}%
  \fi%
&
  \ifpdf%
    \tikzsetnextfilename{cache/quasi4d}%
    \tikz[x=0.2\textwidth, y=0.2\textwidth, ]{\input{pics/quasi4d.tikz}}%
  \else%
    \pgfimage[width=0.2\textwidth]{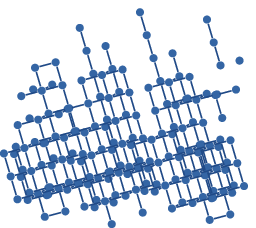}%
  \fi%
&
  \ifpdf%
    \tikzsetnextfilename{cache/quasi5d}%
    \tikz[x=0.2\textwidth, y=0.2\textwidth, ]{\input{pics/quasi5d.tikz}}%
  \else%
    \pgfimage[width=0.2\textwidth]{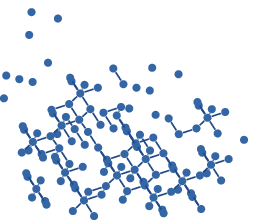}%
  \fi%
&
  \ifpdf%
    \tikzsetnextfilename{cache/quasi6d}%
    \tikz[x=0.2\textwidth, y=0.2\textwidth, ]{\input{pics/quasi6d.tikz}}%
  \else%
    \pgfimage[width=0.2\textwidth]{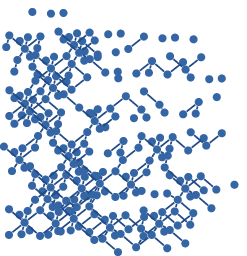}%
  \fi%
\\[2ex]
      \enum{QUASI3D} & \enum{QUASI4D} & \enum{QUASI5D} & \enum{QUASI6D}
    \end{tabularx}
    \par\vspace{1cm}
    \begin{tabularx}{\linewidth}{%
        >{\centering\arraybackslash}X
        >{\centering\arraybackslash}X
        >{\centering\arraybackslash}X
      }
  \ifpdf%
    \tikzsetnextfilename{cache/lindenmayer}%
    \tikz[x=0.25\textwidth, y=0.25\textwidth, ]{\input{pics/lindenmayer.tikz}}%
  \else%
    \pgfimage[width=0.25\textwidth]{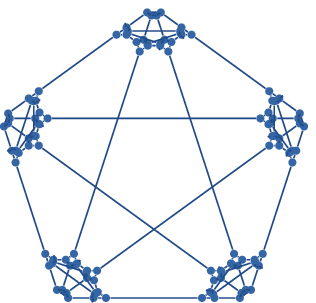}%
  \fi%
&
  \ifpdf%
    \tikzsetnextfilename{cache/mosaic1}%
    \tikz[x=0.25\textwidth, y=0.25\textwidth, ]{
\iftikzgraphpreamble
\def\aspectratio{0.8660254038}
\else
\node[vertex] (v0) at (0.0468750000, 0.6607954253) {};
\node[vertex] (v1) at (0.5468750000, -0.2052299785) {};
\node[vertex] (v2) at (-0.4531250000, -0.2052299785) {};
\node[vertex] (v3) at (0.2968750000, 0.2277827234) {};
\node[vertex] (v4) at (0.0468750000, -0.2052299785) {};
\node[vertex] (v5) at (-0.2031250000, 0.2277827234) {};
\node[vertex] (v6) at (0.2968750000, -0.0608924112) {};
\node[vertex] (v7) at (0.2968750000, -0.1571174561) {};
\node[vertex] (v8) at (-0.3281250000, 0.0112763724) {};
\node[vertex] (v9) at (-0.2031250000, -0.2052299785) {};
\node[vertex] (v10) at (-0.0781250000, 0.0112763724) {};
\node[vertex] (v11) at (0.3802083333, -0.0127798888) {};
\node[vertex] (v12) at (0.4218750000, -0.1330611949) {};
\node[vertex] (v13) at (0.2968750000, 0.0834451561) {};
\node[vertex] (v14) at (0.4218750000, 0.0112763724) {};
\node[vertex] (v15) at (-0.3281250000, -0.2052299785) {};
\node[vertex] (v16) at (-0.3906250000, -0.0969768030) {};
\node[vertex] (v17) at (-0.2656250000, -0.0969768030) {};
\node[vertex] (v18) at (0.1718750000, 0.4442890743) {};
\node[vertex] (v19) at (-0.0781250000, 0.4442890743) {};
\node[vertex] (v20) at (0.0468750000, 0.2277827234) {};
\node[vertex] (v21) at (-0.0156250000, -0.0969768030) {};
\node[vertex] (v22) at (-0.0781250000, -0.2052299785) {};
\node[vertex] (v23) at (-0.1406250000, -0.0969768030) {};
\node[vertex] (v24) at (0.0468750000, 0.3721202907) {};
\node[vertex] (v25) at (0.2135416667, -0.0127798888) {};
\node[vertex] (v26) at (-0.2031250000, -0.0608924112) {};
\node[vertex] (v27) at (-0.2031250000, 0.0112763724) {};
\node[vertex] (v28) at (-0.1406250000, -0.1691455867) {};
\node[vertex] (v29) at (-0.1718750000, -0.1511033908) {};
\node[vertex] (v30) at (-0.1406250000, -0.2052299785) {};
\node[vertex] (v31) at (-0.1093750000, -0.1511033908) {};
\draw[edge] (v1) -- (v4);
\draw[edge] (v3) -- (v4);
\draw[edge] (v4) -- (v6);
\draw[edge] (v4) -- (v7);
\draw[edge] (v6) -- (v7);
\draw[edge] (v1) -- (v7);
\draw[edge] (v5) -- (v8);
\draw[edge] (v5) -- (v10);
\draw[edge] (v1) -- (v12);
\draw[edge] (v6) -- (v12);
\draw[edge] (v6) -- (v13);
\draw[edge] (v3) -- (v13);
\draw[edge] (v3) -- (v14);
\draw[edge] (v1) -- (v14);
\draw[edge] (v11) -- (v12);
\draw[edge] (v11) -- (v13);
\draw[edge] (v11) -- (v14);
\draw[edge] (v9) -- (v15);
\draw[edge] (v2) -- (v15);
\draw[edge] (v2) -- (v16);
\draw[edge] (v8) -- (v16);
\draw[edge] (v8) -- (v17);
\draw[edge] (v9) -- (v17);
\draw[edge] (v15) -- (v17);
\draw[edge] (v15) -- (v16);
\draw[edge] (v16) -- (v17);
\draw[edge] (v3) -- (v18);
\draw[edge] (v0) -- (v18);
\draw[edge] (v0) -- (v19);
\draw[edge] (v5) -- (v19);
\draw[edge] (v5) -- (v20);
\draw[edge] (v3) -- (v20);
\draw[edge] (v18) -- (v20);
\draw[edge] (v18) -- (v19);
\draw[edge] (v19) -- (v20);
\draw[edge] (v10) -- (v21);
\draw[edge] (v4) -- (v21);
\draw[edge] (v4) -- (v22);
\draw[edge] (v10) -- (v23);
\draw[edge] (v21) -- (v23);
\draw[edge] (v21) -- (v22);
\draw[edge] (v18) -- (v24);
\draw[edge] (v19) -- (v24);
\draw[edge] (v20) -- (v24);
\draw[edge] (v3) -- (v25);
\draw[edge] (v6) -- (v25);
\draw[edge] (v4) -- (v25);
\draw[edge] (v10) -- (v27);
\draw[edge] (v8) -- (v27);
\draw[edge] (v17) -- (v26);
\draw[edge] (v23) -- (v26);
\draw[edge] (v26) -- (v27);
\draw[edge] (v23) -- (v29);
\draw[edge] (v9) -- (v29);
\draw[edge] (v9) -- (v30);
\draw[edge] (v22) -- (v30);
\draw[edge] (v22) -- (v31);
\draw[edge] (v23) -- (v31);
\draw[edge] (v28) -- (v29);
\draw[edge] (v28) -- (v30);
\draw[edge] (v28) -- (v31);
\fi}%
  \else%
    \pgfimage[width=0.25\textwidth]{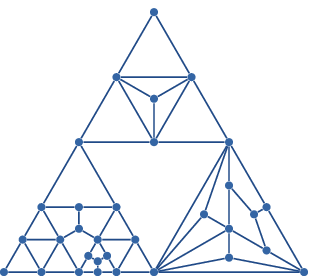}%
  \fi%
&
  \ifpdf%
    \tikzsetnextfilename{cache/mosaic2}%
    \tikz[x=0.25\textwidth, y=0.25\textwidth, ]{
\iftikzgraphpreamble
\def\aspectratio{0.8660254038}
\else
\node[vertex] (v0) at (-0.0000000000, 0.5773502692) {};
\node[vertex] (v1) at (0.5000000000, -0.2886751346) {};
\node[vertex] (v2) at (-0.5000000000, -0.2886751346) {};
\node[vertex] (v3) at (-0.0000000000, 0.0000000000) {};
\node[vertex] (v4) at (0.2500000000, 0.1443375673) {};
\node[vertex] (v5) at (0.0000000000, -0.2886751346) {};
\node[vertex] (v6) at (-0.2500000000, 0.1443375673) {};
\node[vertex] (v7) at (0.1250000000, 0.3608439182) {};
\node[vertex] (v8) at (-0.1250000000, 0.3608439182) {};
\node[vertex] (v9) at (-0.1250000000, 0.0721687836) {};
\node[vertex] (v10) at (0.1250000000, 0.0721687836) {};
\node[vertex] (v11) at (0.2500000000, -0.2886751346) {};
\node[vertex] (v12) at (0.3750000000, -0.0721687836) {};
\node[vertex] (v13) at (0.0000000000, -0.1443375673) {};
\node[vertex] (v14) at (-0.3750000000, -0.0721687836) {};
\node[vertex] (v15) at (-0.2500000000, -0.2886751346) {};
\draw[edge] (v4) -- (v7);
\draw[edge] (v0) -- (v7);
\draw[edge] (v0) -- (v8);
\draw[edge] (v6) -- (v8);
\draw[edge] (v6) -- (v9);
\draw[edge] (v3) -- (v9);
\draw[edge] (v3) -- (v10);
\draw[edge] (v4) -- (v10);
\draw[edge] (v7) -- (v10);
\draw[edge] (v7) -- (v8);
\draw[edge] (v8) -- (v9);
\draw[edge] (v9) -- (v10);
\draw[edge] (v5) -- (v11);
\draw[edge] (v1) -- (v11);
\draw[edge] (v1) -- (v12);
\draw[edge] (v4) -- (v12);
\draw[edge] (v3) -- (v13);
\draw[edge] (v5) -- (v13);
\draw[edge] (v11) -- (v13);
\draw[edge] (v11) -- (v12);
\draw[edge] (v10) -- (v12);
\draw[edge] (v10) -- (v13);
\draw[edge] (v6) -- (v14);
\draw[edge] (v2) -- (v14);
\draw[edge] (v2) -- (v15);
\draw[edge] (v5) -- (v15);
\draw[edge] (v9) -- (v14);
\draw[edge] (v14) -- (v15);
\draw[edge] (v13) -- (v15);
\draw[edge] (v9) -- (v13);
\fi}%
  \else%
    \pgfimage[width=0.25\textwidth]{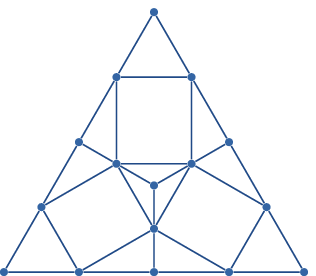}%
  \fi%
\\[2ex]
      \enum{LINDENMAYER} & \enum{MOSAIC1} & \enum{MOSAIC2}
    \end{tabularx}
  \end{center}
  \caption{%
    Examples of generated graphs labeled by the respective generators.  \enum{GRID}, \enum{LINDENMAYER}, \enum{QUASI\meta{$n$}D}, \enum{MOSAIC1}, \enum{MOSAIC2} and \enum{BOTTLE} layouts are native.  \enum{TORUS1} and \enum{TORUS2} are visualized with the stress-minimization algorithm.
  }
  \label{app:fig:generators}
\end{figure}

\begin{figure}[p]
  \begin{center}
    \begin{tabular}{c@{\qquad\qquad}c}

\begin{tikzpicture}[scale = 0.5]

  \path[use as bounding box] (-4.5, -4.5) rectangle (4.5, 4.5);

  \node[Vertex] (v) at (0, 0) {$v$};
  \node[Vertex] (u1) at (  0:4) {$u_1$};
  \node[Vertex] (u2) at ( 90:4) {$u_2$};
  \node[Vertex] (u3) at (180:4) {$u_3$};
  \node[Vertex] (u4) at (270:4) {$u_4$};

  \draw[edge] (v)  -- (u1);
  \draw[edge] (v)  -- (u2);
  \draw[edge] (v)  -- (u3);
  \draw[edge] (v) -- (u4);

\end{tikzpicture} & 

\begin{tikzpicture}[scale = 0.5]

  \path[use as bounding box] (-4.5, -4.5) rectangle (4.5, 4.5);

  \node[Vertex] (v)  at (0, 0)  {$v$};
  \node[Vertex] (u1) at (  0:4) {$u_1$};
  \node[Vertex] (u2) at ( 90:4) {$u_2$};
  \node[Vertex] (u3) at (180:4) {$u_3$};
  \node[Vertex] (u4) at (270:4) {$u_4$};

  \node[vertex] (w1)  at (  0:2) {};
  \node[vertex] (w2)  at ( 30:2) {};
  \node[vertex] (w3)  at ( 60:2) {};
  \node[vertex] (w4)  at ( 90:2) {};
  \node[vertex] (w5)  at (120:2) {};
  \node[vertex] (w6)  at (150:2) {};
  \node[vertex] (w7)  at (180:2) {};
  \node[vertex] (w8)  at (210:2) {};
  \node[vertex] (w9)  at (240:2) {};
  \node[vertex] (w10) at (270:2) {};
  \node[vertex] (w11) at (300:2) {};
  \node[vertex] (w12) at (330:2) {};

  \foreach \i in {1, ..., 12} {
    \draw[edge] (v) -- (w\i);
  }

  \draw[edge] (w1)  -- (u1);
  \draw[edge] (w4)  -- (u2);
  \draw[edge] (w7)  -- (u3);
  \draw[edge] (w10) -- (u4);

\end{tikzpicture}\\
      $\menum{L/SINGLETON}$ & $\menum{L/STAR}_3(v)$\\[2ex]

\begin{tikzpicture}[scale = 0.5]

  \path[use as bounding box] (-4.5, -4.5) rectangle (4.5, 4.5);

  \node[Vertex] (v)  at (0, 0)  {$v$};
  \node[Vertex] (u1) at (  0:4) {$u_1$};
  \node[Vertex] (u2) at ( 90:4) {$u_2$};
  \node[Vertex] (u3) at (180:4) {$u_3$};
  \node[Vertex] (u4) at (270:4) {$u_4$};

  \node[vertex] (w1)  at (  0:2) {};
  \node[vertex] (w2)  at ( 30:2) {};
  \node[vertex] (w3)  at ( 60:2) {};
  \node[vertex] (w4)  at ( 90:2) {};
  \node[vertex] (w5)  at (120:2) {};
  \node[vertex] (w6)  at (150:2) {};
  \node[vertex] (w7)  at (180:2) {};
  \node[vertex] (w8)  at (210:2) {};
  \node[vertex] (w9)  at (240:2) {};
  \node[vertex] (w10) at (270:2) {};
  \node[vertex] (w11) at (300:2) {};
  \node[vertex] (w12) at (330:2) {};

  \draw[edge] (w1) -- (w2) -- (w3) -- (w4) -- (w5) -- (w6) -- (w7) -- (w8) -- (w9) -- (w10) -- (w11) -- (w12) -- (w1);

  \foreach \i in {1, ..., 12} {
    \draw[edge] (v) -- (w\i);
  }

  \draw[edge] (w1)  -- (u1);
  \draw[edge] (w4)  -- (u2);
  \draw[edge] (w7)  -- (u3);
  \draw[edge] (w10) -- (u4);

\end{tikzpicture} & 

\begin{tikzpicture}[scale = 0.5]

  \path[use as bounding box] (-4.5, -4.5) rectangle (4.5, 4.5);

  \node[Vertex] (u1) at (  0:4) {$u_1$};
  \node[Vertex] (u2) at ( 90:4) {$u_2$};
  \node[Vertex] (u3) at (180:4) {$u_3$};
  \node[Vertex] (u4) at (270:4) {$u_4$};

  \node[vertex] (w1)  at (  0:2) {};
  \node[vertex] (w2)  at ( 30:2) {};
  \node[vertex] (w3)  at ( 60:2) {};
  \node[vertex] (w4)  at ( 90:2) {};
  \node[vertex] (w5)  at (120:2) {};
  \node[vertex] (w6)  at (150:2) {};
  \node[vertex] (w7)  at (180:2) {};
  \node[vertex] (w8)  at (210:2) {};
  \node[vertex] (w9)  at (240:2) {};
  \node[vertex] (w10) at (270:2) {};
  \node[vertex] (w11) at (300:2) {};
  \node[vertex] (w12) at (330:2) {};

  \draw[edge] (w1) -- (w2) -- (w3) -- (w4) -- (w5) -- (w6) -- (w7) -- (w8) -- (w9) -- (w10) -- (w11) -- (w12) -- (w1);

  \draw[edge] (w1)  -- (u1);
  \draw[edge] (w4)  -- (u2);
  \draw[edge] (w7)  -- (u3);
  \draw[edge] (w10) -- (u4);

\end{tikzpicture}\\
      $\menum{L/WHEEL}_3(v)$ & $\menum{L/RING}_3(v)$\\[2ex]

\begin{tikzpicture}[scale = 0.5]

  \path[use as bounding box] (-4.5, -4.5) rectangle (4.5, 4.5);

  \node[Vertex] (u1)  at (  0:4) {$u_1$};
  \node[Vertex] (u2)  at ( 90:4) {$u_2$};
  \node[Vertex] (u3)  at (180:4) {$u_3$};
  \node[Vertex] (u4)  at (270:4) {$u_4$};

  \node[vertex] (w1)  at (  0:2) {};
  \node[vertex] (w2)  at ( 45:2) {};
  \node[vertex] (w3)  at ( 90:2) {};
  \node[vertex] (w4)  at (135:2) {};
  \node[vertex] (w5)  at (180:2) {};
  \node[vertex] (w6)  at (225:2) {};
  \node[vertex] (w7)  at (270:2) {};
  \node[vertex] (w8)  at (315:2) {};

  \draw[edge] (w1) -- (u1);
  \draw[edge] (w3) -- (u2);
  \draw[edge] (w5) -- (u3);
  \draw[edge] (w7) -- (u4);

  \foreach \i in {1, ..., 8} {
    \foreach \j in {\i, ..., 8} {
      \ifnumequal{\i}{\j}{}{%
        \draw[edge] (w\i) -- (w\j);
      }
    }
  }

\end{tikzpicture} & 

\begin{tikzpicture}[scale = 0.5]

  \path[use as bounding box] (-4.5, -4.5) rectangle (4.5, 4.5);

  \begin{scope}[scale = 1.5]
    \draw[edge] (-2, -1) grid[step = 1] (+2, +1);
    \foreach \i in {-1, 0, +1} {
      \foreach \j in {-2, -1, 0, +1, +2} {
        \node[vertex] at (\j, \i) {};
      }
    }
    \node[Vertex, fill = white] (v) at (0, 0) {$v$};
  \end{scope}

\end{tikzpicture}\\
      $\menum{L/CLIQUE}_2(v)$ & $\menum{L/GRID}_{3,5}(v)$
    \end{tabular}
  \end{center}
  \caption{%
    Illustration of the \enum{LINDENMAYER} generator operations.  A degree four vertex may be replaced by any of the above subgraphs, except for the bottom right subgraph which replaces a degree zero vertex.
  }
  \label{app:fig:lindenmayer-subgens}
\end{figure}
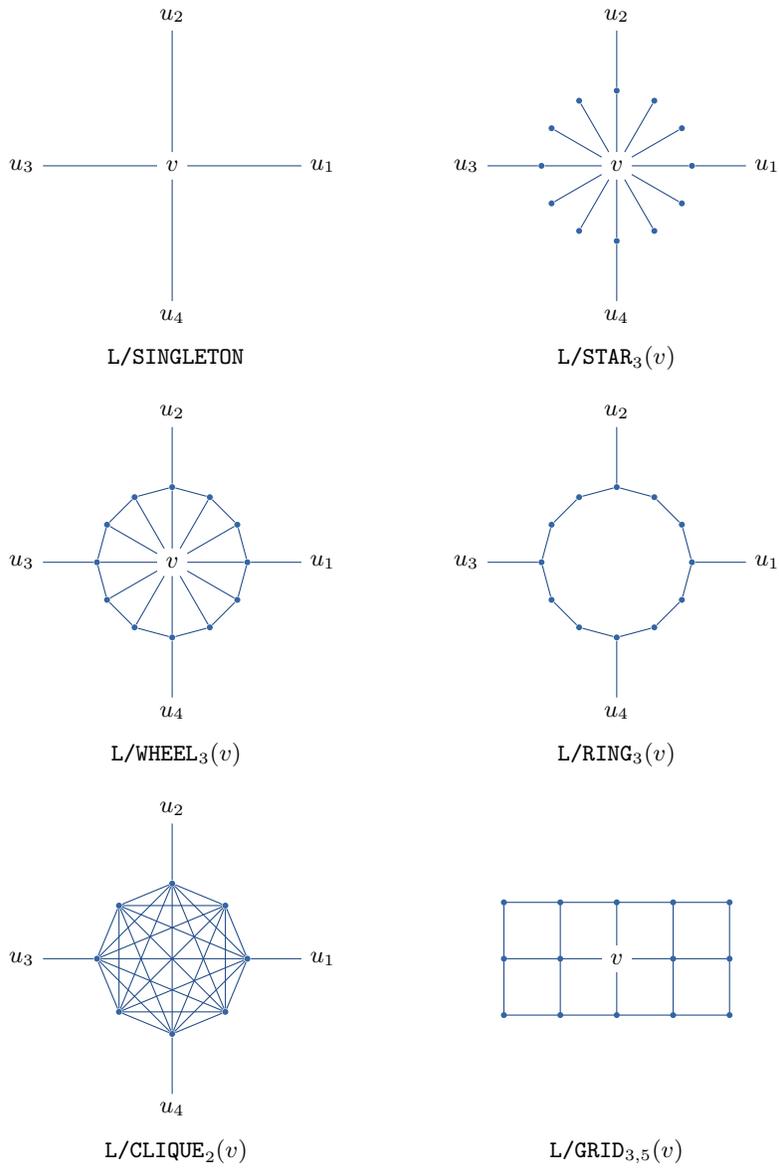

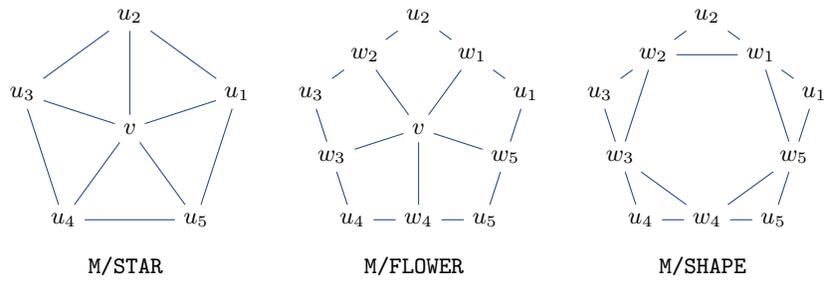
\begin{figure}[p]
  \begin{center}
    \newcommand*{\GenMosaicScale}{0.5}
    \begin{tabular}{c@{\quad}c@{\quad}c}

\providecommand*{\GenMosaicScale}{0.6}
\begin{tikzpicture}[scale = \GenMosaicScale, rotate = 18]

  \node[Vertex] (u1) at (  0:3) {$u_1$};
  \node[Vertex] (u2) at ( 72:3) {$u_2$};
  \node[Vertex] (u3) at (144:3) {$u_3$};
  \node[Vertex] (u4) at (216:3) {$u_4$};
  \node[Vertex] (u5) at (288:3) {$u_5$};

  \node[Vertex] (v) at (0, 0) {$v$};

  \draw[edge] (u1) -- (u2) -- (u3) -- (u4) -- (u5) -- (u1);
  \draw[edge] (v) -- (u1);
  \draw[edge] (v) -- (u2);
  \draw[edge] (v) -- (u3);
  \draw[edge] (v) -- (u4);
  \draw[edge] (v) -- (u5);

\end{tikzpicture}&

\providecommand*{\GenMosaicScale}{0.6}
\begin{tikzpicture}[scale = \GenMosaicScale, rotate = 18]

  \node[Vertex] (u1) at (  0:3) {$u_1$};
  \node[Vertex] (u2) at ( 72:3) {$u_2$};
  \node[Vertex] (u3) at (144:3) {$u_3$};
  \node[Vertex] (u4) at (216:3) {$u_4$};
  \node[Vertex] (u5) at (288:3) {$u_5$};

  \node[Vertex] (w1) at ($(u1)!0.5!(u2)$) {$w_1$};
  \node[Vertex] (w2) at ($(u2)!0.5!(u3)$) {$w_2$};
  \node[Vertex] (w3) at ($(u3)!0.5!(u4)$) {$w_3$};
  \node[Vertex] (w4) at ($(u4)!0.5!(u5)$) {$w_4$};
  \node[Vertex] (w5) at ($(u5)!0.5!(u1)$) {$w_5$};

  \node[Vertex] (v) at (0, 0) {$v$};

  \draw[edge] (u1) -- (w1) -- (u2) -- (w2) -- (u3) -- (w3) -- (u4) -- (w4) -- (u5) -- (w5) -- (u1);

  \draw[edge] (v) -- (w1);
  \draw[edge] (v) -- (w2);
  \draw[edge] (v) -- (w3);
  \draw[edge] (v) -- (w4);
  \draw[edge] (v) -- (w5);

\end{tikzpicture}&

\providecommand*{\GenMosaicScale}{0.6}
\begin{tikzpicture}[scale = \GenMosaicScale, rotate = 18]

  \node[Vertex] (u1) at (  0:3) {$u_1$};
  \node[Vertex] (u2) at ( 72:3) {$u_2$};
  \node[Vertex] (u3) at (144:3) {$u_3$};
  \node[Vertex] (u4) at (216:3) {$u_4$};
  \node[Vertex] (u5) at (288:3) {$u_5$};

  \node[Vertex] (w1) at ($(u1)!0.5!(u2)$) {$w_1$};
  \node[Vertex] (w2) at ($(u2)!0.5!(u3)$) {$w_2$};
  \node[Vertex] (w3) at ($(u3)!0.5!(u4)$) {$w_3$};
  \node[Vertex] (w4) at ($(u4)!0.5!(u5)$) {$w_4$};
  \node[Vertex] (w5) at ($(u5)!0.5!(u1)$) {$w_5$};

  \draw[edge] (u1) -- (w1) -- (u2) -- (w2) -- (u3) -- (w3) -- (u4) -- (w4) -- (u5) -- (w5) -- (u1);
  \draw[edge] (w1) -- (w2) -- (w3) -- (w4) -- (w5) -- (w1);

\end{tikzpicture}\\[1ex]
      \enum{M/STAR} & \enum{M/FLOWER} & \enum{M/SHAPE}
    \end{tabular}
  \end{center}
  \caption{%
    Operations of the \enum{MOSAIC} generator on a pentagonal facet $\{u_1,\ldots,u_5\}$.
  }
  \label{app:fig:mosaic-subgens}
\end{figure}

\begin{figure}[p]
  \begin{center}
    \begin{tabularx}{\linewidth}{%
        >{\centering\arraybackslash}X
        >{\centering\arraybackslash}X
        >{\centering\arraybackslash}X
      }
  \ifpdf%
    \tikzsetnextfilename{cache/native}%
    \tikz[x=0.2\textwidth, y=0.2\textwidth, ]{
\iftikzgraphpreamble
\def\aspectratio{0.8660254038}
\else
\node[vertex] (v0) at (-0.0000000000, 0.5773502692) {};
\node[vertex] (v1) at (0.5000000000, -0.2886751346) {};
\node[vertex] (v2) at (-0.5000000000, -0.2886751346) {};
\node[vertex] (v3) at (-0.0000000000, 0.0000000000) {};
\node[vertex] (v4) at (0.2500000000, 0.1443375673) {};
\node[vertex] (v5) at (0.0000000000, -0.2886751346) {};
\node[vertex] (v6) at (-0.2500000000, 0.1443375673) {};
\node[vertex] (v7) at (-0.0000000000, 0.2165063509) {};
\node[vertex] (v8) at (0.1875000000, -0.1082531755) {};
\node[vertex] (v9) at (-0.1875000000, -0.1082531755) {};
\node[vertex] (v10) at (0.0833333333, 0.3127313958) {};
\node[vertex] (v11) at (0.1250000000, 0.1804219591) {};
\node[vertex] (v12) at (-0.0000000000, 0.3969283101) {};
\node[vertex] (v13) at (0.1250000000, 0.3608439182) {};
\node[vertex] (v14) at (-0.0833333333, 0.3127313958) {};
\node[vertex] (v15) at (-0.1250000000, 0.1804219591) {};
\node[vertex] (v16) at (-0.1250000000, 0.3608439182) {};
\node[vertex] (v17) at (-0.0833333333, 0.1202813061) {};
\node[vertex] (v18) at (-0.0000000000, 0.1082531755) {};
\node[vertex] (v19) at (-0.1250000000, 0.0721687836) {};
\node[vertex] (v20) at (0.0833333333, 0.1202813061) {};
\node[vertex] (v21) at (0.1250000000, 0.0721687836) {};
\node[vertex] (v22) at (0.2291666667, -0.2285344816) {};
\node[vertex] (v23) at (0.0937500000, -0.1984641550) {};
\node[vertex] (v24) at (0.3437500000, -0.1984641550) {};
\node[vertex] (v25) at (0.2500000000, -0.2886751346) {};
\node[vertex] (v26) at (0.3125000000, -0.0841969143) {};
\node[vertex] (v27) at (0.2187500000, 0.0180421959) {};
\node[vertex] (v28) at (0.3750000000, -0.0721687836) {};
\node[vertex] (v29) at (0.1458333333, 0.0120281306) {};
\node[vertex] (v30) at (0.0937500000, -0.0541265877) {};
\node[vertex] (v31) at (0.0625000000, -0.1323094367) {};
\node[vertex] (v32) at (0.0000000000, -0.1443375673) {};
\node[vertex] (v33) at (-0.3125000000, -0.0841969143) {};
\node[vertex] (v34) at (-0.2187500000, 0.0180421959) {};
\node[vertex] (v35) at (-0.3437500000, -0.1984641550) {};
\node[vertex] (v36) at (-0.3750000000, -0.0721687836) {};
\node[vertex] (v37) at (-0.2291666667, -0.2285344816) {};
\node[vertex] (v38) at (-0.0937500000, -0.1984641550) {};
\node[vertex] (v39) at (-0.2500000000, -0.2886751346) {};
\node[vertex] (v40) at (-0.0625000000, -0.1323094367) {};
\node[vertex] (v41) at (-0.0937500000, -0.0541265877) {};
\node[vertex] (v42) at (-0.1458333333, 0.0120281306) {};
\draw[edge] (v4) -- (v11);
\draw[edge] (v7) -- (v11);
\draw[edge] (v7) -- (v12);
\draw[edge] (v0) -- (v12);
\draw[edge] (v0) -- (v13);
\draw[edge] (v4) -- (v13);
\draw[edge] (v10) -- (v11);
\draw[edge] (v10) -- (v12);
\draw[edge] (v10) -- (v13);
\draw[edge] (v7) -- (v15);
\draw[edge] (v6) -- (v15);
\draw[edge] (v6) -- (v16);
\draw[edge] (v0) -- (v16);
\draw[edge] (v12) -- (v14);
\draw[edge] (v14) -- (v15);
\draw[edge] (v14) -- (v16);
\draw[edge] (v7) -- (v18);
\draw[edge] (v3) -- (v18);
\draw[edge] (v3) -- (v19);
\draw[edge] (v6) -- (v19);
\draw[edge] (v15) -- (v17);
\draw[edge] (v17) -- (v18);
\draw[edge] (v17) -- (v19);
\draw[edge] (v4) -- (v21);
\draw[edge] (v3) -- (v21);
\draw[edge] (v18) -- (v20);
\draw[edge] (v11) -- (v20);
\draw[edge] (v20) -- (v21);
\draw[edge] (v5) -- (v23);
\draw[edge] (v8) -- (v23);
\draw[edge] (v8) -- (v24);
\draw[edge] (v1) -- (v24);
\draw[edge] (v1) -- (v25);
\draw[edge] (v5) -- (v25);
\draw[edge] (v22) -- (v23);
\draw[edge] (v22) -- (v24);
\draw[edge] (v22) -- (v25);
\draw[edge] (v8) -- (v27);
\draw[edge] (v4) -- (v27);
\draw[edge] (v4) -- (v28);
\draw[edge] (v1) -- (v28);
\draw[edge] (v24) -- (v26);
\draw[edge] (v26) -- (v27);
\draw[edge] (v26) -- (v28);
\draw[edge] (v8) -- (v30);
\draw[edge] (v3) -- (v30);
\draw[edge] (v27) -- (v29);
\draw[edge] (v29) -- (v30);
\draw[edge] (v21) -- (v29);
\draw[edge] (v5) -- (v32);
\draw[edge] (v3) -- (v32);
\draw[edge] (v30) -- (v31);
\draw[edge] (v23) -- (v31);
\draw[edge] (v31) -- (v32);
\draw[edge] (v6) -- (v34);
\draw[edge] (v9) -- (v34);
\draw[edge] (v9) -- (v35);
\draw[edge] (v2) -- (v35);
\draw[edge] (v2) -- (v36);
\draw[edge] (v6) -- (v36);
\draw[edge] (v33) -- (v34);
\draw[edge] (v33) -- (v35);
\draw[edge] (v33) -- (v36);
\draw[edge] (v9) -- (v38);
\draw[edge] (v5) -- (v38);
\draw[edge] (v5) -- (v39);
\draw[edge] (v2) -- (v39);
\draw[edge] (v35) -- (v37);
\draw[edge] (v37) -- (v38);
\draw[edge] (v37) -- (v39);
\draw[edge] (v9) -- (v41);
\draw[edge] (v3) -- (v41);
\draw[edge] (v38) -- (v40);
\draw[edge] (v40) -- (v41);
\draw[edge] (v32) -- (v40);
\draw[edge] (v41) -- (v42);
\draw[edge] (v34) -- (v42);
\draw[edge] (v19) -- (v42);
\fi}%
  \else%
    \pgfimage[width=0.2\textwidth]{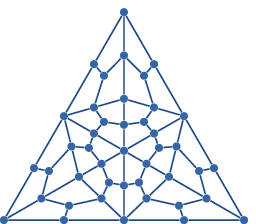}%
  \fi%
&
  \ifpdf%
    \tikzsetnextfilename{cache/fmmm}%
    \tikz[x=0.2\textwidth, y=0.2\textwidth, ]{
\iftikzgraphpreamble
\def\aspectratio{0.9963898917}
\else
\node[vertex] (v0) at (-0.4152464109, -0.0996557804) {};
\node[vertex] (v1) at (0.0757283184, 0.4165897070) {};
\node[vertex] (v2) at (0.2706741667, -0.3198723869) {};
\node[vertex] (v3) at (-0.0000839560, 0.0050373604) {};
\node[vertex] (v4) at (-0.2636218621, 0.2613550500) {};
\node[vertex] (v5) at (0.3392662245, 0.0916799597) {};
\node[vertex] (v6) at (-0.0975568802, -0.3559734699) {};
\node[vertex] (v7) at (-0.3538745697, -0.1068759970) {};
\node[vertex] (v8) at (0.1190496180, 0.3660481907) {};
\node[vertex] (v9) at (0.2851145999, -0.2548904374) {};
\node[vertex] (v10) at (-0.5235496600, 0.0122575770) {};
\node[vertex] (v11) at (-0.3791453278, 0.0916799597) {};
\node[vertex] (v12) at (-0.5199395517, -0.1610276215) {};
\node[vertex] (v13) at (-0.4188565192, 0.1169507178) {};
\node[vertex] (v14) at (-0.4405171690, -0.3090420620) {};
\node[vertex] (v15) at (-0.2816724037, -0.2909915204) {};
\node[vertex] (v16) at (-0.3105532701, -0.2909915204) {};
\node[vertex] (v17) at (-0.2058601293, -0.1971287046) {};
\node[vertex] (v18) at (-0.1914196961, -0.0382839392) {};
\node[vertex] (v19) at (-0.0542355806, -0.1826882713) {};
\node[vertex] (v20) at (-0.2780622954, 0.1061203929) {};
\node[vertex] (v21) at (-0.1372680715, 0.1386113676) {};
\node[vertex] (v22) at (0.2851145999, 0.4454705734) {};
\node[vertex] (v23) at (0.3067752498, 0.2938460247) {};
\node[vertex] (v24) at (0.1587608093, 0.5285030644) {};
\node[vertex] (v25) at (0.2490135169, 0.2938460247) {};
\node[vertex] (v26) at (-0.0145243892, 0.5248929561) {};
\node[vertex] (v27) at (-0.0831164470, 0.3732684074) {};
\node[vertex] (v28) at (-0.1300478549, 0.4093694904) {};
\node[vertex] (v29) at (-0.0831164470, 0.2721853749) {};
\node[vertex] (v30) at (0.0648979935, 0.1963731005) {};
\node[vertex] (v31) at (0.2345730837, 0.1927629922) {};
\node[vertex] (v32) at (0.1804214592, 0.0519687684) {};
\node[vertex] (v33) at (0.2490135169, -0.4678868273) {};
\node[vertex] (v34) at (0.1154395097, -0.3704139031) {};
\node[vertex] (v35) at (0.3898077407, -0.3848543363) {};
\node[vertex] (v36) at (0.0937788599, -0.4173453111) {};
\node[vertex] (v37) at (0.4764503400, -0.2585005457) {};
\node[vertex] (v38) at (0.4114683906, -0.0996557804) {};
\node[vertex] (v39) at (0.3681470909, -0.1321467551) {};
\node[vertex] (v40) at (0.2959449249, -0.0491142641) {};
\node[vertex] (v41) at (0.1515405927, -0.1285366468) {};
\node[vertex] (v42) at (0.0612878851, -0.2729409789) {};
\draw[edge] (v4) -- (v11);
\draw[edge] (v7) -- (v11);
\draw[edge] (v7) -- (v12);
\draw[edge] (v0) -- (v12);
\draw[edge] (v0) -- (v13);
\draw[edge] (v4) -- (v13);
\draw[edge] (v10) -- (v11);
\draw[edge] (v10) -- (v12);
\draw[edge] (v10) -- (v13);
\draw[edge] (v7) -- (v15);
\draw[edge] (v6) -- (v15);
\draw[edge] (v6) -- (v16);
\draw[edge] (v0) -- (v16);
\draw[edge] (v12) -- (v14);
\draw[edge] (v14) -- (v15);
\draw[edge] (v14) -- (v16);
\draw[edge] (v7) -- (v18);
\draw[edge] (v3) -- (v18);
\draw[edge] (v3) -- (v19);
\draw[edge] (v6) -- (v19);
\draw[edge] (v15) -- (v17);
\draw[edge] (v17) -- (v18);
\draw[edge] (v17) -- (v19);
\draw[edge] (v4) -- (v21);
\draw[edge] (v3) -- (v21);
\draw[edge] (v18) -- (v20);
\draw[edge] (v11) -- (v20);
\draw[edge] (v20) -- (v21);
\draw[edge] (v5) -- (v23);
\draw[edge] (v8) -- (v23);
\draw[edge] (v8) -- (v24);
\draw[edge] (v1) -- (v24);
\draw[edge] (v1) -- (v25);
\draw[edge] (v5) -- (v25);
\draw[edge] (v22) -- (v23);
\draw[edge] (v22) -- (v24);
\draw[edge] (v22) -- (v25);
\draw[edge] (v8) -- (v27);
\draw[edge] (v4) -- (v27);
\draw[edge] (v4) -- (v28);
\draw[edge] (v1) -- (v28);
\draw[edge] (v24) -- (v26);
\draw[edge] (v26) -- (v27);
\draw[edge] (v26) -- (v28);
\draw[edge] (v8) -- (v30);
\draw[edge] (v3) -- (v30);
\draw[edge] (v27) -- (v29);
\draw[edge] (v29) -- (v30);
\draw[edge] (v21) -- (v29);
\draw[edge] (v5) -- (v32);
\draw[edge] (v3) -- (v32);
\draw[edge] (v30) -- (v31);
\draw[edge] (v23) -- (v31);
\draw[edge] (v31) -- (v32);
\draw[edge] (v6) -- (v34);
\draw[edge] (v9) -- (v34);
\draw[edge] (v9) -- (v35);
\draw[edge] (v2) -- (v35);
\draw[edge] (v2) -- (v36);
\draw[edge] (v6) -- (v36);
\draw[edge] (v33) -- (v34);
\draw[edge] (v33) -- (v35);
\draw[edge] (v33) -- (v36);
\draw[edge] (v9) -- (v38);
\draw[edge] (v5) -- (v38);
\draw[edge] (v5) -- (v39);
\draw[edge] (v2) -- (v39);
\draw[edge] (v35) -- (v37);
\draw[edge] (v37) -- (v38);
\draw[edge] (v37) -- (v39);
\draw[edge] (v9) -- (v41);
\draw[edge] (v3) -- (v41);
\draw[edge] (v38) -- (v40);
\draw[edge] (v40) -- (v41);
\draw[edge] (v32) -- (v40);
\draw[edge] (v41) -- (v42);
\draw[edge] (v34) -- (v42);
\draw[edge] (v19) -- (v42);
\fi}%
  \else%
    \pgfimage[width=0.2\textwidth]{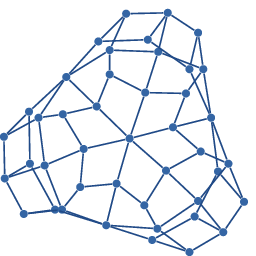}%
  \fi%
&
  \ifpdf%
    \tikzsetnextfilename{cache/stress}%
    \tikz[x=0.2\textwidth, y=0.2\textwidth, ]{
\iftikzgraphpreamble
\def\aspectratio{0.8838609335}
\else
\node[vertex] (v0) at (-0.3374868077, 0.1926256633) {};
\node[vertex] (v1) at (0.3355617860, 0.1959598921) {};
\node[vertex] (v2) at (0.0019247210, -0.3885849802) {};
\node[vertex] (v3) at (0.0000000065, -0.0000000124) {};
\node[vertex] (v4) at (-0.0016198577, 0.3269031014) {};
\node[vertex] (v5) at (0.2839163385, -0.1620487240) {};
\node[vertex] (v6) at (-0.2822964716, -0.1648543956) {};
\node[vertex] (v7) at (-0.2673895672, 0.1526153145) {};
\node[vertex] (v8) at (0.2658637228, 0.1552581688) {};
\node[vertex] (v9) at (0.0015260238, -0.3078738261) {};
\node[vertex] (v10) at (-0.3177880314, 0.3934402881) {};
\node[vertex] (v11) at (-0.1688686685, 0.3187044631) {};
\node[vertex] (v12) at (-0.4232134427, 0.2415541103) {};
\node[vertex] (v13) at (-0.1810750557, 0.3651841595) {};
\node[vertex] (v14) at (-0.5003766799, 0.0735372690) {};
\node[vertex] (v15) at (-0.3602930047, -0.0166801912) {};
\node[vertex] (v16) at (-0.4065206823, -0.0298052194) {};
\node[vertex] (v17) at (-0.2625947157, -0.0121349871) {};
\node[vertex] (v18) at (-0.1406891632, 0.0802999204) {};
\node[vertex] (v19) at (-0.1458714824, -0.0851854802) {};
\node[vertex] (v20) at (-0.1230961257, 0.2322725116) {};
\node[vertex] (v21) at (-0.0008370331, 0.1689211449) {};
\node[vertex] (v22) at (0.4996233201, 0.0784923856) {};
\node[vertex] (v23) at (0.3604405453, -0.0131081548) {};
\node[vertex] (v24) at (0.4207987710, 0.2457364516) {};
\node[vertex] (v25) at (0.4067962083, -0.0257759274) {};
\node[vertex] (v26) at (0.3138734525, 0.3965702801) {};
\node[vertex] (v27) at (0.1657014245, 0.3203627204) {};
\node[vertex] (v28) at (0.1774478152, 0.3669601938) {};
\node[vertex] (v29) at (0.1207881856, 0.2334811232) {};
\node[vertex] (v30) at (0.1398864084, 0.0816903575) {};
\node[vertex] (v31) at (0.2627020137, -0.0095319322) {};
\node[vertex] (v32) at (0.1467085262, -0.0837356853) {};
\node[vertex] (v33) at (-0.1818352732, -0.4719326547) {};
\node[vertex] (v34) at (-0.1915718577, -0.3055967818) {};
\node[vertex] (v35) at (0.0024147202, -0.4872906534) {};
\node[vertex] (v36) at (-0.2257212130, -0.3394076869) {};
\node[vertex] (v37) at (0.1865032026, -0.4701075503) {};
\node[vertex] (v38) at (0.1945919566, -0.3036828137) {};
\node[vertex] (v39) at (0.2290724549, -0.3371546150) {};
\node[vertex] (v40) at (0.1418065581, -0.2213462222) {};
\node[vertex] (v41) at (0.0008028057, -0.1619903752) {};
\node[vertex] (v42) at (-0.1396058340, -0.2227406498) {};
\draw[edge] (v4) -- (v11);
\draw[edge] (v7) -- (v11);
\draw[edge] (v7) -- (v12);
\draw[edge] (v0) -- (v12);
\draw[edge] (v0) -- (v13);
\draw[edge] (v4) -- (v13);
\draw[edge] (v10) -- (v11);
\draw[edge] (v10) -- (v12);
\draw[edge] (v10) -- (v13);
\draw[edge] (v7) -- (v15);
\draw[edge] (v6) -- (v15);
\draw[edge] (v6) -- (v16);
\draw[edge] (v0) -- (v16);
\draw[edge] (v12) -- (v14);
\draw[edge] (v14) -- (v15);
\draw[edge] (v14) -- (v16);
\draw[edge] (v7) -- (v18);
\draw[edge] (v3) -- (v18);
\draw[edge] (v3) -- (v19);
\draw[edge] (v6) -- (v19);
\draw[edge] (v15) -- (v17);
\draw[edge] (v17) -- (v18);
\draw[edge] (v17) -- (v19);
\draw[edge] (v4) -- (v21);
\draw[edge] (v3) -- (v21);
\draw[edge] (v18) -- (v20);
\draw[edge] (v11) -- (v20);
\draw[edge] (v20) -- (v21);
\draw[edge] (v5) -- (v23);
\draw[edge] (v8) -- (v23);
\draw[edge] (v8) -- (v24);
\draw[edge] (v1) -- (v24);
\draw[edge] (v1) -- (v25);
\draw[edge] (v5) -- (v25);
\draw[edge] (v22) -- (v23);
\draw[edge] (v22) -- (v24);
\draw[edge] (v22) -- (v25);
\draw[edge] (v8) -- (v27);
\draw[edge] (v4) -- (v27);
\draw[edge] (v4) -- (v28);
\draw[edge] (v1) -- (v28);
\draw[edge] (v24) -- (v26);
\draw[edge] (v26) -- (v27);
\draw[edge] (v26) -- (v28);
\draw[edge] (v8) -- (v30);
\draw[edge] (v3) -- (v30);
\draw[edge] (v27) -- (v29);
\draw[edge] (v29) -- (v30);
\draw[edge] (v21) -- (v29);
\draw[edge] (v5) -- (v32);
\draw[edge] (v3) -- (v32);
\draw[edge] (v30) -- (v31);
\draw[edge] (v23) -- (v31);
\draw[edge] (v31) -- (v32);
\draw[edge] (v6) -- (v34);
\draw[edge] (v9) -- (v34);
\draw[edge] (v9) -- (v35);
\draw[edge] (v2) -- (v35);
\draw[edge] (v2) -- (v36);
\draw[edge] (v6) -- (v36);
\draw[edge] (v33) -- (v34);
\draw[edge] (v33) -- (v35);
\draw[edge] (v33) -- (v36);
\draw[edge] (v9) -- (v38);
\draw[edge] (v5) -- (v38);
\draw[edge] (v5) -- (v39);
\draw[edge] (v2) -- (v39);
\draw[edge] (v35) -- (v37);
\draw[edge] (v37) -- (v38);
\draw[edge] (v37) -- (v39);
\draw[edge] (v9) -- (v41);
\draw[edge] (v3) -- (v41);
\draw[edge] (v38) -- (v40);
\draw[edge] (v40) -- (v41);
\draw[edge] (v32) -- (v40);
\draw[edge] (v41) -- (v42);
\draw[edge] (v34) -- (v42);
\draw[edge] (v19) -- (v42);
\fi}%
  \else%
    \pgfimage[width=0.2\textwidth]{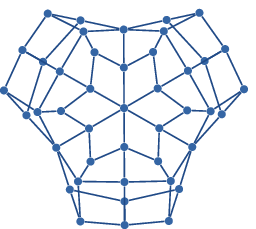}%
  \fi%
\\[2ex]
      \enum{NATIVE} & \enum{FMMM} & \enum{STRESS}
    \end{tabularx}
    \par\vspace{1cm}
    \begin{tabularx}{\linewidth}{%
        >{\centering\arraybackslash}X
        >{\centering\arraybackslash}X
        >{\centering\arraybackslash}X
      }
  \ifpdf%
    \tikzsetnextfilename{cache/random-uniform}%
    \tikz[x=0.2\textwidth, y=0.2\textwidth, ]{
\iftikzgraphpreamble
\def\aspectratio{1.0132018315}
\else
\node[vertex] (v0) at (-0.2574845341, -0.1639597483) {};
\node[vertex] (v1) at (0.3636291939, 0.1600308741) {};
\node[vertex] (v2) at (-0.0420672742, -0.4243047077) {};
\node[vertex] (v3) at (-0.2342433890, -0.2749267576) {};
\node[vertex] (v4) at (-0.2596414303, -0.0985573776) {};
\node[vertex] (v5) at (0.0544069147, -0.2782034463) {};
\node[vertex] (v6) at (-0.3930477870, -0.3900469307) {};
\node[vertex] (v7) at (0.2003842542, 0.2104234729) {};
\node[vertex] (v8) at (-0.2952447661, -0.4434189896) {};
\node[vertex] (v9) at (0.4458302200, 0.1729147697) {};
\node[vertex] (v10) at (0.4228967274, -0.2754732196) {};
\node[vertex] (v11) at (-0.4831563451, -0.0601653185) {};
\node[vertex] (v12) at (0.2077023720, -0.1233712293) {};
\node[vertex] (v13) at (-0.0585425464, -0.0872838999) {};
\node[vertex] (v14) at (-0.2298234286, 0.2403517196) {};
\node[vertex] (v15) at (-0.2603858225, 0.5565810104) {};
\node[vertex] (v16) at (0.4299452001, -0.2295791251) {};
\node[vertex] (v17) at (0.1595723482, 0.2422687500) {};
\node[vertex] (v18) at (-0.1552688108, 0.0642946262) {};
\node[vertex] (v19) at (0.3235075857, -0.0213231308) {};
\node[vertex] (v20) at (-0.4404574695, -0.0954309395) {};
\node[vertex] (v21) at (-0.1426704052, -0.1519997880) {};
\node[vertex] (v22) at (-0.0215455344, -0.1931521461) {};
\node[vertex] (v23) at (-0.3392566290, 0.1356431166) {};
\node[vertex] (v24) at (0.1672058725, -0.1415643614) {};
\node[vertex] (v25) at (0.5038138408, 0.5513267261) {};
\node[vertex] (v26) at (-0.2301853330, 0.1054755233) {};
\node[vertex] (v27) at (-0.4660578674, -0.2867343638) {};
\node[vertex] (v28) at (0.0556343647, 0.5523070312) {};
\node[vertex] (v29) at (0.1612342969, 0.3325429571) {};
\node[vertex] (v30) at (0.4048489097, 0.1069293254) {};
\node[vertex] (v31) at (-0.1167038906, 0.5175797967) {};
\node[vertex] (v32) at (-0.1916313327, -0.1949564613) {};
\node[vertex] (v33) at (-0.4113858427, 0.2279647128) {};
\node[vertex] (v34) at (0.2892394748, -0.3740967911) {};
\node[vertex] (v35) at (-0.0868587768, -0.3889070917) {};
\node[vertex] (v36) at (0.4156663408, -0.1758601063) {};
\node[vertex] (v37) at (0.0874605938, -0.4136863488) {};
\node[vertex] (v38) at (0.3136616954, -0.3459675509) {};
\node[vertex] (v39) at (0.1295237942, 0.0186343115) {};
\node[vertex] (v40) at (0.0499649344, 0.5130895887) {};
\node[vertex] (v41) at (-0.0885956539, 0.5007485535) {};
\node[vertex] (v42) at (0.0181259350, 0.4238629638) {};
\draw[edge] (v4) -- (v11);
\draw[edge] (v7) -- (v11);
\draw[edge] (v7) -- (v12);
\draw[edge] (v0) -- (v12);
\draw[edge] (v0) -- (v13);
\draw[edge] (v4) -- (v13);
\draw[edge] (v10) -- (v11);
\draw[edge] (v10) -- (v12);
\draw[edge] (v10) -- (v13);
\draw[edge] (v7) -- (v15);
\draw[edge] (v6) -- (v15);
\draw[edge] (v6) -- (v16);
\draw[edge] (v0) -- (v16);
\draw[edge] (v12) -- (v14);
\draw[edge] (v14) -- (v15);
\draw[edge] (v14) -- (v16);
\draw[edge] (v7) -- (v18);
\draw[edge] (v3) -- (v18);
\draw[edge] (v3) -- (v19);
\draw[edge] (v6) -- (v19);
\draw[edge] (v15) -- (v17);
\draw[edge] (v17) -- (v18);
\draw[edge] (v17) -- (v19);
\draw[edge] (v4) -- (v21);
\draw[edge] (v3) -- (v21);
\draw[edge] (v18) -- (v20);
\draw[edge] (v11) -- (v20);
\draw[edge] (v20) -- (v21);
\draw[edge] (v5) -- (v23);
\draw[edge] (v8) -- (v23);
\draw[edge] (v8) -- (v24);
\draw[edge] (v1) -- (v24);
\draw[edge] (v1) -- (v25);
\draw[edge] (v5) -- (v25);
\draw[edge] (v22) -- (v23);
\draw[edge] (v22) -- (v24);
\draw[edge] (v22) -- (v25);
\draw[edge] (v8) -- (v27);
\draw[edge] (v4) -- (v27);
\draw[edge] (v4) -- (v28);
\draw[edge] (v1) -- (v28);
\draw[edge] (v24) -- (v26);
\draw[edge] (v26) -- (v27);
\draw[edge] (v26) -- (v28);
\draw[edge] (v8) -- (v30);
\draw[edge] (v3) -- (v30);
\draw[edge] (v27) -- (v29);
\draw[edge] (v29) -- (v30);
\draw[edge] (v21) -- (v29);
\draw[edge] (v5) -- (v32);
\draw[edge] (v3) -- (v32);
\draw[edge] (v30) -- (v31);
\draw[edge] (v23) -- (v31);
\draw[edge] (v31) -- (v32);
\draw[edge] (v6) -- (v34);
\draw[edge] (v9) -- (v34);
\draw[edge] (v9) -- (v35);
\draw[edge] (v2) -- (v35);
\draw[edge] (v2) -- (v36);
\draw[edge] (v6) -- (v36);
\draw[edge] (v33) -- (v34);
\draw[edge] (v33) -- (v35);
\draw[edge] (v33) -- (v36);
\draw[edge] (v9) -- (v38);
\draw[edge] (v5) -- (v38);
\draw[edge] (v5) -- (v39);
\draw[edge] (v2) -- (v39);
\draw[edge] (v35) -- (v37);
\draw[edge] (v37) -- (v38);
\draw[edge] (v37) -- (v39);
\draw[edge] (v9) -- (v41);
\draw[edge] (v3) -- (v41);
\draw[edge] (v38) -- (v40);
\draw[edge] (v40) -- (v41);
\draw[edge] (v32) -- (v40);
\draw[edge] (v41) -- (v42);
\draw[edge] (v34) -- (v42);
\draw[edge] (v19) -- (v42);
\fi}%
  \else%
    \pgfimage[width=0.2\textwidth]{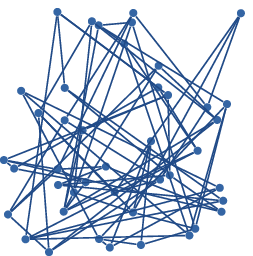}%
  \fi%
&
  \ifpdf%
    \tikzsetnextfilename{cache/random-normal}%
    \tikz[x=0.2\textwidth, y=0.2\textwidth, ]{
\iftikzgraphpreamble
\def\aspectratio{1.1715726826}
\else
\node[vertex] (v0) at (-0.1219495656, -0.1838475180) {};
\node[vertex] (v1) at (0.0972024919, 0.2435146133) {};
\node[vertex] (v2) at (-0.0804280636, -0.0096878133) {};
\node[vertex] (v3) at (-0.1177371091, -0.1064294083) {};
\node[vertex] (v4) at (-0.1032209735, -0.2313895017) {};
\node[vertex] (v5) at (-0.2407420883, 0.0246252210) {};
\node[vertex] (v6) at (0.2634822154, 0.2689691744) {};
\node[vertex] (v7) at (0.0684765074, 0.1691143501) {};
\node[vertex] (v8) at (-0.0053508178, 0.0382989089) {};
\node[vertex] (v9) at (-0.2174879751, 0.2441025249) {};
\node[vertex] (v10) at (-0.3349379769, -0.2184498435) {};
\node[vertex] (v11) at (0.1820015115, -0.2185478290) {};
\node[vertex] (v12) at (-0.0326642829, 0.0825978453) {};
\node[vertex] (v13) at (0.3184865793, 0.2095927525) {};
\node[vertex] (v14) at (0.0428787128, -0.3950274263) {};
\node[vertex] (v15) at (-0.0510873051, 0.2876418294) {};
\node[vertex] (v16) at (-0.0043369629, -0.0766925106) {};
\node[vertex] (v17) at (-0.2237586792, -0.2002570644) {};
\node[vertex] (v18) at (-0.3066535520, -0.0633455074) {};
\node[vertex] (v19) at (0.0695516609, -0.2113734232) {};
\node[vertex] (v20) at (-0.2644786496, 0.2052919074) {};
\node[vertex] (v21) at (0.0840486848, -0.3124271402) {};
\node[vertex] (v22) at (0.1009215250, 0.0098204938) {};
\node[vertex] (v23) at (0.2817392731, 0.1279269828) {};
\node[vertex] (v24) at (0.0528549462, 0.2268192170) {};
\node[vertex] (v25) at (0.1183843858, -0.0248341130) {};
\node[vertex] (v26) at (-0.1743639839, -0.1689928159) {};
\node[vertex] (v27) at (0.0649487662, -0.1037702363) {};
\node[vertex] (v28) at (-0.0416971278, 0.0442489800) {};
\node[vertex] (v29) at (-0.1121981761, -0.0296839072) {};
\node[vertex] (v30) at (-0.0555006416, 0.1352592929) {};
\node[vertex] (v31) at (-0.0991887927, 0.0205552934) {};
\node[vertex] (v32) at (-0.0544834486, 0.0598609509) {};
\node[vertex] (v33) at (-0.1491788920, 0.4765616028) {};
\node[vertex] (v34) at (0.1604620093, 0.0116949972) {};
\node[vertex] (v35) at (0.1523814827, -0.0290008474) {};
\node[vertex] (v36) at (0.2527077521, -0.0019579246) {};
\node[vertex] (v37) at (-0.0178198035, -0.5234383972) {};
\node[vertex] (v38) at (-0.1143486056, 0.1973458280) {};
\node[vertex] (v39) at (0.0936626053, 0.0591866153) {};
\node[vertex] (v40) at (0.5186155540, -0.0505779715) {};
\node[vertex] (v41) at (0.0610032896, -0.1472953556) {};
\node[vertex] (v42) at (-0.0601964798, 0.1639971733) {};
\draw[edge] (v4) -- (v11);
\draw[edge] (v7) -- (v11);
\draw[edge] (v7) -- (v12);
\draw[edge] (v0) -- (v12);
\draw[edge] (v0) -- (v13);
\draw[edge] (v4) -- (v13);
\draw[edge] (v10) -- (v11);
\draw[edge] (v10) -- (v12);
\draw[edge] (v10) -- (v13);
\draw[edge] (v7) -- (v15);
\draw[edge] (v6) -- (v15);
\draw[edge] (v6) -- (v16);
\draw[edge] (v0) -- (v16);
\draw[edge] (v12) -- (v14);
\draw[edge] (v14) -- (v15);
\draw[edge] (v14) -- (v16);
\draw[edge] (v7) -- (v18);
\draw[edge] (v3) -- (v18);
\draw[edge] (v3) -- (v19);
\draw[edge] (v6) -- (v19);
\draw[edge] (v15) -- (v17);
\draw[edge] (v17) -- (v18);
\draw[edge] (v17) -- (v19);
\draw[edge] (v4) -- (v21);
\draw[edge] (v3) -- (v21);
\draw[edge] (v18) -- (v20);
\draw[edge] (v11) -- (v20);
\draw[edge] (v20) -- (v21);
\draw[edge] (v5) -- (v23);
\draw[edge] (v8) -- (v23);
\draw[edge] (v8) -- (v24);
\draw[edge] (v1) -- (v24);
\draw[edge] (v1) -- (v25);
\draw[edge] (v5) -- (v25);
\draw[edge] (v22) -- (v23);
\draw[edge] (v22) -- (v24);
\draw[edge] (v22) -- (v25);
\draw[edge] (v8) -- (v27);
\draw[edge] (v4) -- (v27);
\draw[edge] (v4) -- (v28);
\draw[edge] (v1) -- (v28);
\draw[edge] (v24) -- (v26);
\draw[edge] (v26) -- (v27);
\draw[edge] (v26) -- (v28);
\draw[edge] (v8) -- (v30);
\draw[edge] (v3) -- (v30);
\draw[edge] (v27) -- (v29);
\draw[edge] (v29) -- (v30);
\draw[edge] (v21) -- (v29);
\draw[edge] (v5) -- (v32);
\draw[edge] (v3) -- (v32);
\draw[edge] (v30) -- (v31);
\draw[edge] (v23) -- (v31);
\draw[edge] (v31) -- (v32);
\draw[edge] (v6) -- (v34);
\draw[edge] (v9) -- (v34);
\draw[edge] (v9) -- (v35);
\draw[edge] (v2) -- (v35);
\draw[edge] (v2) -- (v36);
\draw[edge] (v6) -- (v36);
\draw[edge] (v33) -- (v34);
\draw[edge] (v33) -- (v35);
\draw[edge] (v33) -- (v36);
\draw[edge] (v9) -- (v38);
\draw[edge] (v5) -- (v38);
\draw[edge] (v5) -- (v39);
\draw[edge] (v2) -- (v39);
\draw[edge] (v35) -- (v37);
\draw[edge] (v37) -- (v38);
\draw[edge] (v37) -- (v39);
\draw[edge] (v9) -- (v41);
\draw[edge] (v3) -- (v41);
\draw[edge] (v38) -- (v40);
\draw[edge] (v40) -- (v41);
\draw[edge] (v32) -- (v40);
\draw[edge] (v41) -- (v42);
\draw[edge] (v34) -- (v42);
\draw[edge] (v19) -- (v42);
\fi}%
  \else%
    \pgfimage[width=0.2\textwidth]{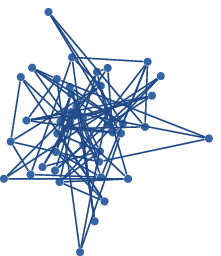}%
  \fi%
&
  \ifpdf%
    \tikzsetnextfilename{cache/phantom}%
    \tikz[x=0.2\textwidth, y=0.2\textwidth, ]{
\iftikzgraphpreamble
\def\aspectratio{0.9691780822}
\else
\node[vertex] (v0) at (-0.2499203568, -0.3725708824) {};
\node[vertex] (v1) at (0.2911755336, 0.1000318573) {};
\node[vertex] (v2) at (-0.0307422746, -0.1944886907) {};
\node[vertex] (v3) at (0.4555590953, -0.1157215674) {};
\node[vertex] (v4) at (-0.2053998089, -0.2321599235) {};
\node[vertex] (v5) at (0.5651481363, -0.0266804715) {};
\node[vertex] (v6) at (0.0172029309, -0.1842147181) {};
\node[vertex] (v7) at (-0.2875915897, 0.1924976107) {};
\node[vertex] (v8) at (-0.2088244664, 0.0555113093) {};
\node[vertex] (v9) at (-0.0649888500, -0.0129818414) {};
\node[vertex] (v10) at (0.2432303281, 0.0760592545) {};
\node[vertex] (v11) at (-0.0410162472, -0.0780503345) {};
\node[vertex] (v12) at (-0.3629340554, 0.0555113093) {};
\node[vertex] (v13) at (-0.1060847404, 0.1548263778) {};
\node[vertex] (v14) at (-0.0478655623, 0.1514017203) {};
\node[vertex] (v15) at (-0.1334820006, -0.2355845811) {};
\node[vertex] (v16) at (0.2295316980, -0.0951736222) {};
\node[vertex] (v17) at (0.0514495062, 0.0794839121) {};
\node[vertex] (v18) at (0.0000796432, -0.3759955400) {};
\node[vertex] (v19) at (-0.3800573431, -0.0198311564) {};
\node[vertex] (v20) at (-0.1129340554, 0.1034565148) {};
\node[vertex] (v21) at (-0.1232080280, -0.0575023893) {};
\node[vertex] (v22) at (0.2398056706, -0.0301051290) {};
\node[vertex] (v23) at (-0.2567696719, 0.1171551450) {};
\node[vertex] (v24) at (0.1199426569, 0.2130455559) {};
\node[vertex] (v25) at (0.4452851227, 0.0383880217) {};
\node[vertex] (v26) at (-0.2807422746, -0.1465434852) {};
\node[vertex] (v27) at (0.0137782733, -0.0746256770) {};
\node[vertex] (v28) at (0.4144632048, 0.0178400765) {};
\node[vertex] (v29) at (-0.3834820006, 0.3294839121) {};
\node[vertex] (v30) at (-0.3766326856, -0.4684612934) {};
\node[vertex] (v31) at (0.1062440268, 0.0349633641) {};
\node[vertex] (v32) at (-0.0136189869, 0.3020866518) {};
\node[vertex] (v33) at (-0.0273176171, 0.0349633641) {};
\node[vertex] (v34) at (0.5891207391, 0.1274291176) {};
\node[vertex] (v35) at (0.3014495062, -0.1670914304) {};
\node[vertex] (v36) at (-0.2122491239, 0.1377030902) {};
\node[vertex] (v37) at (-0.1403313157, -0.1499681427) {};
\node[vertex] (v38) at (0.1267919720, -0.0746256770) {};
\node[vertex] (v39) at (-0.3834820006, 0.5007167888) {};
\node[vertex] (v40) at (0.6165179994, 0.0623606244) {};
\node[vertex] (v41) at (-0.2190984390, -0.0472284167) {};
\node[vertex] (v42) at (-0.1780025486, 0.2746893915) {};
\draw[edge] (v4) -- (v11);
\draw[edge] (v7) -- (v11);
\draw[edge] (v7) -- (v12);
\draw[edge] (v0) -- (v12);
\draw[edge] (v0) -- (v13);
\draw[edge] (v4) -- (v13);
\draw[edge] (v10) -- (v11);
\draw[edge] (v10) -- (v12);
\draw[edge] (v10) -- (v13);
\draw[edge] (v7) -- (v15);
\draw[edge] (v6) -- (v15);
\draw[edge] (v6) -- (v16);
\draw[edge] (v0) -- (v16);
\draw[edge] (v12) -- (v14);
\draw[edge] (v14) -- (v15);
\draw[edge] (v14) -- (v16);
\draw[edge] (v7) -- (v18);
\draw[edge] (v3) -- (v18);
\draw[edge] (v3) -- (v19);
\draw[edge] (v6) -- (v19);
\draw[edge] (v15) -- (v17);
\draw[edge] (v17) -- (v18);
\draw[edge] (v17) -- (v19);
\draw[edge] (v4) -- (v21);
\draw[edge] (v3) -- (v21);
\draw[edge] (v18) -- (v20);
\draw[edge] (v11) -- (v20);
\draw[edge] (v20) -- (v21);
\draw[edge] (v5) -- (v23);
\draw[edge] (v8) -- (v23);
\draw[edge] (v8) -- (v24);
\draw[edge] (v1) -- (v24);
\draw[edge] (v1) -- (v25);
\draw[edge] (v5) -- (v25);
\draw[edge] (v22) -- (v23);
\draw[edge] (v22) -- (v24);
\draw[edge] (v22) -- (v25);
\draw[edge] (v8) -- (v27);
\draw[edge] (v4) -- (v27);
\draw[edge] (v4) -- (v28);
\draw[edge] (v1) -- (v28);
\draw[edge] (v24) -- (v26);
\draw[edge] (v26) -- (v27);
\draw[edge] (v26) -- (v28);
\draw[edge] (v8) -- (v30);
\draw[edge] (v3) -- (v30);
\draw[edge] (v27) -- (v29);
\draw[edge] (v29) -- (v30);
\draw[edge] (v21) -- (v29);
\draw[edge] (v5) -- (v32);
\draw[edge] (v3) -- (v32);
\draw[edge] (v30) -- (v31);
\draw[edge] (v23) -- (v31);
\draw[edge] (v31) -- (v32);
\draw[edge] (v6) -- (v34);
\draw[edge] (v9) -- (v34);
\draw[edge] (v9) -- (v35);
\draw[edge] (v2) -- (v35);
\draw[edge] (v2) -- (v36);
\draw[edge] (v6) -- (v36);
\draw[edge] (v33) -- (v34);
\draw[edge] (v33) -- (v35);
\draw[edge] (v33) -- (v36);
\draw[edge] (v9) -- (v38);
\draw[edge] (v5) -- (v38);
\draw[edge] (v5) -- (v39);
\draw[edge] (v2) -- (v39);
\draw[edge] (v35) -- (v37);
\draw[edge] (v37) -- (v38);
\draw[edge] (v37) -- (v39);
\draw[edge] (v9) -- (v41);
\draw[edge] (v3) -- (v41);
\draw[edge] (v38) -- (v40);
\draw[edge] (v40) -- (v41);
\draw[edge] (v32) -- (v40);
\draw[edge] (v41) -- (v42);
\draw[edge] (v34) -- (v42);
\draw[edge] (v19) -- (v42);
\fi}%
  \else%
    \pgfimage[width=0.2\textwidth]{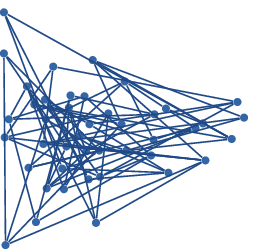}%
  \fi%
\\[2ex]
      \enum{RANDOM\_UNIFORM} & \enum{RANDOM\_NORMAL} & \enum{PHANTOM}
    \end{tabularx}
  \end{center}
  \caption{%
    Examples of different layouts for the same graph.  \enum{RANDOM\_UNIFORM}, \enum{RANDOM\_NORMAL} are random layouts where vertex positions are sampled from the uniform and the normal distributions, respectively.
  }
  \label{app:fig:layouts}
\end{figure}

\begin{figure}[p]
  \begin{center}
    \begin{tabularx}{\linewidth}{%
        l
        >{\centering\arraybackslash}X
        >{\centering\arraybackslash}X
        >{\centering\arraybackslash}X
        >{\centering\arraybackslash}X
      }
      \rotatebox{90}{\enum{PERTURB}}&
  \ifpdf%
    \tikzsetnextfilename{cache/perturb-00000}%
    \tikz[x=0.2\textwidth, y=0.2\textwidth, ]{\input{pics/perturb-00000.tikz}}%
  \else%
    \pgfimage[width=0.2\textwidth]{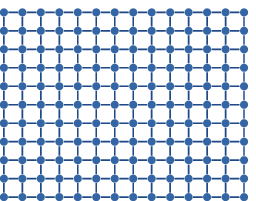}%
  \fi%
&
  \ifpdf%
    \tikzsetnextfilename{cache/perturb-01500}%
    \tikz[x=0.2\textwidth, y=0.2\textwidth, ]{\input{pics/perturb-01500.tikz}}%
  \else%
    \pgfimage[width=0.2\textwidth]{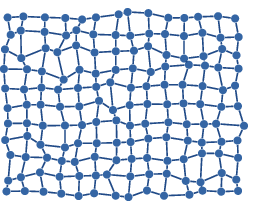}%
  \fi%
&
  \ifpdf%
    \tikzsetnextfilename{cache/perturb-05000}%
    \tikz[x=0.2\textwidth, y=0.2\textwidth, ]{\input{pics/perturb-05000.tikz}}%
  \else%
    \pgfimage[width=0.2\textwidth]{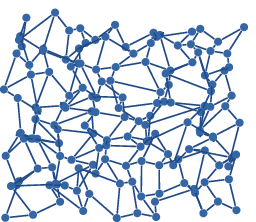}%
  \fi%
&
  \ifpdf%
    \tikzsetnextfilename{cache/perturb-10000}%
    \tikz[x=0.2\textwidth, y=0.2\textwidth, ]{\input{pics/perturb-10000.tikz}}%
  \else%
    \pgfimage[width=0.2\textwidth]{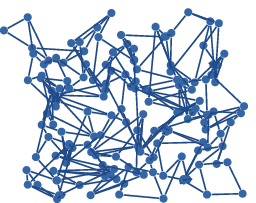}%
  \fi%
\\[2ex]
      \rotatebox{90}{\enum{FLIP\_NODES}}&
  \ifpdf%
    \tikzsetnextfilename{cache/flip-nodes-00000}%
    \tikz[x=0.2\textwidth, y=0.2\textwidth, ]{\input{pics/flip-nodes-00000.tikz}}%
  \else%
    \pgfimage[width=0.2\textwidth]{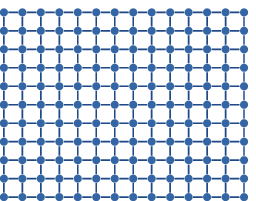}%
  \fi%
&
  \ifpdf%
    \tikzsetnextfilename{cache/flip-nodes-01500}%
    \tikz[x=0.2\textwidth, y=0.2\textwidth, ]{\input{pics/flip-nodes-01500.tikz}}%
  \else%
    \pgfimage[width=0.2\textwidth]{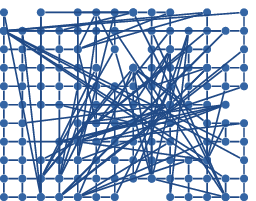}%
  \fi%
&
  \ifpdf%
    \tikzsetnextfilename{cache/flip-nodes-05000}%
    \tikz[x=0.2\textwidth, y=0.2\textwidth, ]{\input{pics/flip-nodes-05000.tikz}}%
  \else%
    \pgfimage[width=0.2\textwidth]{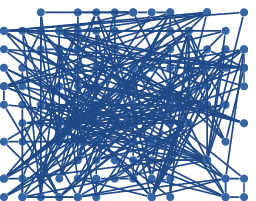}%
  \fi%
&
  \ifpdf%
    \tikzsetnextfilename{cache/flip-nodes-10000}%
    \tikz[x=0.2\textwidth, y=0.2\textwidth, ]{\input{pics/flip-nodes-10000.tikz}}%
  \else%
    \pgfimage[width=0.2\textwidth]{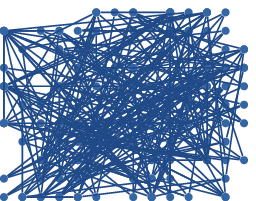}%
  \fi%
\\[2ex]
      \rotatebox{90}{\enum{FLIP\_EDGES}}&
  \ifpdf%
    \tikzsetnextfilename{cache/flip-edges-00000}%
    \tikz[x=0.2\textwidth, y=0.2\textwidth, ]{\input{pics/flip-edges-00000.tikz}}%
  \else%
    \pgfimage[width=0.2\textwidth]{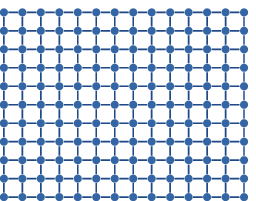}%
  \fi%
&
  \ifpdf%
    \tikzsetnextfilename{cache/flip-edges-01500}%
    \tikz[x=0.2\textwidth, y=0.2\textwidth, ]{\input{pics/flip-edges-01500.tikz}}%
  \else%
    \pgfimage[width=0.2\textwidth]{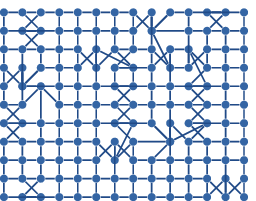}%
  \fi%
&
  \ifpdf%
    \tikzsetnextfilename{cache/flip-edges-05000}%
    \tikz[x=0.2\textwidth, y=0.2\textwidth, ]{\input{pics/flip-edges-05000.tikz}}%
  \else%
    \pgfimage[width=0.2\textwidth]{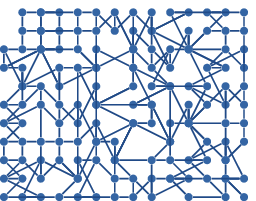}%
  \fi%
&
  \ifpdf%
    \tikzsetnextfilename{cache/flip-edges-10000}%
    \tikz[x=0.2\textwidth, y=0.2\textwidth, ]{\input{pics/flip-edges-10000.tikz}}%
  \else%
    \pgfimage[width=0.2\textwidth]{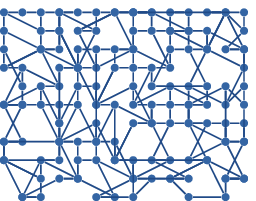}%
  \fi%
\\[2ex]
      \rotatebox{90}{\enum{MOVLSQ}}&
  \ifpdf%
    \tikzsetnextfilename{cache/movlsq-00000}%
    \tikz[x=0.2\textwidth, y=0.2\textwidth, ]{\input{pics/movlsq-00000.tikz}}%
  \else%
    \pgfimage[width=0.2\textwidth]{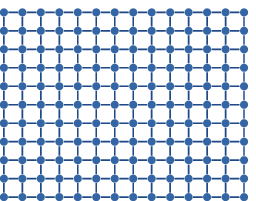}%
  \fi%
&
  \ifpdf%
    \tikzsetnextfilename{cache/movlsq-01500}%
    \tikz[x=0.2\textwidth, y=0.2\textwidth, ]{\input{pics/movlsq-01500.tikz}}%
  \else%
    \pgfimage[width=0.2\textwidth]{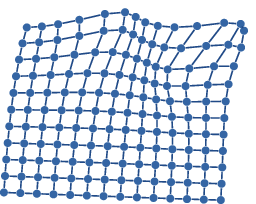}%
  \fi%
&
  \ifpdf%
    \tikzsetnextfilename{cache/movlsq-05000}%
    \tikz[x=0.2\textwidth, y=0.2\textwidth, ]{\input{pics/movlsq-05000.tikz}}%
  \else%
    \pgfimage[width=0.2\textwidth]{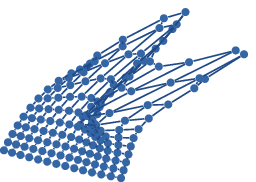}%
  \fi%
&
  \ifpdf%
    \tikzsetnextfilename{cache/movlsq-10000}%
    \tikz[x=0.2\textwidth, y=0.2\textwidth, ]{\input{pics/movlsq-10000.tikz}}%
  \else%
    \pgfimage[width=0.2\textwidth]{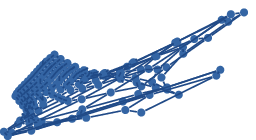}%
  \fi%
\\[2ex]
      & $r=0\percent$ & $r=15\percent$ & $r=50\percent$ & $r=100\percent$
    \end{tabularx}
  \end{center}
  \caption{%
    Examples of applying  different layout worsening techniques at different rates.
  }
  \label{app:fig:worsening}
\end{figure}

\begin{figure}[p]
  \begin{center}
    \begin{tabularx}{\linewidth}{%
        >{\centering\arraybackslash}X
        >{\centering\arraybackslash}X
        >{\centering\arraybackslash}X
        >{\centering\arraybackslash}X
      }
  \ifpdf%
    \tikzsetnextfilename{cache/linear-00000}%
    \tikz[x=0.2\textwidth, y=0.2\textwidth, ]{\input{pics/linear-00000.tikz}}%
  \else%
    \pgfimage[width=0.2\textwidth]{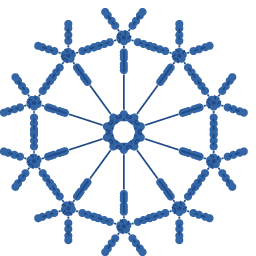}%
  \fi%
&
  \ifpdf%
    \tikzsetnextfilename{cache/linear-02500}%
    \tikz[x=0.2\textwidth, y=0.2\textwidth, ]{\input{pics/linear-02500.tikz}}%
  \else%
    \pgfimage[width=0.2\textwidth]{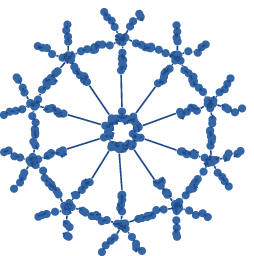}%
  \fi%
&
  \ifpdf%
    \tikzsetnextfilename{cache/linear-07500}%
    \tikz[x=0.2\textwidth, y=0.2\textwidth, ]{\input{pics/linear-07500.tikz}}%
  \else%
    \pgfimage[width=0.2\textwidth]{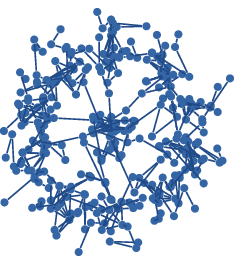}%
  \fi%
&
  \ifpdf%
    \tikzsetnextfilename{cache/linear-10000}%
    \tikz[x=0.2\textwidth, y=0.2\textwidth, ]{\input{pics/linear-10000.tikz}}%
  \else%
    \pgfimage[width=0.2\textwidth]{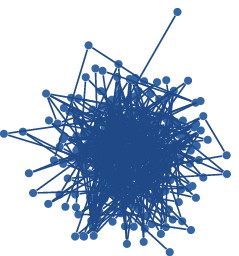}%
  \fi%
\\[2ex]
      $r=0\percent$ & $r=25\percent$ & $r=75\percent$ & $r=100\percent$
    \end{tabularx}
  \end{center}
  \caption{%
    Example of linear interpolation between a proper and a garbage layout.
  }
  \label{app:fig:interpolating}
\end{figure}

\fi

\end{document}